\documentclass{aastex62}

\newcommand{\Eexc}{$E_{\rm exc}$}
\newcommand{\Teff}{$T_{\rm eff}$}  
\newcommand{\kms}{km\,s$^{-1}$}
\def\ione{\,{\sc i}}
\def\ii{\,{\sc ii}}
\def\iii{\,{\sc iii}}
\def\iv{\,{\sc iv}}

\received{}
\revised{}
\accepted{}
\submitjournal{ApJ}

\shorttitle{The NLTE Analyses of Carbon Emission Lines in the Atmospheres of O and B type Stars }
\shortauthors{Alexeeva et al.}

\begin{document}

\title{The NLTE Analyses of Carbon Emission Lines in the Atmospheres of O and B type Stars }

\author{Sofya Alexeeva}
\affiliation{Shandong Provincial Key Laboratory of Optical Astronomy and Solar-Terrestrial Environment, Institute of Space Sciences, Shandong University, Weihai 264209, China \\}
\email{alexeeva@sdu.edu.cn}
\nocollaboration

\author{Kozo Sadakane}
\affiliation{Astronomical Institute, Osaka Kyoiku University, Asahigaoka, Kashiwara-shi, Osaka 582-8582, Japan\\}
\nocollaboration

\author{Masayoshi Nishimura}
\affiliation{2-6, Nishiyama-Maruo, Yawata-shi, Kyoto 614-8353, Japan \\}
\nocollaboration

\author{Junju Du}
\affiliation{Shandong Provincial Key Laboratory of Optical Astronomy and Solar-Terrestrial Environment, Institute of Space Sciences, Shandong University, Weihai 264209, China  \\}
\nocollaboration

\author{Shaoming Hu} 
\affiliation{Shandong Provincial Key Laboratory of Optical Astronomy and Solar-Terrestrial Environment, Institute of Space Sciences, Shandong University, Weihai 264209, China \\}
\email{husm@sdu.edu.cn}
\nocollaboration

\begin{abstract}

  We present a model atom for C\ione--C\ii--C\iii--C\iv\ using the most up-to-date atomic
  data and evaluated the non-local thermodynamic equilibrium (NLTE) line formation
  in classical 1D atmospheric models of O-B-type stars.
  Our models predict the emission lines of C\ii\ 9903~\AA\ and 18\,535~\AA\ to appear at effective temperature \Teff~$\geq$~17\,500~K,
  those of C\ii\ 6151~\AA\ and 6461~\AA\ to appear at \Teff~$>$~25\,000~K, and those of C\iii\ 5695, 6728--44, 9701--17~\AA\ to appear at \Teff~$\geq$~35\,000~K (log~$g$=4.0). 
  Emission occurs in the lines of minority species, where the photoionization-recombination mechanism provides a depopulation of the lower levels to a greater extent than the upper levels.
  For C\ii\ 9903 and 18\,535~\AA, the upper levels are mainly populated from C\iii\ reservoir through the Rydberg states. 
  For C\iii\ 5695 and 6728--44~\AA, the lower levels are depopulated due to photon losses in UV transitions at 885, 1308, and 1426--28~\AA\ which become optically thin in the photosphere.   
 
  We analysed the lines of C\ione, C\ii, C\iii, and C\iv\ for twenty-two O-B-type stars with temperature range between 15\,800 $\leq$~\Teff~$\leq$ 38\,000~K. 
   Abundances from emission lines of C\ione, C\ii\ and C\iii\ are in agreement with those from absorption ones for most of the stars.  
  We obtained log~$\epsilon_{\rm C}$=8.36$\pm$0.08 from twenty B-type stars, that is in line with the present-day Cosmic Abundance Standard.
  The obtained carbon abundances in 15~Mon and HD~42088 from emission and absorption lines are 8.27$\pm$0.11 and 8.31$\pm$0.11, respectively.

\end{abstract}


\section{Introduction} \label{sec:intro}

  The spectra of the majority of main sequence B-type stars have absorption lines of different chemical elements, which mainly form in the photosphere of a star. 
  Rapid development of observational techniques has resulted in dramatic improvement of the quality of spectral observations. 
  When spectrographs of much higher resolution (R $>$ 30\,000) became available, 
  the phenomenon of sharp and weak emission lines (WELs) of metals in optical spectra of B-type stars was evident. 
  WELs of C\ione, Mg\ii, Al\ii, Si\ii, P\ii, Ca\ii, Cr\ii,  Mn\ii, Fe\ii, Ni\ii, Cu\ii, and Hg\ii\ were detected in 
  the visible and near IR spectral regions in the spectra of B-type stars 
  \citep{2000ApJ...530L..89S, 2000A&A...362L..13W, 2001PASJ...53.1223S, 2001A&A...377L..27S, 2001ApJ...546L.115S, 2004A&A...425..263C, 2004A&A...418.1073W, 2007A&A...466.1083H,  2007A&A...475.1041C, 2016MNRAS.462.1123A, 2017PASJ...69...48S}. 
   Recently, \citet{2019PASJ...71...45S} published a detailed list of WELs observed in $\iota$~Her (B3IV). Their list contains numerous previously unknown very weak WELs
   originating from highly excited levels of Fe\ii\ in the visual region, and an emission line of C\ii\ at 9903.46~\AA. 
  
  The WELs originated from high-excitation states, are detected over a range of element abundance and are found among both chemically-normal and chemically-peculiar stars. 
  What is the physical nature of WELs and whether they could be used for abundance determination? Which processes do produce emission component in these lines?
  These enigmatic features are still not well understood.
 
  In the literature, there are some examples of reproducing the observed emission lines using standard plane-parallel non-local thermodynamic equilibrium (NLTE) modeling. For example, Mg\ione\ 12~$\mu$m in the Sun \citep{1992A&A...253..567C} and Procyon \citep{2004ApJ...617..551R},
  Mg\ione\ 12 and 18~$\mu$m in three K giants \citep{2008A&A...486..985S}, Mg\ione\ 7 and 12~$\mu$m in the Sun and Arcturus \citep{2015AA...579A..53O},
  Mg\ione\ 7 and 12~$\mu$m in the Sun, Procyon and three K giants \citep{2018ApJ...866..153A},
  Mn\ii\ 6122--6132~\AA\ in the three late-type B stars \citep{2001A&A...377L..27S}, C\ii\ 6151, 6462~\AA\ in $\tau$ Sco (B0V) and C\ii\ 6462~\AA\ in HR 1861 (B1V), HR~3055 (B0III), and HR~2928 (B2II)  
  \citep{2006ApJ...639L..39N, 2008A&A...481..199N},
  C\ione\ 8335, 9405, 9061--9111, and 9603--58~\AA\ in 21~Peg (B9.5V), $\pi$~Cet (B7IV), $\iota$~Her (B3IV) \citep{2016MNRAS.462.1123A}, Ca\ii\ 8912--27 and 9890~\AA\ in $\iota$~Her (B3IV) \citep{2018MNRAS.477.3343S}. 
  
  One of the pioneer works regarding the interpretation of emission lines of C\iii\ in the spectra of O-type stars belongs to \citet{1982SvA....26..563S}. 
  They performed theoretical analyses of the C\iii\ 9710~\AA\ line by using plane-pararell model atmospheres and NLTE approach and predicted the line to appear in emission as a results of departures from LTE (among O-type stars). 
  After decades, the emission lines of C\iii\ at 4647--50--51 and 5696~\AA\ were predicted theoretically in O-type stars in \citet{2012A&A...545A..95M} with using the code \textsc{cmfgen} \citep{1998ApJ...496..407H} to compute NLTE models including winds, line-blanketing and spherical geometry. \citet{2012A&A...545A..95M} found that a tight coupling of the C\iii\ 4647--50--51 and C\iii\ 5696~\AA\ lines to UV-transitions regulates the population of the associated levels.
  \citet{2018A&A...615A...4C} derived realistic carbon abundances by means of the NLTE atmosphere code \textsc{fastwind} \citep{2005A&A...435..669P} with recently developed model 
  atom C\ii--C\iii--C\iv. They investigated the spectra of a sample of six O-type dwarfs and supergiants, and found in most cases a moderate depletion compared to the solar value.
   In the literature, we can find no theoretical studies in which emission line of the C\ii\ 9903~\AA, observed in B-type stars, and C\ione\ 8335, 9405, 9061--9111, and 9603--58~\AA\ emission lines, observed in hotter B-type stars
   (with \Teff$>$17\,500~K), are well reproduced. 
      
  Carbon is one of the most abundant elements in the Universe and the main element in thermonuclear reactions. The determination of carbon abundances in O-B-type stars is important for stellar and galactochemical evolution. 
   There are many NLTE studies of carbon in OB stars available in the literature from the last decades, such as 
  \citet{1992ApJ...387..673G,1994ApJ...426..170C,1999A&A...350..598A,1999ApJ...522..950D, 2001ApJ...563..325D, 2001ApJ...552..309D, 2004ApJ...604..362D, 2004ApJ...606..514D, 2009A&A...496..841H, 2012AA...539A.143N,2015A&A...578A.109M, 2015A&A...575A..34M,  2017A&A...599A..30M}. 
  Several model atoms of carbon were constructed and several independent codes  were used to perform the analyses. 

  Carbon lines in the visible spectral region can be measured over a wide range of effective temperatures up to 55\,000 K.
  In the low-resolution spectroscopy, the strongest features in the metal line spectra are demanded for study.
  In the case of carbon, the C\ii\ 4267 and C\iii\ 4647~\AA\ lines are strong in B-type stars, and C\iii\ 5695 and C\iv\ 5801, 5811~\AA\ in O-type stars.
  These lines can be used as abundance indicators even in low-resolution spectroscopy.
   It should be noticed that the C\iii\ 5695~\AA\ line can be a good gravity indicator, since it strongly depends on luminosity classes (and surface gravity), that was pointed out in the spectroscopic atlas of O-B stars of \citet{1980ApJS...44..535W} and theoretical analyses of \citet{2018A&A...615A...4C}. 
  However, these lines are highly sensitive not only to the choice of stellar atmospheric parameters but also to the NLTE effects. 
  For example, \citet{2013MNRAS.428.3497L} found that for stars with \Teff\ between 20\,000 and 24\,000~K the C\ii\ 4267~\AA\ line to give lower log~$\epsilon_{\rm C}$ 
  values, and the underestimation amounts to 0.20--0.70~dex. A tendency of the 4267~\AA\ line to yield lower values of log~$\epsilon_{\rm C}$ for B stars has been noted in earlier studies, e.g. \citet{1992ApJ...387..673G, 1992A&A...262..171K, 1996ApJ...473..452S}.
   \citet{2006ApJ...639L..39N} and \citet{2008A&A...481..199N} presented a solution to the long-standing discrepancy between NLTE analyses of 
  the C\ii\ 4267 and 6578/82~\AA\ multiplets in six slowly-rotating early B-type dwarfs and giants, which cover a wide parameter range.
  Their comprehensive NLTE model atom of C\ii/\iii/\iv\ was constructed with critically selected atomic data and calibrated with high-SN spectra.  

  The aims of our work are $(1)$ to treat the method of accurate abundance determination in O-B stars from different carbon lines
  of four ionization stages, C\ione, C\ii, C\iii, and C\iv\ based on NLTE line formation; $(2)$ to present the modeling of C\ione, C\ii\ and C\iii\ emission lines in the spectra of O-B-type stars and an interpretation of these lines in the NLTE framework, including C\ii\ 9903~\AA\ emission lines, which were not reported before; $(3)$ to eliminate the significant discrepancy between C\ii\ 4267 \AA\ and other C\ii\ lines with future possibility applying it in both high- and low-resolution spectroscopy;  $(4)$ and check the reliability of C\ione, C\ii\ and C\iii\ emission lines in carbon abundance determination. 

  We construct a comprehensive model atom for C\ione--C\ii--C\iii--C\iv\ using the most up-to-date atomic data available so far, and analyse lines of C\ione, C\ii, C\iii, and C\iv\ in high-resolution spectra of reference B- and O-type stars with well-known stellar parameters.  We focus on carbon emission lines, C\ione\ 8335, 9405, 9061--9111, and 9603--58~\AA, C\ii\ 9903~\AA, and C\iii\ 5695, 6744, 9710~\AA\ in the spectra of O-B-type stars, trying to reproduce them along with absorption ones using the uniformly presented model atom for C\ione--C\ii--C\iii--C\iv.

  The paper is organized as follows. Section \ref{Sect:atom} describes an updated model atom of C\ione--C\ii--C\iii--C\iv, discusses departures from LTE in the model atmospheres 
  of O and B-type stars, and investigates mechanisms of producing the C\ii--C\iii\ emission lines. In Section \ref{sec:stellar}, 
  we analyse the carbon absorption and emission lines observed in a sample of twenty B-type stars and two O stars. We determine the C abundance of the selected stars, make some discussions and 
   compare our results with other studies from literature. 
  Our conclusions are summarized in Section \ref{Sect:Conclusions}.

\section{NLTE Line formation for C\ii~-- C\iii}\label{Sect:atom}

 \subsection{Model atom and atomic data}\label{subSect:atom}
 
\begin{figure*}
 \begin{center}
 \includegraphics[scale=0.6,angle=-90]{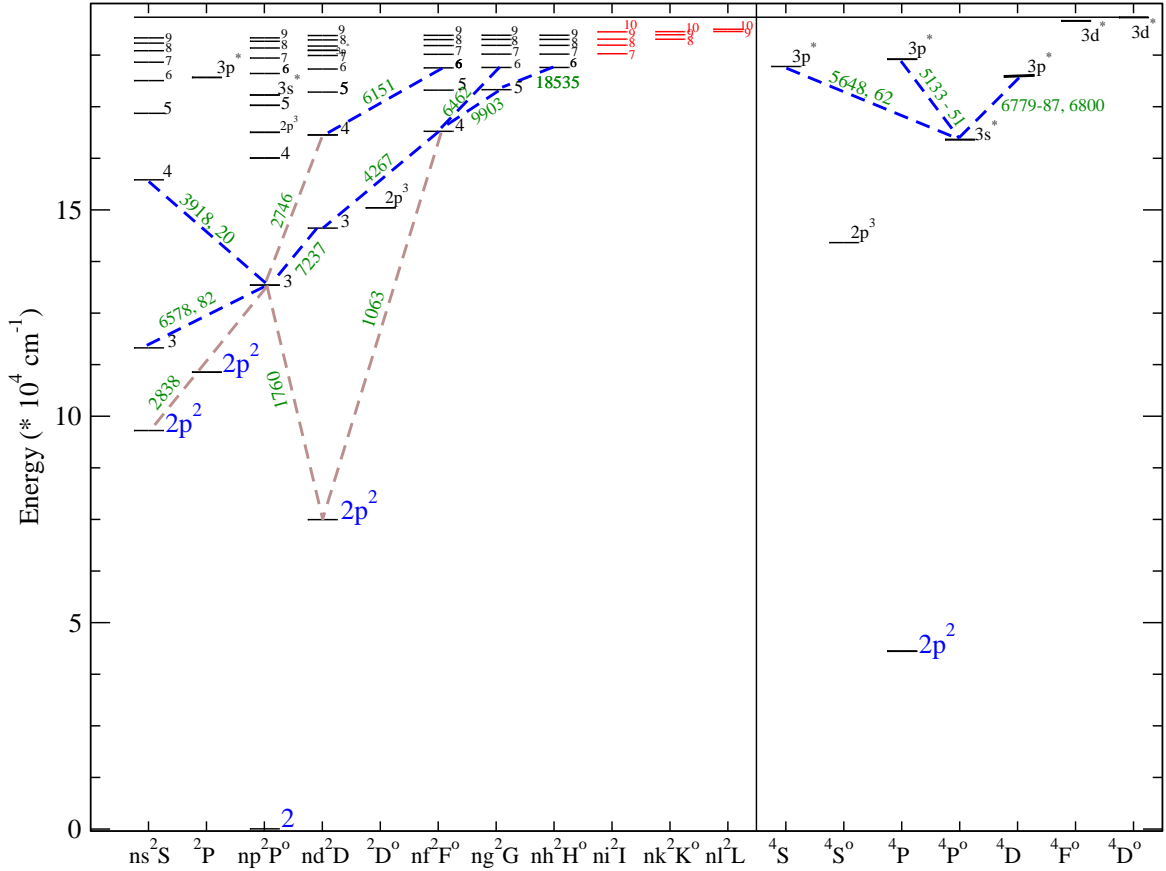}
 \caption{Term diagram for C\ii\ with extended levels.  The dashed lines indicate the transitions, from which the investigated spectral lines arise. }
 \label{Grot_C2}
 \end{center}
 \end{figure*}

\begin{figure*}
 \begin{center}
 \includegraphics[scale=0.6]{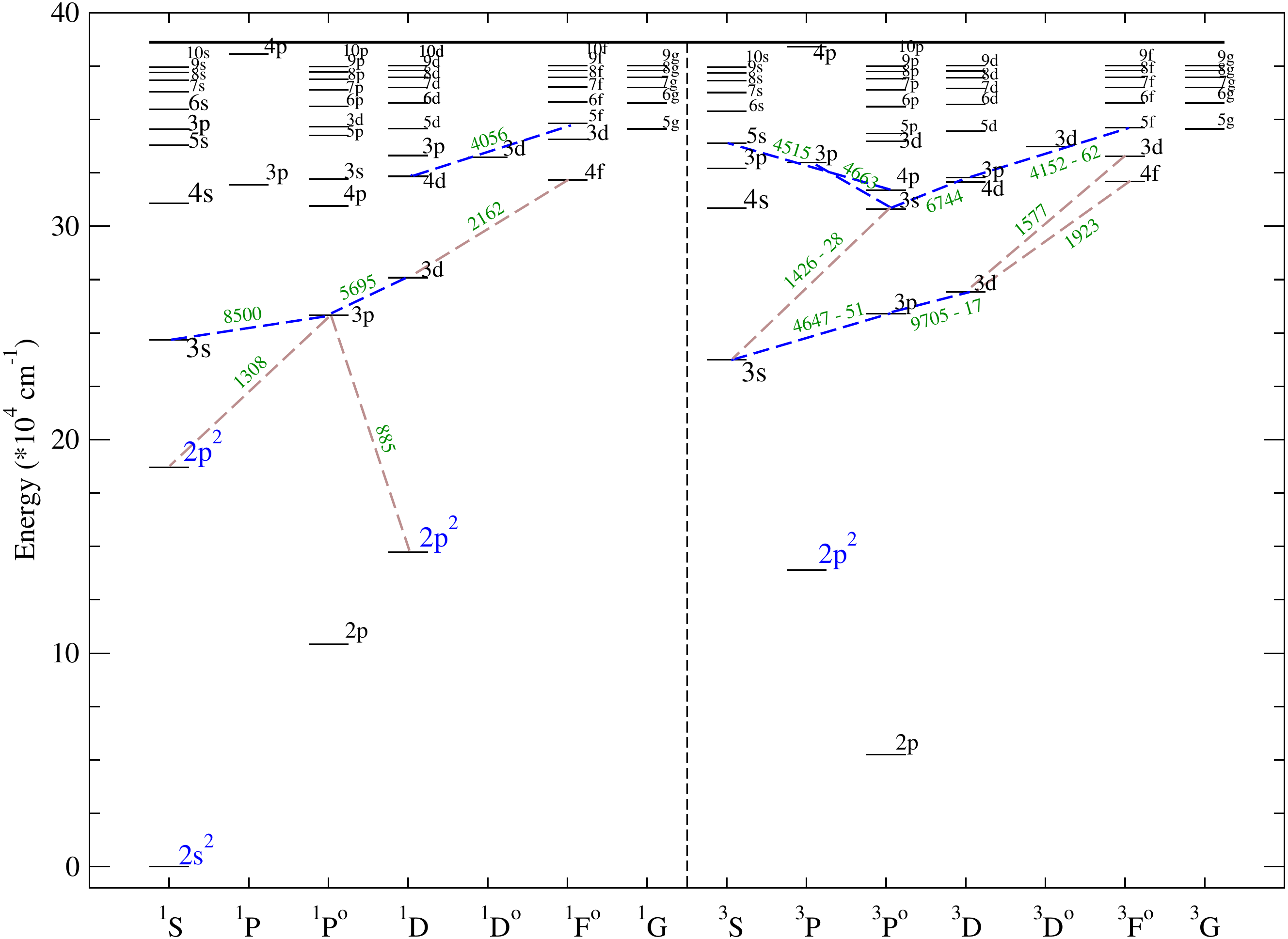}
 \caption{Term diagram for C\iii. The dashed lines indicate the transitions, from which the investigated spectral lines arise. }
 \label{Grot_C3}
 \end{center}
 \end{figure*}

 {\bf Energy levels.} The term structures of C\ione\ and C\ii\ were described in detail in \citet{2015MNRAS.453.1619A} and \citet{2016MNRAS.462.1123A}, respectively.
 Here, the model atom was updated by extending levels of C\ii\ with $n$ = 7--10, $l$=6--8, C\iii\ levels belonging to singlet and triplet terms of the 1s$^2$2p $nl$ ($n$ = 3--4, $l$=0--2), 1s$^2$2s $nl$ ($n$ = 3--10, $l$=0--2), 1s$^2$2s $n$f ($n$ = 4--10), 1s$^2$2s $n$g ($n$ = 5--10), 1s$^2$2s$^2$, 1s$^2$2p$^2$ electronic configurations, C\iv\ levels belonging to doublet terms of the 1s$^2$ $nl$ ($n$ = 2--6, $l$=0--1), 1s$^2$ $n$d ($n$ = 3--6), 1s$^2$ $n$f ($n$ = 4--6), 1s$^2$ 5g electronic configurations. Fine structure splitting was taken into account everywhere up to $n$ = 5 for C\ione\ and C\ii, but the fine structure was neglected for C\iii\ and C\iv\ levels. 
 Most energy levels were taken from the NIST database \footnote{\url{https://www.nist.gov/pml/atomic-spectra-database}} \citep{NIST_ASD}.
 The term diagrams for C\ii\ and C\iii\ are shown in Fig.\,\ref{Grot_C2} and Fig.\,\ref{Grot_C3}, respectively. 
 
 {\bf Radiative and collisional data.}
 
 {\bf C\ione:} Radiative and collisional data for C\ione\ were described in detail in \citet{2016MNRAS.462.1123A}. 
 
 {\bf C\ii:} The transition probabilities were taken from the NIST and VALD \footnote{\url{http://vald.astro.uu.se/~vald/php/vald.php}} \citep{vald} data bases, where available, and the Opacity Project (OP) data base TOPbase \footnote{\url{http://cdsweb.u-strasbg.fr/topbase/topbase.html}} \citep{1989JPhB...22..389L, 1993BICDS..42...39C,1993AAS...99..179H} except for the
 transitions between levels with $n$ = 7$-$10 and $l$ = 6$-$8, for which oscillator strengths were calculated following \citet{1957ApJS....3...37G}.
 Photo-ionization cross sections for levels with $n \le 10$, $l \le 3$ were taken from TOPbase, and for levels $n$g ($n$ = 5--9) relevant data were adopted from NORAD-Atomic-Data web-page \footnote{\url{http://norad.astronomy.ohio-state.edu/}} \citep{1995ApJS..101..423N}. 
 For C\ii\ levels with $n$ = 7--10, $l$ = 5--8, we adopted the hydrogen-like cross sections. 
 For the transitions connecting the 30 lowest fine-structure levels in C\ii, we used the effective collision strengths from \citet{2005AA...432..731W}. 
 The remaining transitions in both C\ii\ and C\iii\ were treated using an approximation formula of \citet{1962ApJ...136..906V} for the allowed transitions and assuming the effective collision strength $\Omega_{ij}$ = 1 for the forbidden transitions. 
  Threshold photoionization cross sections were adopted from TOPbase and \citet{1995ApJS..101..423N}, with empirical correction of one order of magnitude higher for the 
  6f $^2$F$^o$ and 6g $^2$G levels following by \citet{2008A&A...481..199N}.
 
 {\bf C\iii:} The transition probabilities were taken from TOPbase. Photo-ionization cross sections for levels of C\iii\ were adopted from NORAD-Atomic-Data web-page \citep{1997ApJS..111..339N}. 
 For electron-impact excitation of all the transitions between the lowest 98 energy levels of C\iii\ (up to 1s$^2$2s 7d), we adopted recent accurate collisional data from
 \citet{2014A&A...566A.104F}, which are available in the archives of APAP (Atomic Processes for Astrophysical Plasmas) Network \footnote{\url{http://www.apap-network.org/}}.
 
 {\bf C\iv:} The transition probabilities were adopted from NIST. Effective collision strengths for transitions between the lowest nine terms (up to 1s$^2$ 4f) were adopted from \citet{2000JPhB...33.1013G} and downloaded from APAP Network. Photo-ionization cross sections of ground and excited states of C\iv\ were taken from NORAD-Atomic-Data web-page \citep{1997ApJS..111..339N}. 
 
 Ionization by electronic collisions was everywhere treated through the \citet{Seaton1962} classical path approximation
 with threshold photoionization cross sections from OP and \citet{1995ApJS..101..423N, 1997ApJS..111..339N}. 

 \subsection{Method of calculations}
 
  To solve the radiative transfer and statistical equilibrium equations, we used the code \textsc{detail} \citep{detail} based on the accelerated $\Lambda$-iteration method \citep{rh91}. 
  The \textsc{detail} opacity package was updated by \citet{2011JPhCS.328a2015P} by including bound-free opacities of neutral and ionized species.
  We then used the code \textsc{synthV\_NLTE} \citep{2016MNRAS.456.1221R} to calculate the synthetic NLTE line profiles with the obtained departure coefficients, $b_{\rm{i}}$ = $n_{\rm{NLTE}}$ / $n_{\rm{LTE}}$.
  Here, $n_{\rm{NLTE}}$ and $n_{\rm{LTE}}$ are the statistical equilibrium and thermal (Saha-Boltzmann) number densities, respectively. Comparison with the observed spectrum and spectral line fitting were performed with the \textsc{binmag} code\footnote{\url{http://www.astro.uu.se/~oleg/binmag.html}} \citep{binmag3,2018ascl.soft05015K}.
  For consistency, the calculations were performed with plane-parallel (1D), chemically homogeneous model atmospheres from the Kurucz's grid\footnote{\url{http://www.oact.inaf.it/castelli/castelli/grids.html}} \citep{2003IAUS..210P.A20C} for all stars, including two late O-type stars. 
  We suppose that for O-type stars, the spherical and hydrodynamic NLTE atmospheres would be more physically justified. 
  However, our two O-type stars are unevolved (log~$g$ = 3.84 and 4.00) with the simple photospheric physics. They must be unaffected by strong stellar winds unlike the hotter and more luminous supergiants. 
  As for NLTE models, \citet{2007A&A...467..295N} demonstrated that, at least, in the range of 20\,000 $\leq$ \Teff\ $\leq$ 35\,000 K and 3.0 $\leq$ log~$g$ $\leq$ 4.5, the hybrid NLTE approach (LTE model atmosphere plus NLTE line formation) is equivalent to full hydrostatic NLTE computations. Although, the temperatures of our O-type stars are higher by $\sim$ 3000~K compared to investigated limit of \citet{2007A&A...467..295N}, we believe that 
  it will not affect the final results too much.

  Table \ref{tab1} lists lines of C\ione, C\ii, C\iii, and C\iv\ used in line formation analyses together with the adopted line data.
  Oscillator strengths were taken from the NIST database \citep{NIST_ASD}. For most lines, Stark collisional data are adopted from the VALD data base.

\begin{deluxetable*}{lccr}
\tablecaption{Lines of C\ione, C\ii, C\iii, and C\iv\ used in our analysis. \label{tab1}}
\tabletypesize{\scriptsize}
\tablehead{
\colhead{$\lambda$} & \colhead{Transition} & \colhead{\Eexc} & \colhead{log~$gf$}  \\
\colhead{\AA\,}     & \colhead{}           & \colhead{eV}    & \colhead{}        
}
\colnumbers
\startdata
 C\ione\   &                              &       &          \\
 9078.28    & 3s $^3$P$^o$ -- 3p $^3$P     & 7.48  &  $-$0.581 \\
 9088.51    & 3s $^3$P$^o$ -- 3p $^3$P     & 7.48  &  $-$0.43  \\
 9094.83    & 3s $^3$P$^o$ -- 3p $^3$P     & 7.49  &  ~0.151 \\
 9111.80    & 3s $^3$P$^o$ -- 3p $^3$P     & 7.49  &  $-$0.297 \\
 9405.73    & 3s $^1$P$^o$ -- 3p $^1$D     & 7.49  &  ~0.286 \\
 9658.43    & 3s $^3$P$^o$ -- 3p $^3$S     & 7.49  &  $-$0.280 \\ \hline
 C\ii\     &                              &       &          \\
 3918.96    & 3p $^2$P$^o$ -- 4s $^2$S     & 16.33 & $-$0.533  \\
 3920.68    & 3p $^2$P$^o$ -- 4s $^2$S     & 16.33 & $-$0.232  \\
 4267.00    & 3d $^2$D     -- 4f $^2$F$^o$ & 18.05 & ~0.563   \\
 4267.26    & 3d $^2$D     -- 4f $^2$F$^o$ & 18.05 & ~0.716   \\
 4267.26    & 3d $^2$D     -- 4f $^2$F$^o$ & 18.05 & $-$0.584  \\
 5132.95    & 2p3s $^4$P$^o$ -- 2p3p $^4$P & 20.70 & $-$0.211  \\
 5133.28    & 2p3s $^4$P$^o$ -- 2p3p $^4$P & 20.70 & $-$0.178  \\
 5137.25    & 2p3s $^4$P$^o$ -- 2p3p $^4$P & 20.70 & $-$0.911  \\
 5139.17    & 2p3s $^4$P$^o$ -- 2p3p $^4$P & 20.70 & $-$0.707  \\
 5143.49    & 2p3s $^4$P$^o$ -- 2p3p $^4$P & 20.70 & $-$0.212  \\
 5145.16    & 2p3s $^4$P$^o$ -- 2p3p $^4$P & 20.71 & ~0.189   \\
 5151.09    & 2p3s $^4$P$^o$ -- 2p3p $^4$P & 20.71 & $-$0.179  \\
 5648.07    & 3s $^4$P$^o$  --  3p $^4$S   & 20.70 & $-$0.424  \\
 5662.47    & 3s $^4$P$^o$  --  3p $^4$S   & 20.70 & $-$0.249  \\
 6151.26    & 4d $^2$D -- 6f $^2$F$^o$     & 20.84 & $-$0.164  \\
 6151.53    & 4d $^2$D -- 6f $^2$F$^o$     & 20.84 & $-$1.310  \\
 6151.53    & 4d $^2$D -- 6f $^2$F$^o$     & 20.84 & $-$0.009  \\
 6461.94    & 4f $^2$F$^o$ -- 6g $^2$G     & 20.95 & ~0.161  \\
 6461.94    & 4f $^2$F$^o$ -- 6g $^2$G     & 20.95 & ~0.048  \\
 6461.94    & 4f $^2$F$^o$ -- 6g $^2$G     & 20.95 & $-$1.383  \\
 6578.05    & 3s $^2$S  -- 3p $^2$P$^o$    & 14.45 & $-$0.021  \\
 6582.88    & 3s $^2$S  -- 3p $^2$P$^o$    & 14.45 & $-$0.323  \\
 6779.94    & 3s $^4$P$^o$ -- 3p $^4$D     & 20.70 &  0.025  \\
 6780.59    & 3s $^4$P$^o$ -- 3p $^4$D     & 20.70 & $-$0.377  \\
 6783.90    & 3s $^4$P$^o$ -- 3p $^4$D     & 20.70 & ~0.304  \\
 6800.68    & 3s $^4$P$^o$ -- 3p $^4$D     & 20.70 & $-$0.343  \\                                           
 9903.46    & 4f $^2$F$^o$ -- 5g $^2$G     & 20.95 & $-$0.523  \\
 9903.46    & 4f $^2$F$^o$ -- 5g $^2$G     & 20.95 & ~0.908  \\
 9903.46    & 4f $^2$F$^o$ -- 5g $^2$G     & 20.95 & ~1.021  \\
 18\,535.94   & 5g $^2$G -- 6h $^2$H$^o$     & 22.20 & ~1.217  \\
 18\,535.94   & 5g $^2$G -- 6h $^2$H$^o$     & 22.20 & ~1.129  \\
 18\,535.94   & 5g $^2$G -- 6h $^2$H$^o$     & 22.20 & $-$0.515  \\ \hline
 C\iii\     &                              &       &       \\  
 4056.06    &  4d $^1$D -- 5f $^1$F$^o$    & 40.19 & ~0.267  \\
 4152.51    &  3p $^3$D -- 5f $^3$F$^o$    & 40.05 & $-$0.112  \\
 4162.86    &  3p $^3$D -- 5f $^3$F$^o$    & 40.06 & ~0.218  \\
 4516.00    &  4p $^3$P$^o$ -- 5s$^3$S     & 39.40 & $-$0.058  \\
 4647.42    &  3s $^3$S  -- 3p $^3$P$^o$   & 29.53 & ~0.070  \\ 
 4650.25    &  3s $^3$S  -- 3p $^3$P$^o$   & 29.53 & $-$0.151  \\
 4651.47    &  3s $^3$S -- 3p $^3$P$^o$    & 29.53 & $-$0.629  \\
 4652.05    &  3s $^3$P$^o$ --  3p $^3$P   & 38.21 & $-$0.432  \\
 4663.64    &  3s $^3$P$^o$ -- 3p $^3$P    & 38.21 & $-$0.530  \\
 4665.86    &  3s $^3$P$^o$ -- 3p $^3$P    & 38.22 & ~0.044  \\
 5695.92    &  3p $^1$P$^o$ -- 3d $^1$D    & 32.10 & ~0.017  \\
 6731.04    &  3s $^3$P$^o$ -- 3p $^3$D    & 38.22 & $-$0.293  \\
 6744.39    &  3s $^3$P$^o$ -- 3p $^3$D    & 38.22 & $-$0.022  \\ 
 7037.25    &  3s $^1$P$^o$ -- 4d $^1$D    & 38.44 &  ~0.443 \\
 8500.32    &  3s $^1$S  -- 3p $^1$P$^o$   & 30.64 & $-$0.484  \\ 
 9705.41    &  3p $^3$P$^o$  -- 3d $^3$D   & 32.20 & $-$0.378  \\
 9706.44    &  3p $^3$P$^o$  -- 3d $^3$D   & 32.20 & $-$0.855  \\
 9715.09    &  3p $^3$P$^o$  -- 3d $^3$D   & 32.20 & $-$0.107  \\
 9717.75    &  3p $^3$P$^o$  -- 3d $^3$D   & 32.20 & $-$0.855  \\\hline
  C\iv\     &                              &       &       \\ 
 5801.31    &  3s $^2$S  -- 3p $^2$P$^o$   & 37.55 & $-$0.194 \\
 5811.97    &  3s $^2$S  -- 3p $^2$P$^o$   & 37.55 & $-$0.495 \\ \hline
\enddata
\end{deluxetable*}

 \subsection{Departures from LTE for C\ii\ -- C\iii\ in B-type star}\label{Sect:departure}

  \begin{figure*}
 \begin{minipage}{170mm}
 \begin{center}
 \parbox{0.265\linewidth}{\includegraphics[scale=0.25]{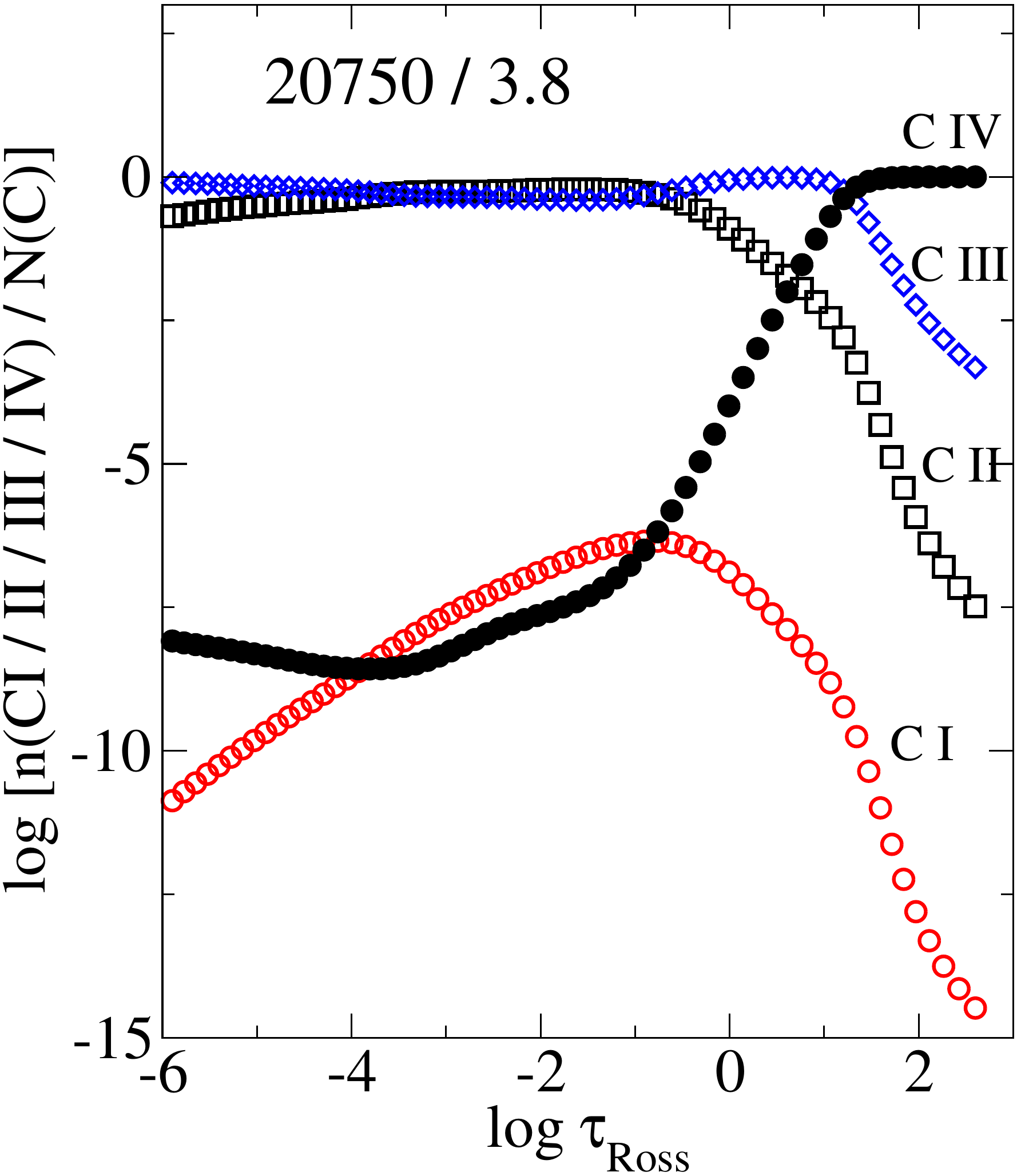}\\
 \centering}
 \parbox{0.23\linewidth}{\includegraphics[scale=0.25]{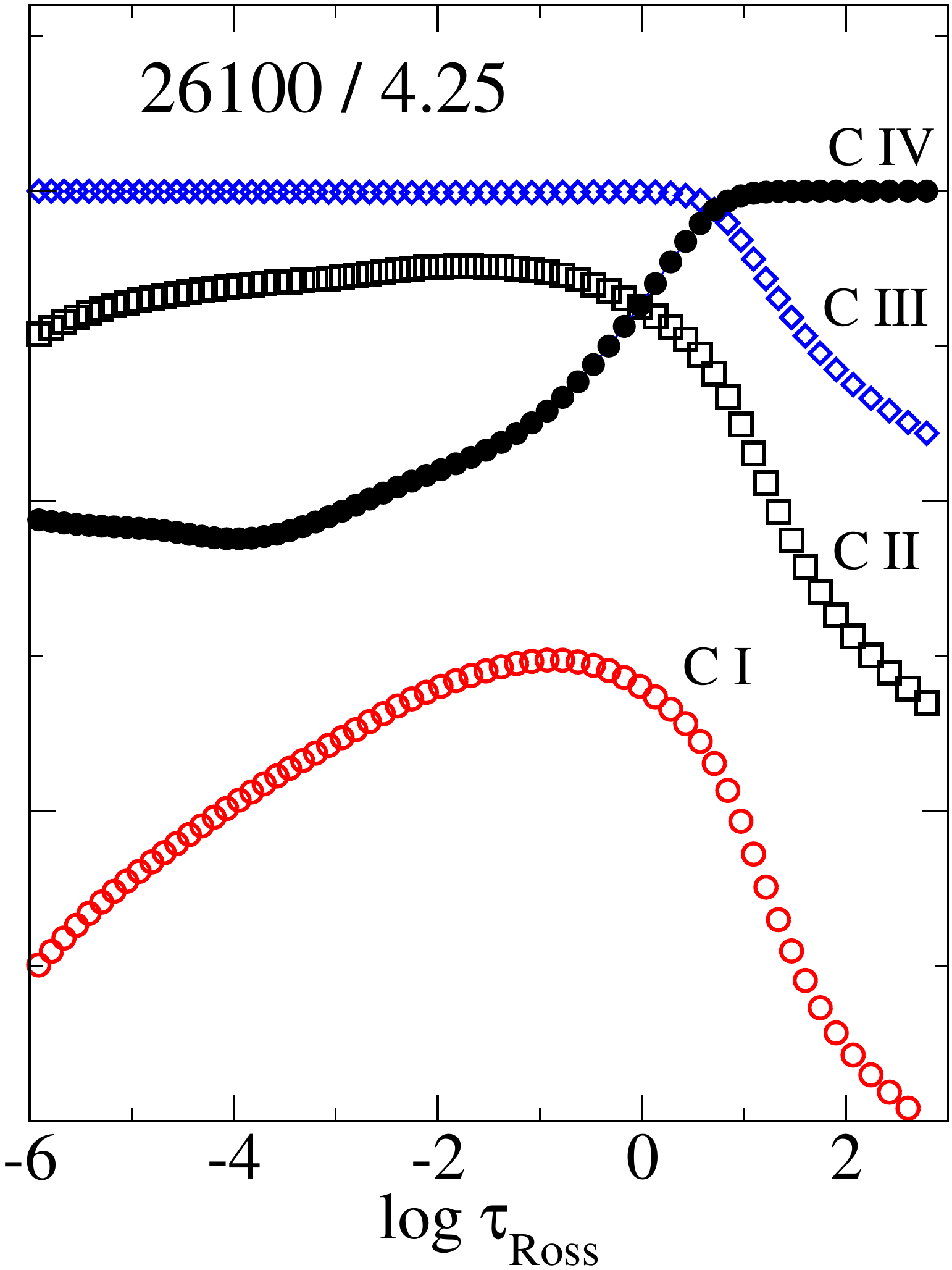}\\
 \centering}
  \parbox{0.23\linewidth}{\includegraphics[scale=0.25]{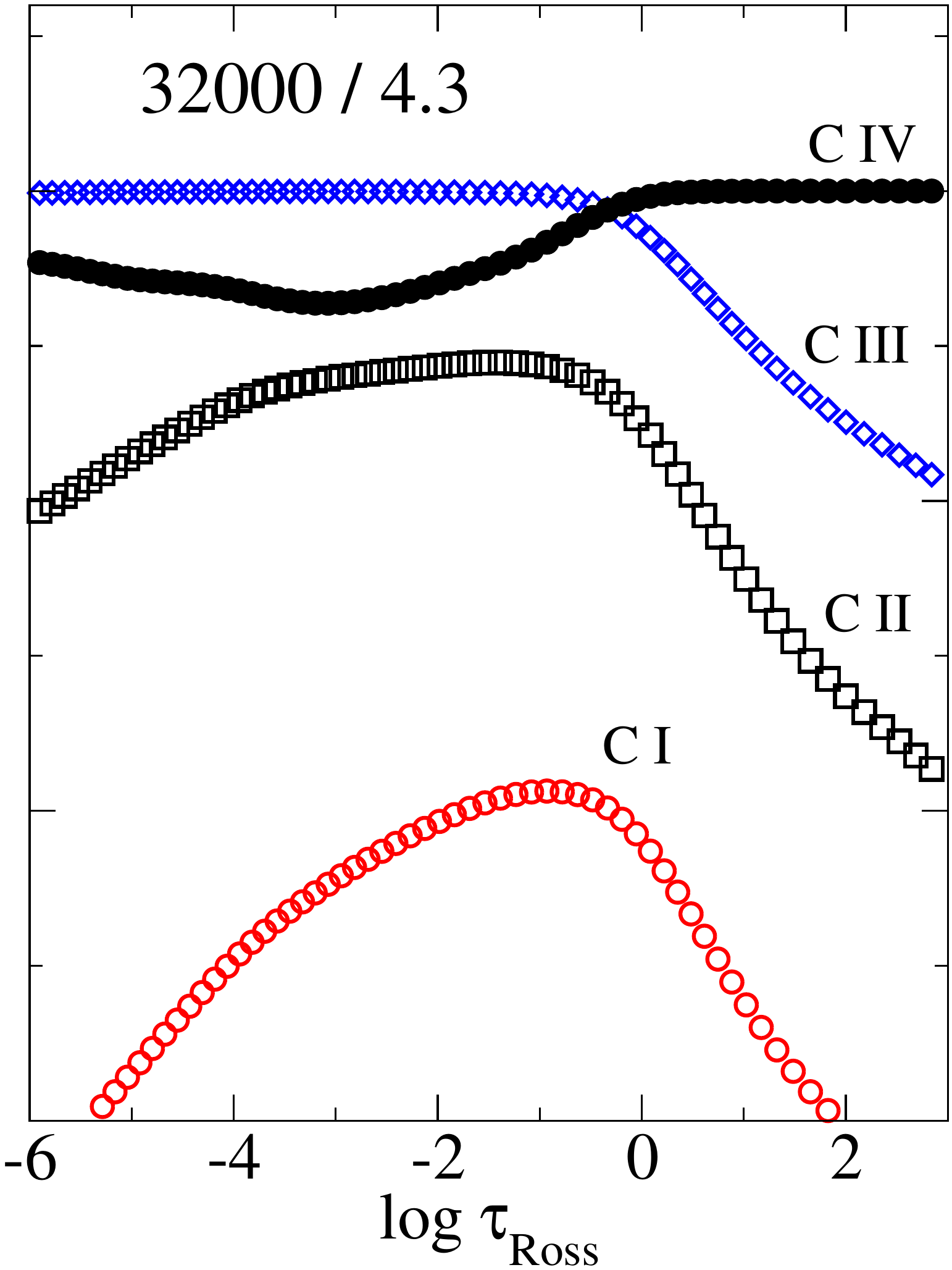}\\ 
 \centering}
 \parbox{0.18\linewidth}{\includegraphics[scale=0.25]{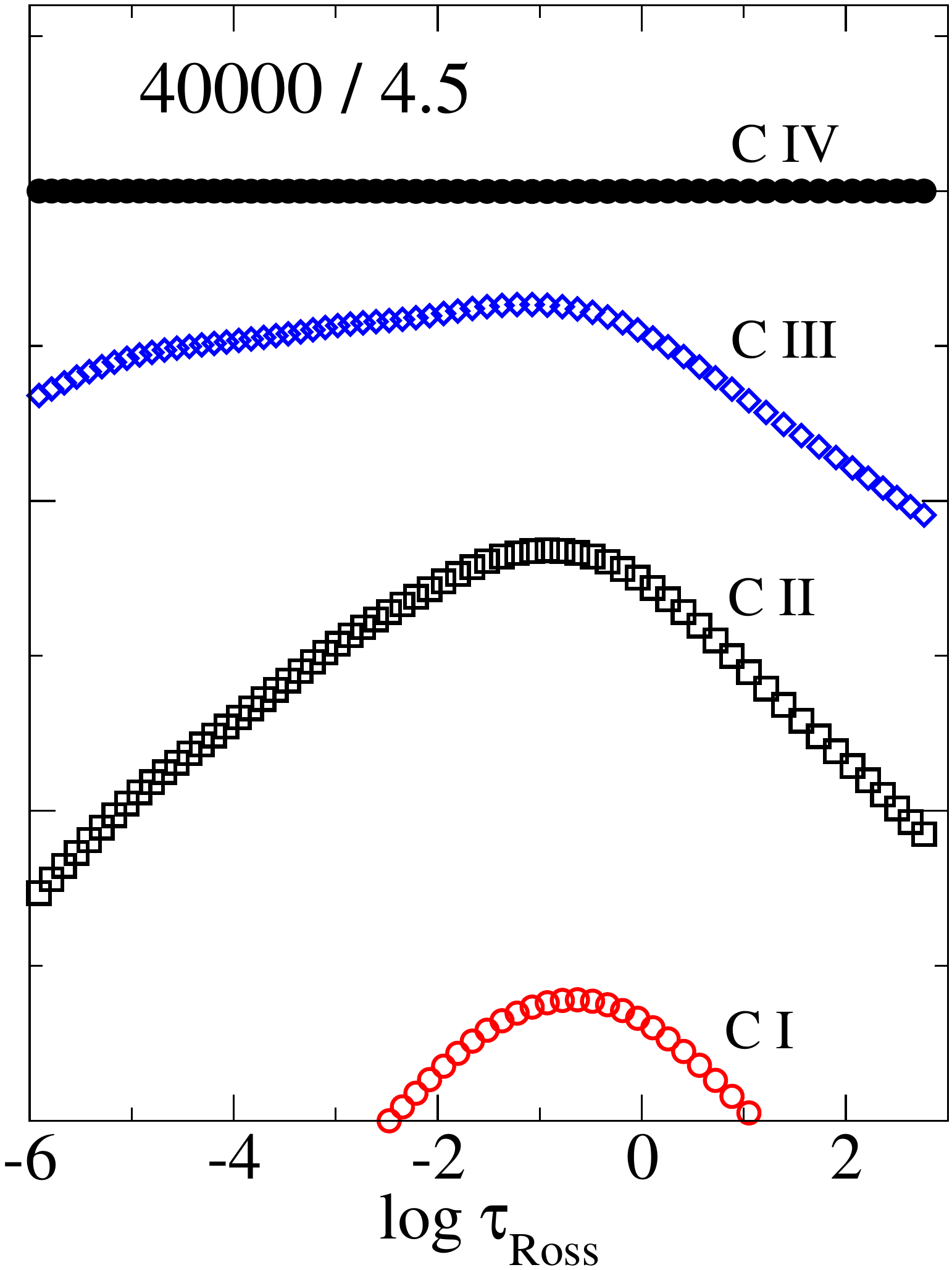}\\
 \centering}
 \hspace{1\linewidth}
 \hfill
 \\[0ex]
 \caption{Ionization fractions of C\ione, C\ii, C\iii, and C\iv\ in the model atmospheres of different effective temperatures. }
 \label{balance}
 \end{center}
 \end{minipage}
 \end{figure*}

 \begin{figure*}
  \begin{minipage}{175mm}
  \parbox{0.45\linewidth}{\includegraphics[scale=0.55]{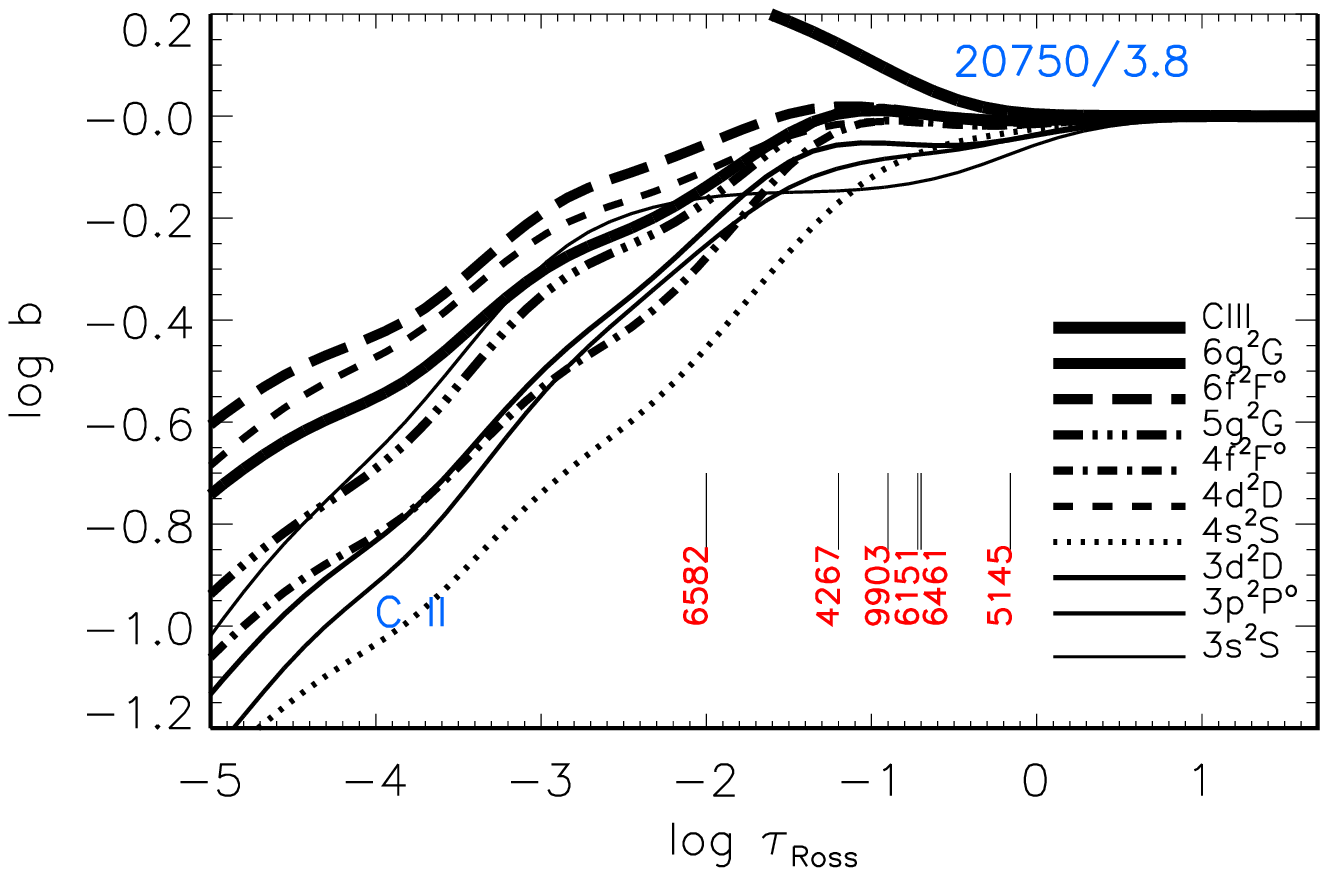}\\
 \centering}
 \parbox{0.45\linewidth}{\includegraphics[scale=0.55]{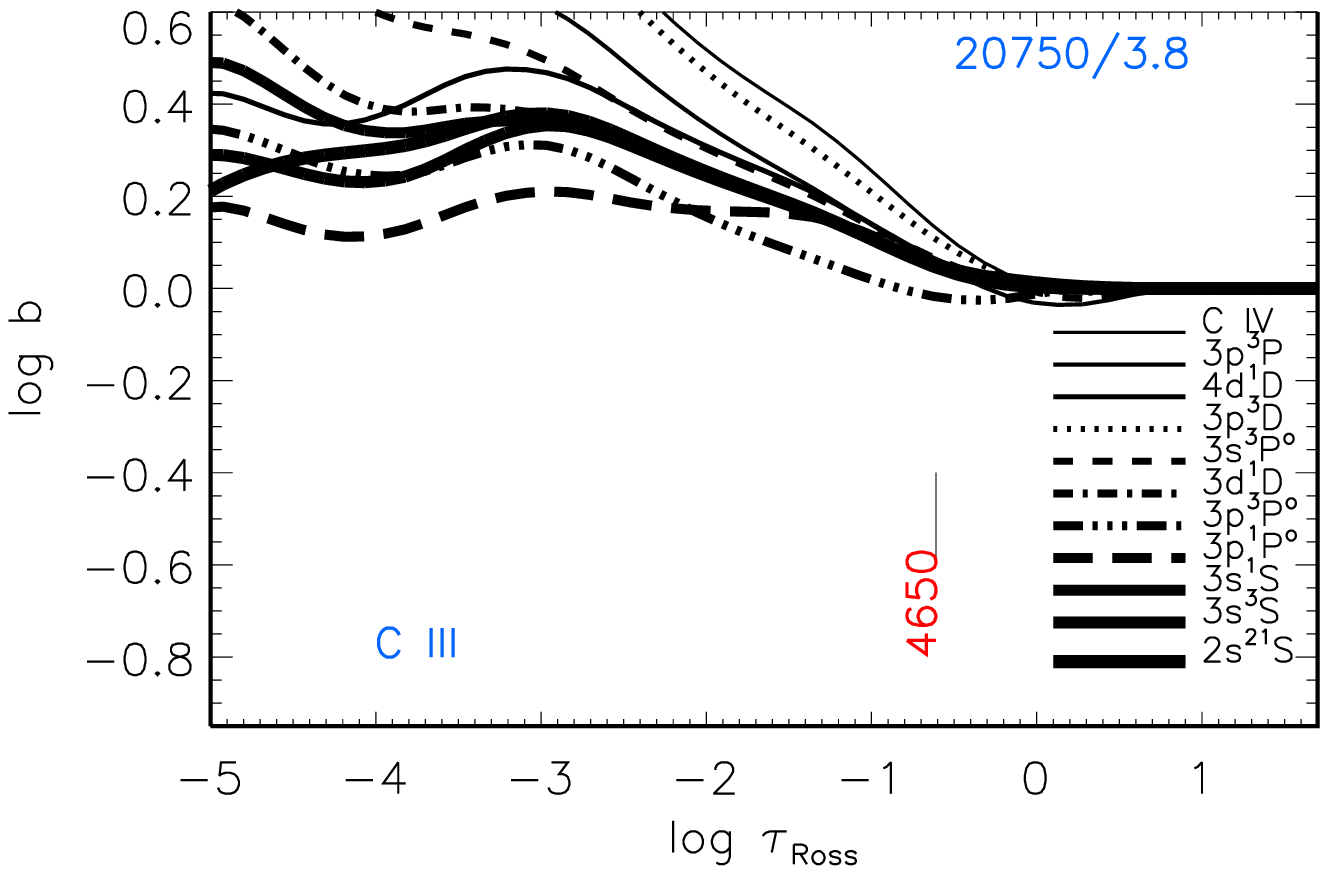}\\
 \centering}
 \hspace{0.00\linewidth}
 \hfill
 \\[0ex]
 \parbox{0.45\linewidth}{\includegraphics[scale=0.55]{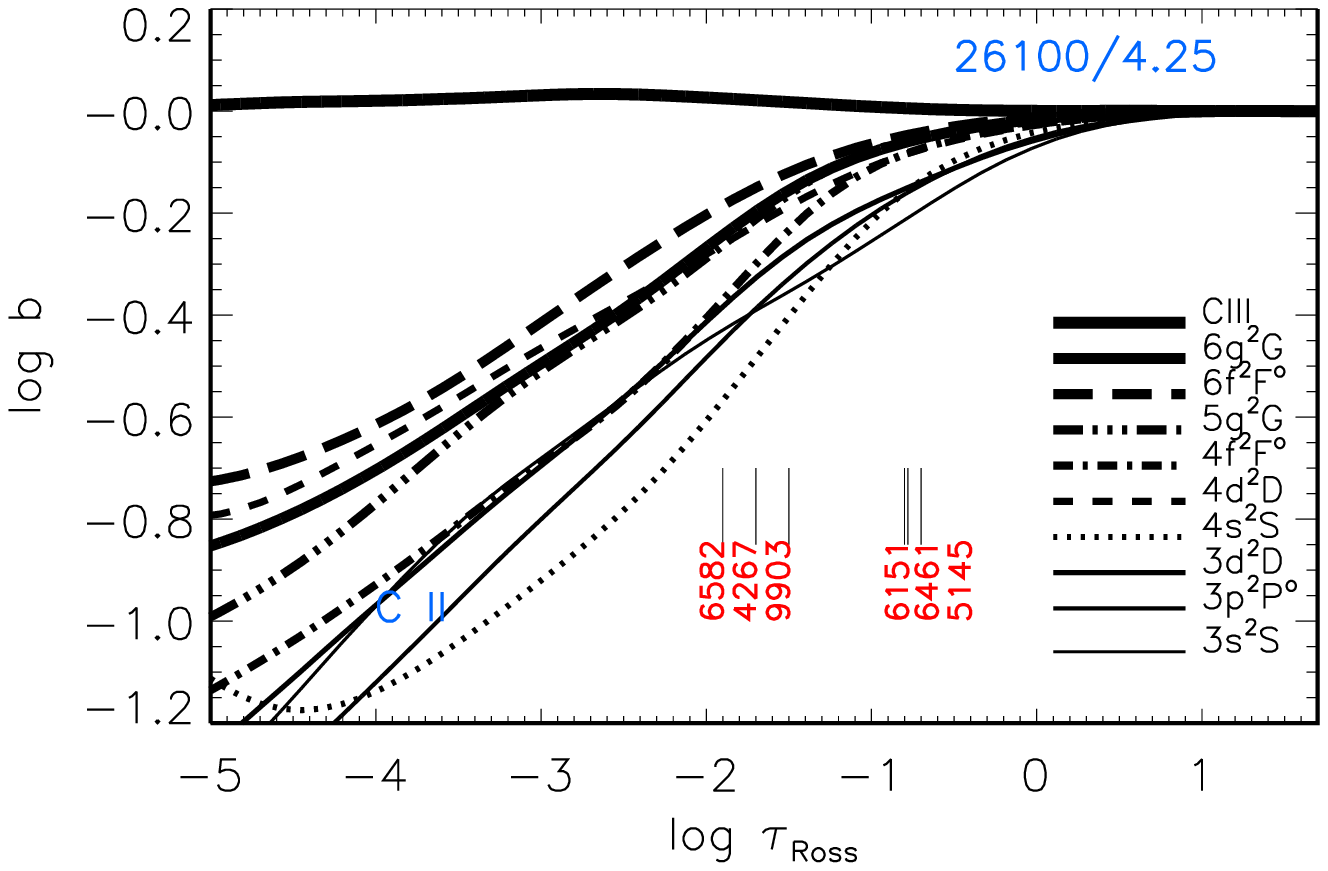}\\
 \centering}
 \parbox{0.45\linewidth}{\includegraphics[scale=0.55]{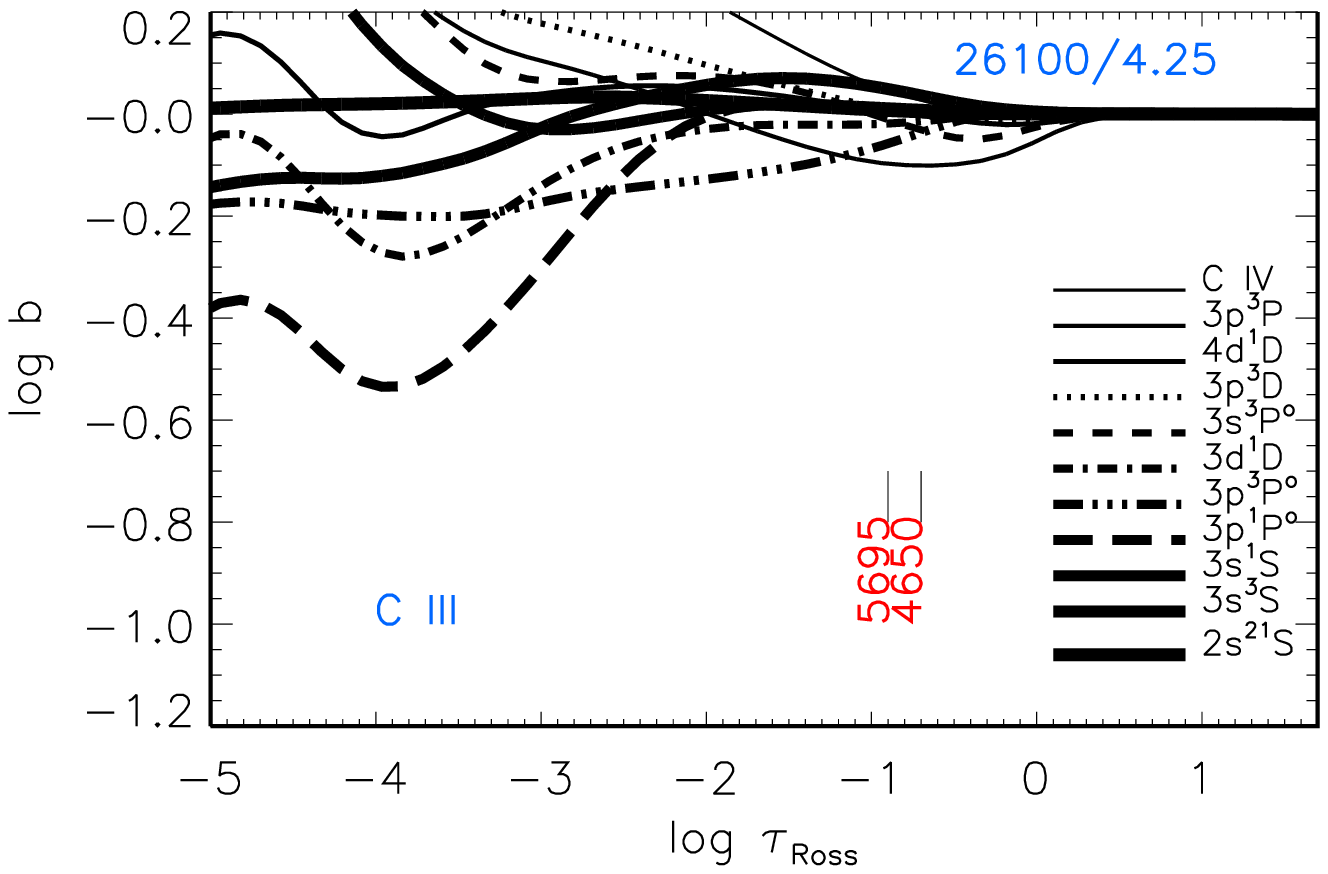}\\
 \centering}
 \hspace{0.00\linewidth}
 \hfill
 \\[0ex]
 \parbox{0.45\linewidth}{\includegraphics[scale=0.55]{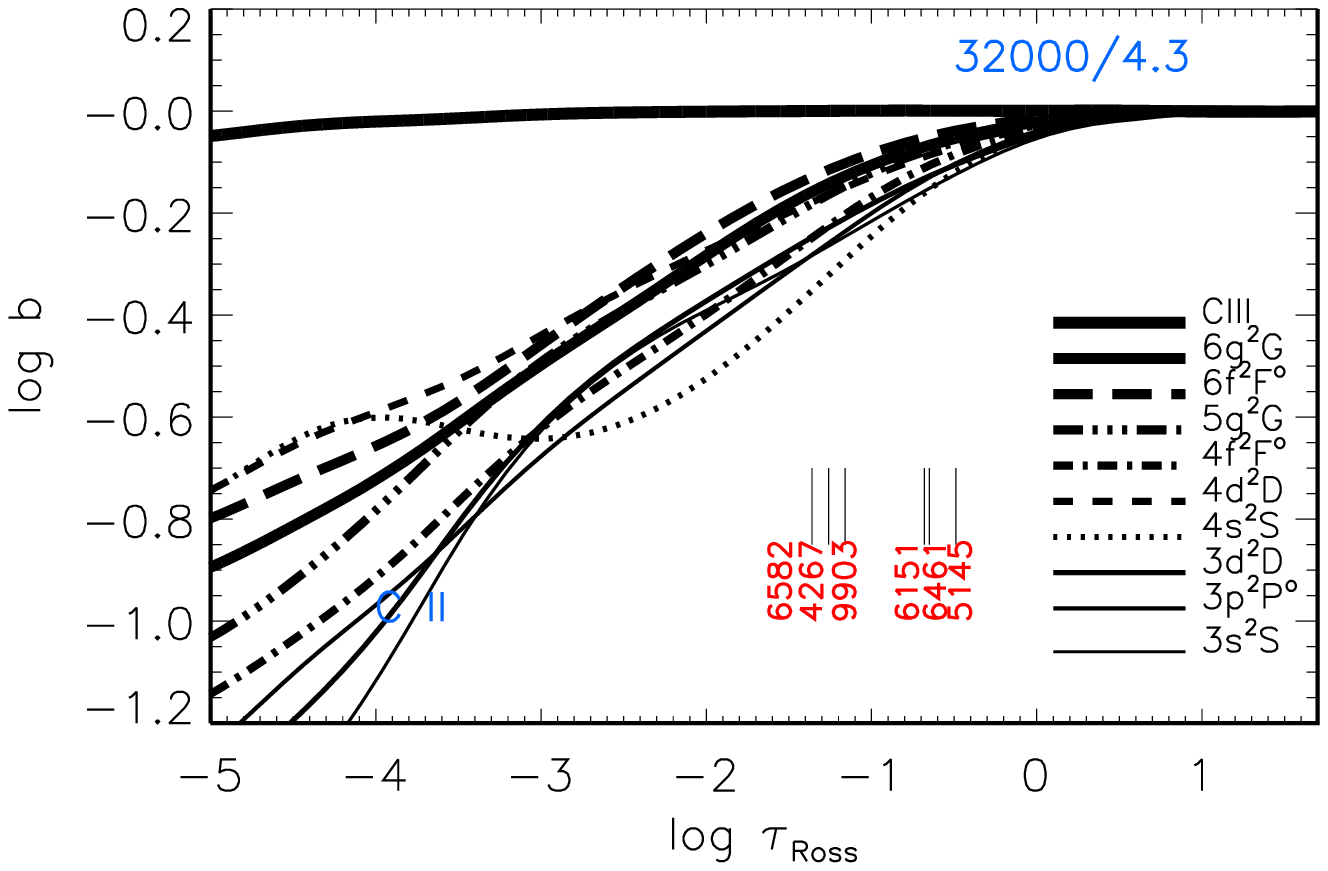}\\
 \centering}
 \parbox{0.45\linewidth}{\includegraphics[scale=0.55]{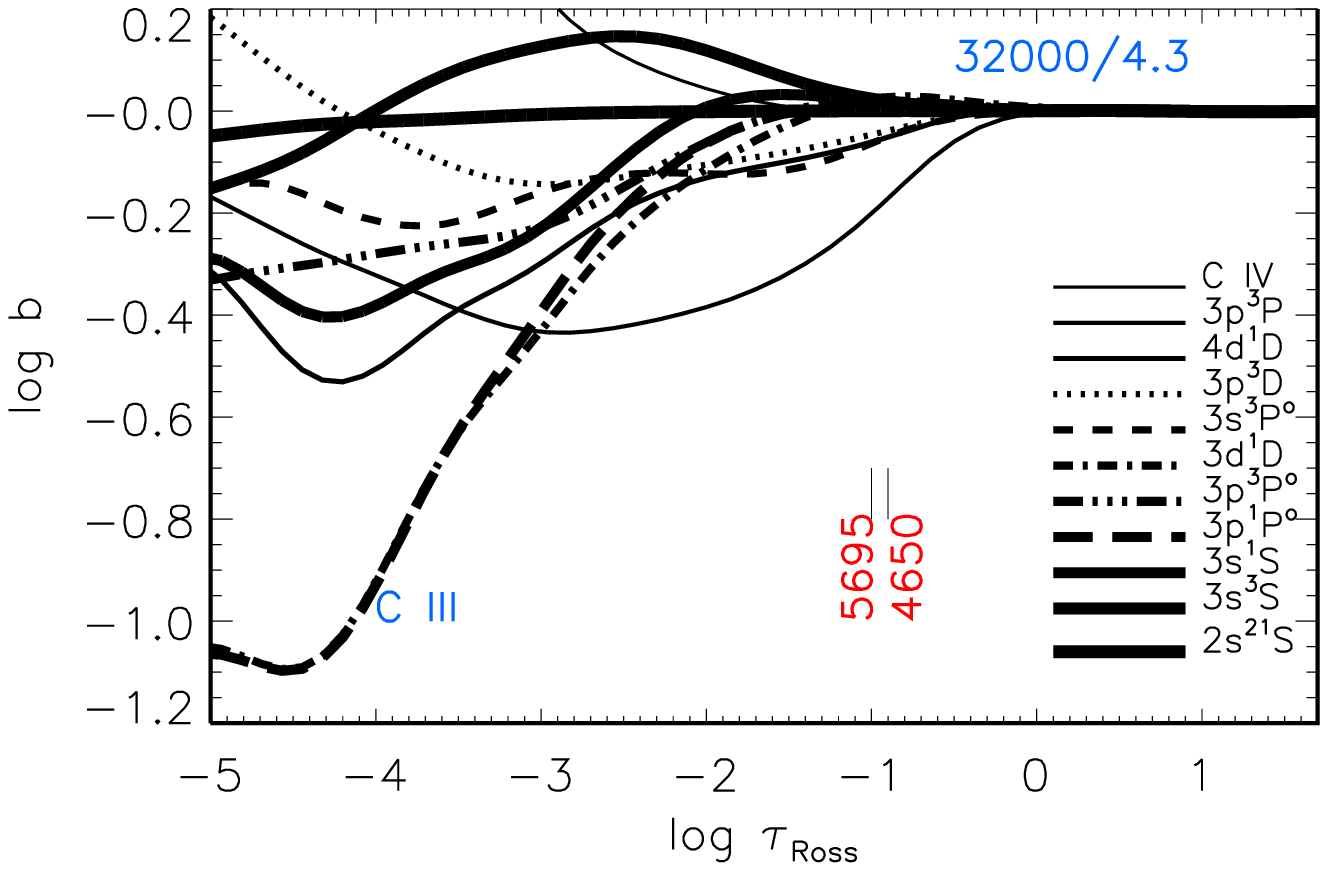}\\
 \centering}
 \hfill
 \\[0ex]
 \parbox{0.45\linewidth}{\includegraphics[scale=0.55]{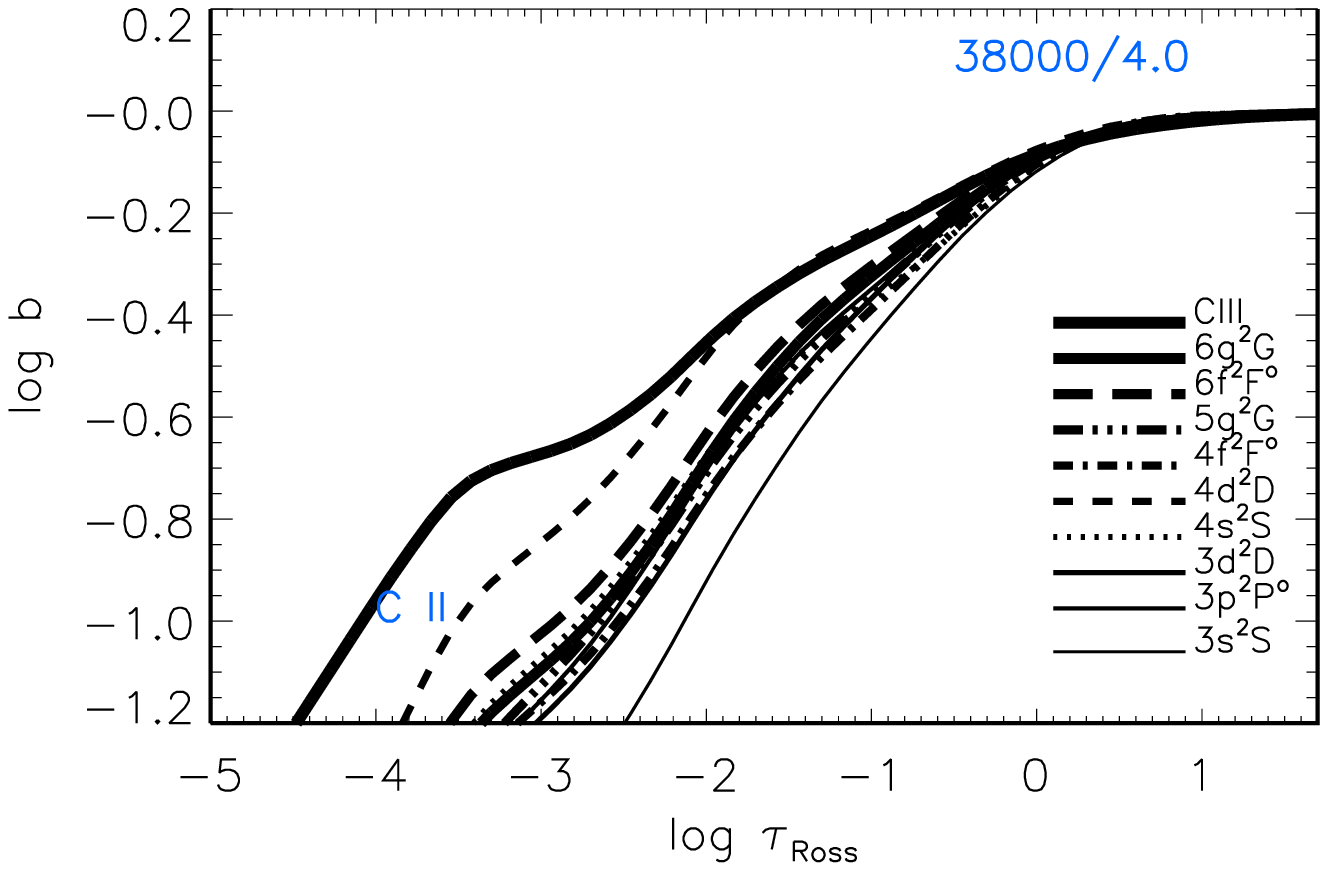}\\
 \centering}
 \parbox{0.45\linewidth}{\includegraphics[scale=0.55]{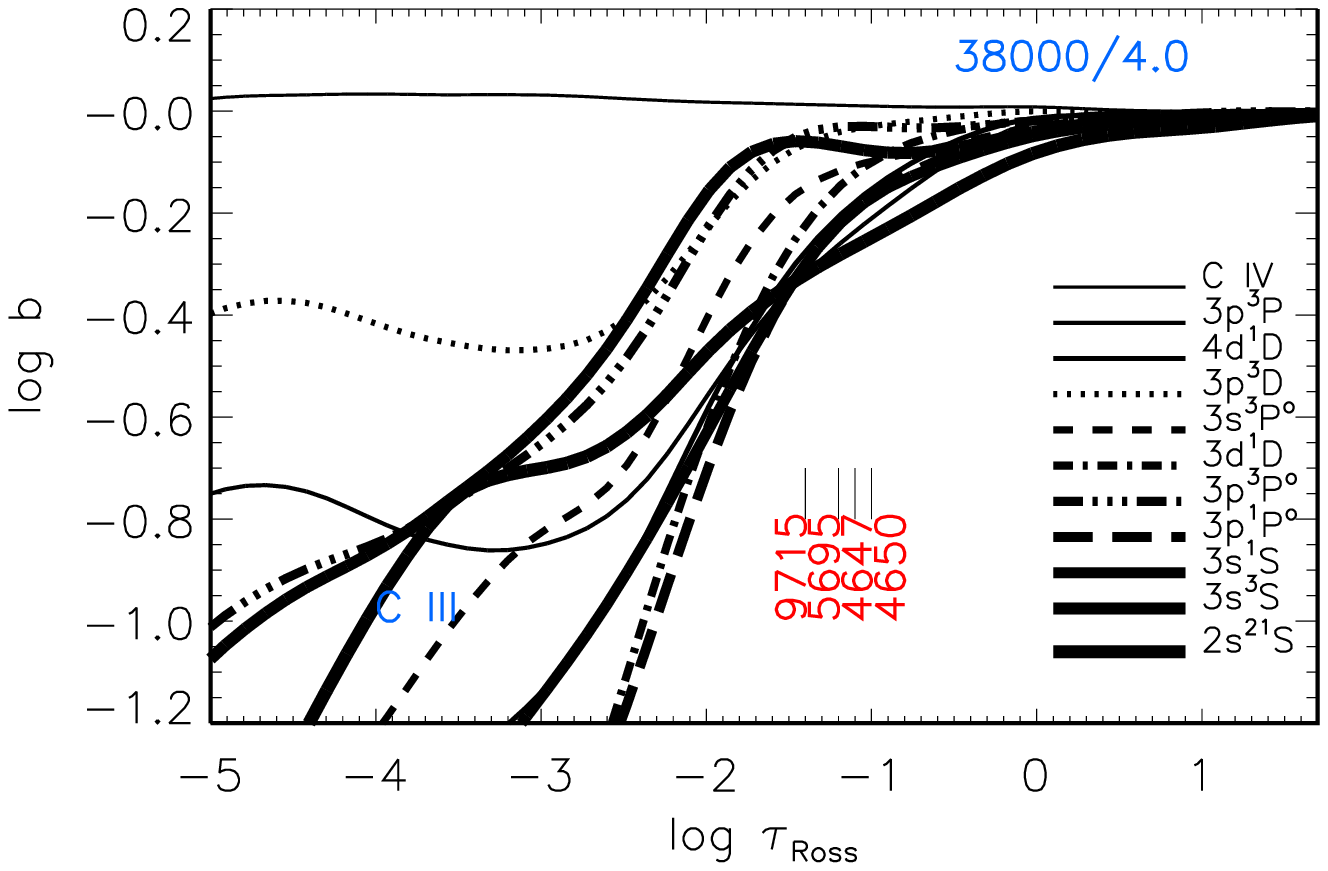}\\
 \centering}
 \hfill
 \caption{Departure coefficients for the C\ii\ and C\iii\ levels as a function of $\log \tau_{Ross}$ in the  model atmospheres with \Teff /log~$g$ = 20750 / 3.8; 26100/ 4.25; 32000 / 4.3; and 38000 / 4.0. 
  The vertical lines indicate the formation depths of cores of individual lines. }
 \label{DC}
 \end{minipage}
 \end{figure*}

 \begin{figure*}
 \begin{minipage}{175mm}
 \begin{center}
 \parbox{0.45\linewidth}{\includegraphics[scale=0.55]{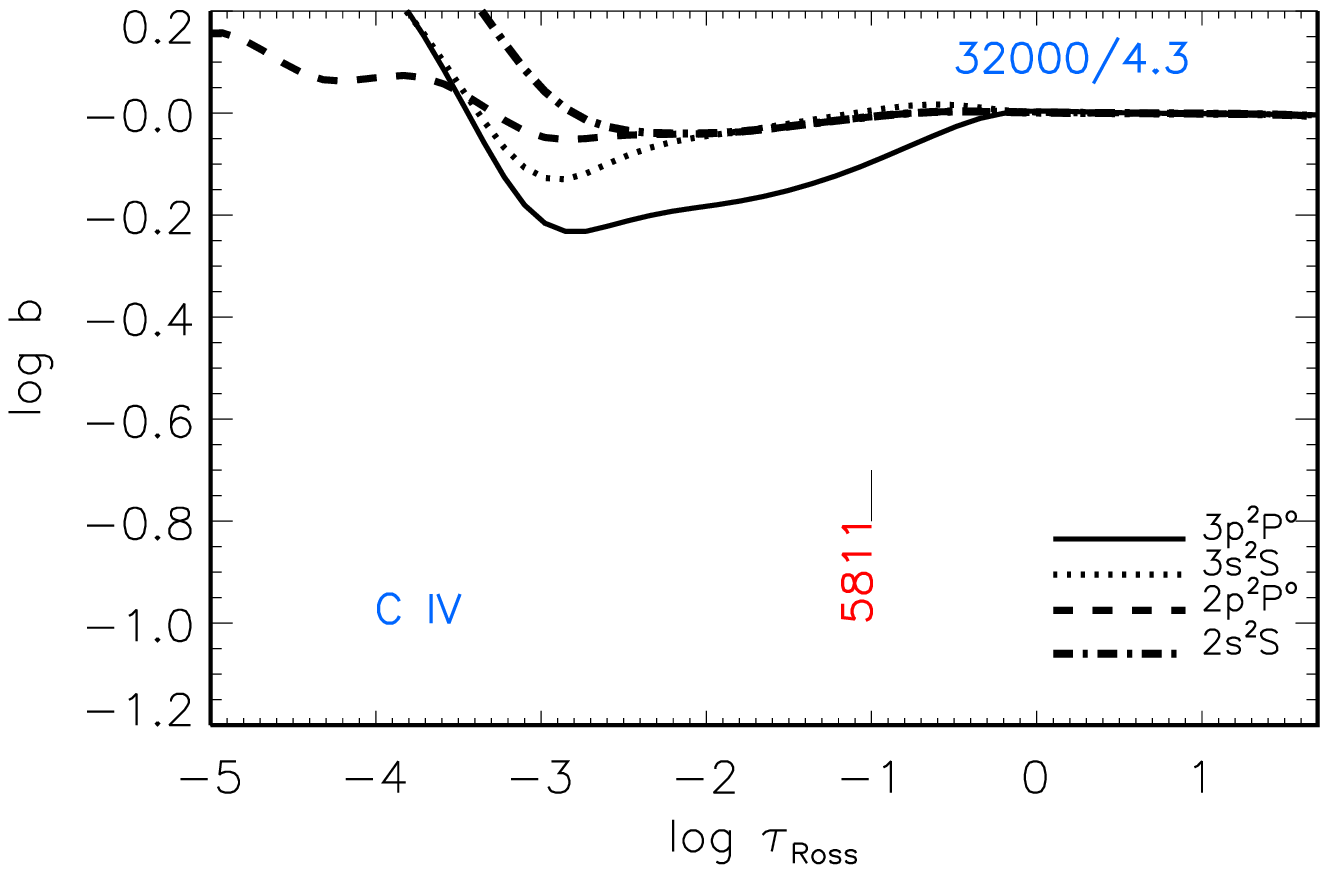}\\
 \centering}
 \parbox{0.45\linewidth}{\includegraphics[scale=0.55]{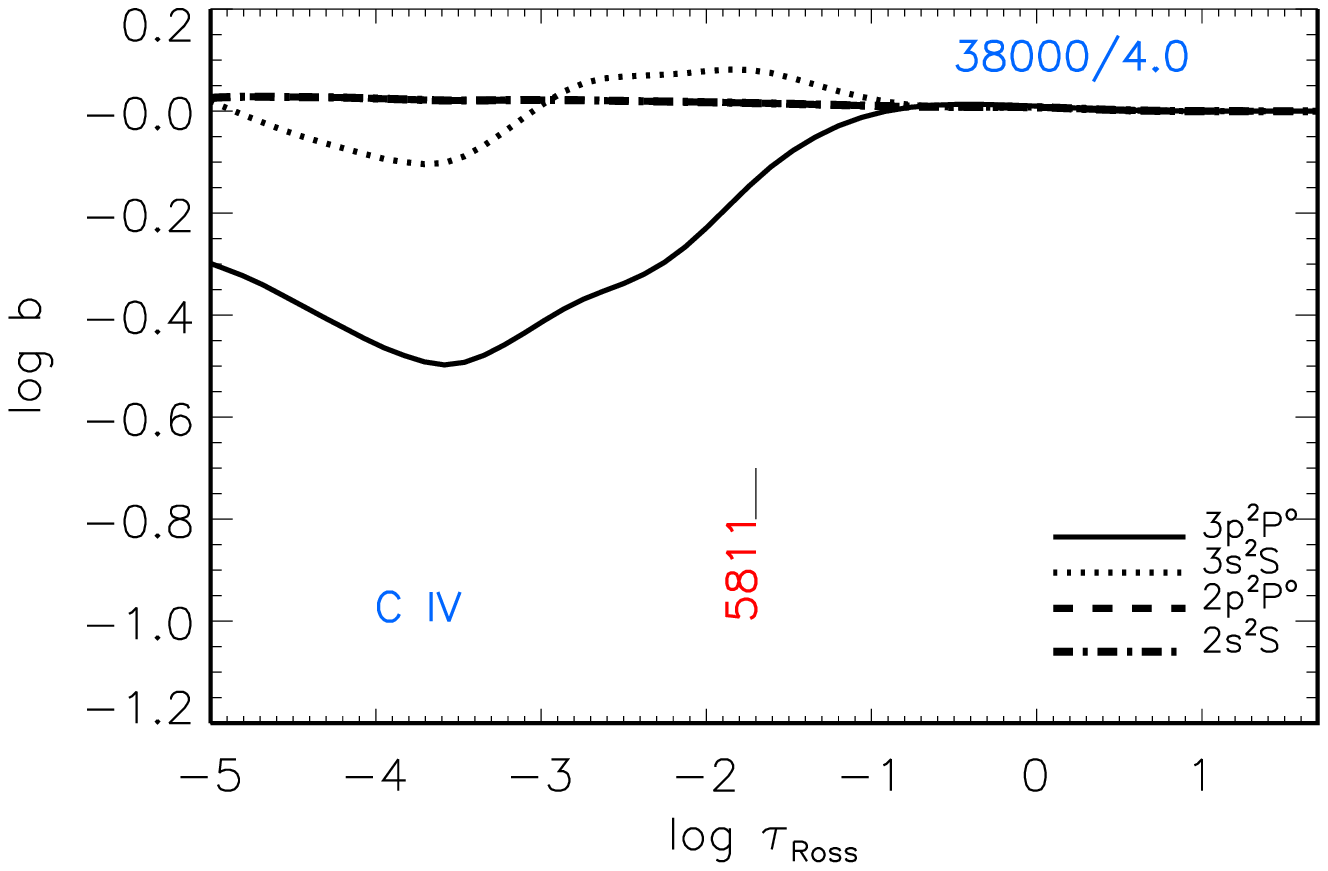}\\
 \centering}
 \hspace{1\linewidth}
 \hfill
 \\[0ex]
 \caption{Departure coefficients for the C\iv\ levels as a function of $\log \tau_{Ross}$ in the  model atmospheres with \Teff /log~$g$ =  32000 / 4.3; and 38000 / 4.0.  }
 \label{DCC4}
 \end{center}
 \end{minipage}
 \end{figure*}

 
 { Figure}\,\ref{balance} displays the ionization fraction of different carbon ions C\ione, C\ii, C\iii, and C\iv\ in stellar atmospheres of different effective temperatures.
 In the atmospheres with \Teff\ $\lesssim$ 32\,000 K, C\iii\ is the dominant ionization stage in the line-formation region, 
 while C\iv\ becomes the major species in the hotter atmospheres. 

 Figures\,\ref{DC} and \ref{DCC4} show the departure coefficients for the selected levels in the model atmospheres with \Teff /log~$g$ = 20\,750 / 3.8, 26\,100 / 4.25, 32\,000 / 4.3, and 38\,000 / 4.0.
 The departure coefficients are equal to unity deep in the atmosphere, where the gas density is large and collisional processes dominate, enforcing the LTE. 
 
 In the models 20\,750 / 3.8, 26\,100 / 4.25, and 32\,000 / 4.3, the NLTE leads to depleted populations of the C\ii\ levels because of UV overionization, while in
 the last two models the C\iii\ ground state is close to the TE population, since doubly ionized carbon dominates the element abundance outwards log~$\tau_{Ross} = 0$.

 In the model 20\,750 / 3.8, outward of log~$\tau_{Ross} = -0.2$, the C\iii\ ground state is overpopulated via photoionization processes C\ii + h$\nu$ $\rightarrow$ C\iii +e$^-$, 
 and the overpopulation is extended to the remaining C\iii\ levels due to their coupling to the ground state. In hotter atmospheres 26\,100 / 4.25 and 32\,000 / 4.3,
 spontaneous transitions dominate, resulting in a depopulation of the C\iii\ levels, while the C\iii\ ground state keeps its LTE populations. 
 
 In the model 38\,000 / 4.0, outward of log~$\tau_{Ross} = 0.3$, NLTE leads to depleted populations of the C\ii\ and C\iii\ levels because of UV overionization, while the C\iv\ ground state almost keeps its LTE populations. 

 Analysis of the departure coefficients of the lower (b$_l$) and upper (b$_u$) levels at the line-formation depths allows to understand NLTE effects for a given spectral line. 
 If b$_l>$ 1 and (or) the line source function is smaller than the Planck function, that is, b$_l>$ b$_u$, a NLTE strengthening of lines occurs. 
 The carbon lines of four ionization stages are listed in Table~\ref{tab1}. 
 
 {\bf C\ii\ 4267~\AA } This line, arising from 3d $^2$D -- 4f $^2$F$^o$, is strong. In the 20\,750 / 3.8 model, its core forms at atmospheric depths ($\log \tau_{Ross} = -1.2$) dominated by overionization of C\ii, resulting in the weakened line and positive abundance correction, $\Delta_{\rm NLTE}$ = log$\epsilon_{\rm NLTE}$ - log$\epsilon_{\rm LTE}$, which equals to +0.25~dex. In the model 26\,100 / 4.25, the line shifts to the outer layers, where the deviations from LTE are strong and abundance correction amounts to +0.77~dex. While in the model 32\,000 / 4.3, the line shifts to the inner layers and $\Delta_{\rm NLTE}$ = +0.33~dex. 
 
 {\bf C\ii\ 6151, 6461~\AA } These lines, arising from 4d $^2$D -- 6f $^2$F$^o$, 4f $^2$F$^o$ -- 6g $^2$G, are weak and their cores form at atmospheric depths around $\log \tau_{Ross} = -0.75$ in the models 20\,750 / 3.8, 26\,100 / 4.25, and 32\,000 / 4.3, where b$_l<$~b$_u$ and b$_l<$1, that leads to weakening of these lines. The abundance corrections for C\ii\ 6151~\AA\ are $\Delta_{\rm NLTE}$ = +0.28, +1.19 dex in the  20\,750 / 3.8, 26\,100 / 4.25 models, respectively, while it appears as weak emission line in 32\,000 / 4.3. For C\ii\ 6461~\AA, $\Delta_{\rm NLTE}$ = +0.38~dex in the  20\,750 / 3.8, while this line appears as emission in the hotter models. 
                                                
 {\bf C\ii\ 6582~\AA } This line, together with 6578, arising from 3s $^2$S  -- 3p $^2$P$^o$, is strong and in the model 20\,750 / 3.8 its core forms around $\log \tau_{Ross} = -2.0$, where the departure coefficient of the upper level 3p $^2$P$^o$ drops rapidly. This results in dropping the line source function below the Planck function and enhancing absorption in the line core. In contrast, in the line wings, absorption is weaker compared with the LTE case due to overall overionization. As a result the abundance correction is slightly negative $-$0.16~dex.
 In the 26\,100 / 4.25 and 32\,000 / 4.3 models, the overionization of C\ii\ leads to weakening of these lines and results in positive abundance corrections, +0.45 and +0.22~dex, respectively. 
   Moreover, these lines are lying in the wing of the H$\alpha$ line at 6563\AA. This makes the formation of these lines very sensitive to modeling of the hydrogen line formation.

 {\bf C\ii\ 9903~\AA } This line, arising from 4f $^2$F$^o$ -- 5g $^2$G, is strong. In the line formation region, $\log \tau_{Ross} \sim -1.25$, b$_l<$~b$_u$ and b$_l<$1, that leads to weakening of the line. It appears as an emission line in the presented models with \Teff\ from 17\,500 to 33\,400 K, and disappears when \Teff\ is higher than 33\,400 K.  
 
 {\bf C\iii\ 4650~\AA } This line, as with 4647~\AA\ and 4651~\AA\ lines, arising from 3s $^3$S  -- 3p $^3$P$^o$, is strong and in the models 20\,750 / 3.8, 26\,100 / 4.25, and 32\,000 / 4.3, b$_l>$ b$_u$ in the line formation region and
 the line source function drops below the Planck function resulting in strengthened lines and negative NLTE abundance corrections of $\Delta_{\rm NLTE}$ = $-$0.23~dex, $-$0.33~dex, $-$0.61~dex respectively. 
 In the model 38\,000 / 4.0 b$_l<$~b$_u$ in the line formation region, which leads to weakening of the line and $\Delta_{\rm NLTE}$ = $+$0.64~dex. 
 
 {\bf C\iii\ 5695~\AA } The line, arising from 3p $^1$P$^o$ -- 3d $^1$D, is strong. Its core forms at atmospheric depths around $\log \tau_{Ross} = -1.0$.
  The line has slightly negative abundance correction, $-$0.12, in the model 26\,100 / 4.25 and slightly positive one, +0.06 dex, in 32\,000 / 4.3. 
  In the model 38\,000 / 4.0, the line formation region moves to inner layers, where b$_l<$~b$_u$, and b$_l<$1, and the line appears in emission. 
 
 {\bf C\iv\ 5811~\AA } This line and 5801~\AA\ arising from 3s $^2$S  -- 3p $^2$P$^o$, appear to be stronger in the hotter atmospheres. In the line formation region b$_l>$ b$_u$ that results in strengthened line and negative NLTE abundance correction of $\Delta_{\rm NLTE}$ = $-$0.37~dex in the model 32\,000 / 4.3. In a hotter model, 38\,000 / 4.0, line formation region moves to
 outer layers (Fig.\,\ref{DCC4}), where the departures from LTE are stronger. So, in the model 38\,000 / 4.0 the NLTE abundance correction is $-$1.29~dex.  
 
 \subsection{C\ione\ emission lines} 
 
 The mechanisms driving the C\ione\ emission were described in details in \citet{2016MNRAS.462.1123A}. 
 Our NLTE calculations predicted that C\ione\ emission lines to appear at effective temperature from 9250 to 10\,500~K depending on log~$g$. 
 They are near IR C\ione\ lines at 8335~\AA\ (3s $^1$P$^o$ -- 3p $^1$S) and 9405~\AA\ (3s $^1$P$^o$ -- 3p $^1$D) singlet lines and at higher temperature, \Teff $>$ 15\,000 K (log~$g$ = 4), 
 in the 9061--9111~\AA\ (3s $^3$P$^o$ -- 3p $^3$P), and 9603--9658~\AA\ (3s $^3$P$^o$ -- 3p $^3$S) triplet lines. 
 Having appeared at specific \Teff/log~$g$, the emission is strengthened towards higher temperature, reaches its maximum and disappears at about \Teff $=$ 22\,000 K (in the case of log~$g$ = 3).
 
  \subsection{C\ii\ emission lines} 
  
  Our NLTE calculations in the grid of model atmospheres with 18\,000~K $\le$ \Teff\ $\le$ 35\,000~K and log~$g$ = 4 predict that C\ii\ lines at 6151~\AA\ (4d $^2$D -- 6f $^2$F$^o$), 
  6461~\AA\ (4f $^2$F$^o$ -- 6g $^2$G), 9903~\AA\ (4f $^2$F$^o$ -- 5g $^2$G), and 18\,535~\AA\ (5g $^2$G -- 6h $^2$H$^o$) may appear as emission lines depending on the atmospheric parameters (Fig. \ref{param34}). 
  The NET rates, that were calculated as NET = $n_{\rm{l}}$($R_{\rm{lu}}$ + $C_{\rm{lu}}$)$-$$n_{\rm{u}}$ ($R_{\rm{ul}}$ + $C_{\rm{ul}}$), allow to know what processes in C\ii\ do drive the emission in the lines.
  Here, $R_{\rm{lu}}$ and $C_{\rm{lu}}$ are radiative and collisional rates, respectively, for the transition from low to upper level, and $R_{\rm{ul}}$ and $C_{\rm{ul}}$ for the inverse transition.
  The upper levels of the investigated transitions, 6f $^2$F$^o$, 5g $^2$G, 6g $^2$G, 6h $^2$H$^o$, are mainly populated by recombination from C\iii\ and 
  cascade transitions from Rydberg states, while the low-excitation levels (4d $^2$D and 4f $^2$F$^o$) have depleted populations due to photon losses in the lines of C\ii\ arising from the transitions 3p $^2$P$^o$ -- 4d $^2$D (2746~\AA) and 3d $^2$D -- 4f $^2$F$^o$ (4267~\AA). 
  Extra depopulation of the level 4f $^2$F$^o$ is caused by overionization. 
  These mechanisms result in a depopulation of the lower levels to a greater extent than the upper levels and are responsible for the emission phenomenon. 
  Test calculations with the model atom, where the Rygberg states ($l>$6) were not included, demonstrate the decrease of the relative flux at the line center of C\ii\ 9903~\AA\ from 1.30 to 1.06  in the model atmosphere 26\,100 / 4.25. It demonstrates a crucial role of these Rygberg states in emission phenomenon at C\ii\ 9903~\AA.

 \subsection{C\iii\ emission lines}
 
\begin{figure*}
   \begin{minipage}{180mm}
 \begin{center}
 \parbox{0.45\linewidth}{\includegraphics[scale=0.25]{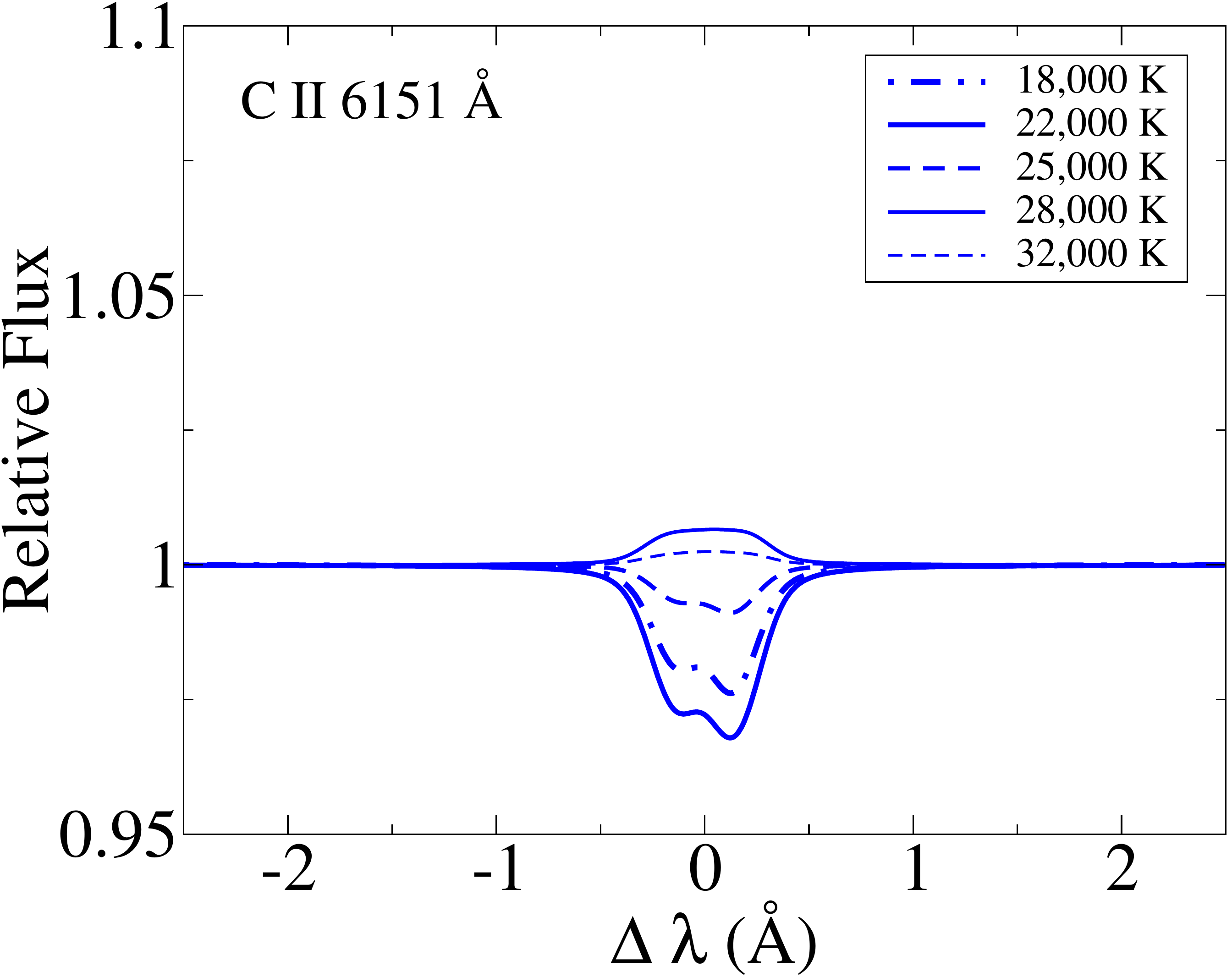}\\
 \centering}
 \parbox{0.45\linewidth}{\includegraphics[scale=0.25]{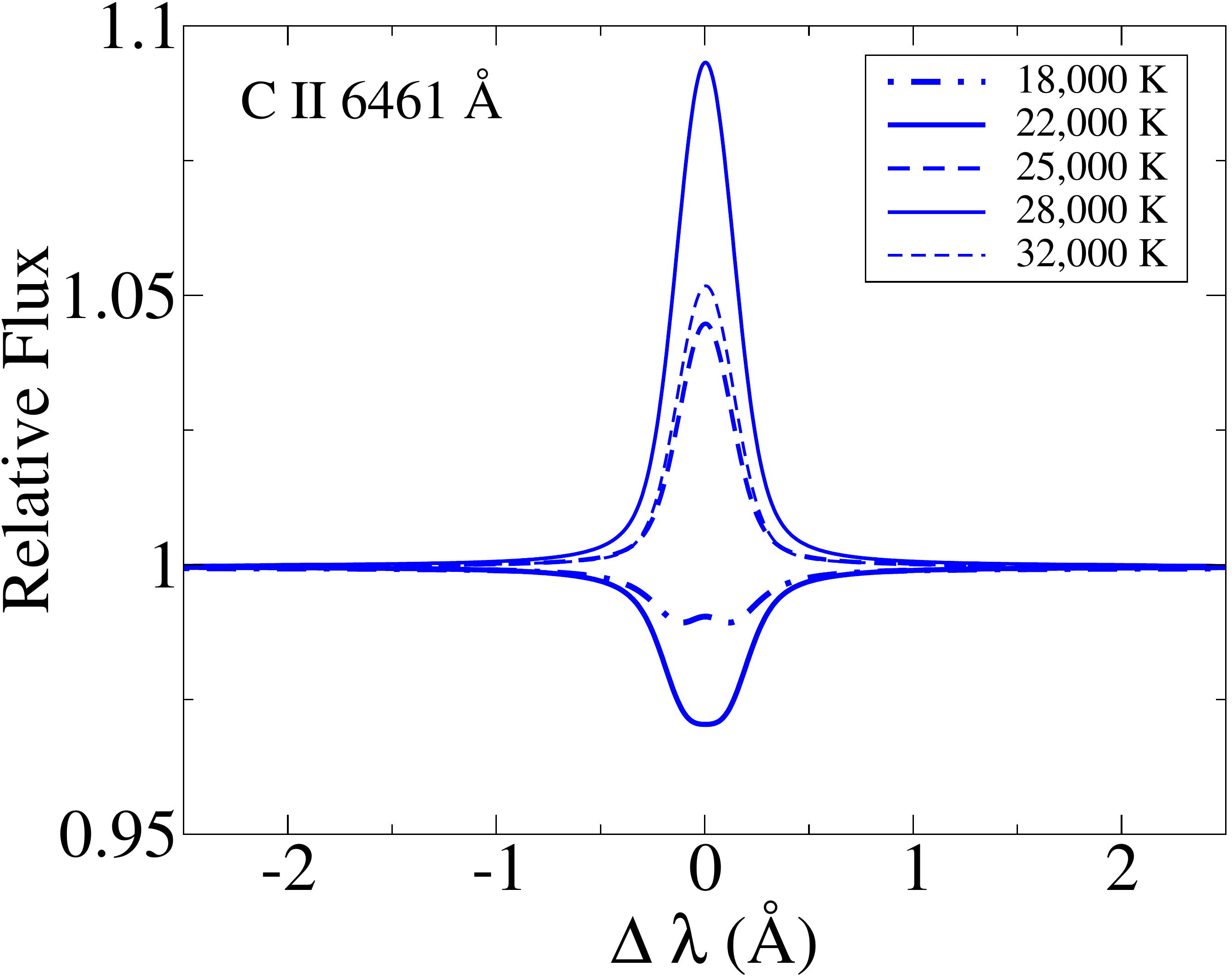}\\
\centering}
 \hspace{1\linewidth}
 \hfill
 \\[0ex]
 \parbox{0.45\linewidth}{\includegraphics[scale=0.25]{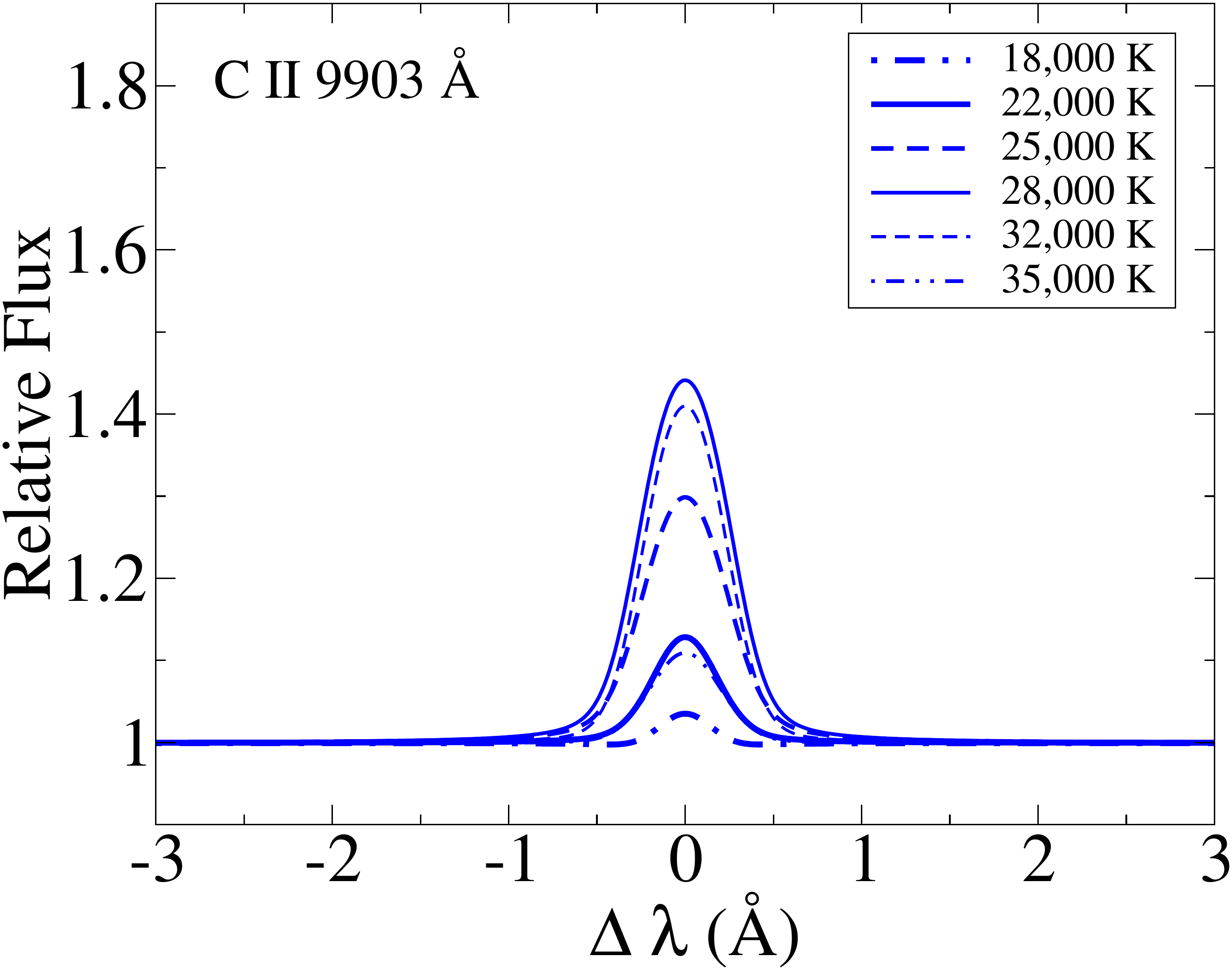}\\
 \centering}
 \parbox{0.45\linewidth}{\includegraphics[scale=0.25]{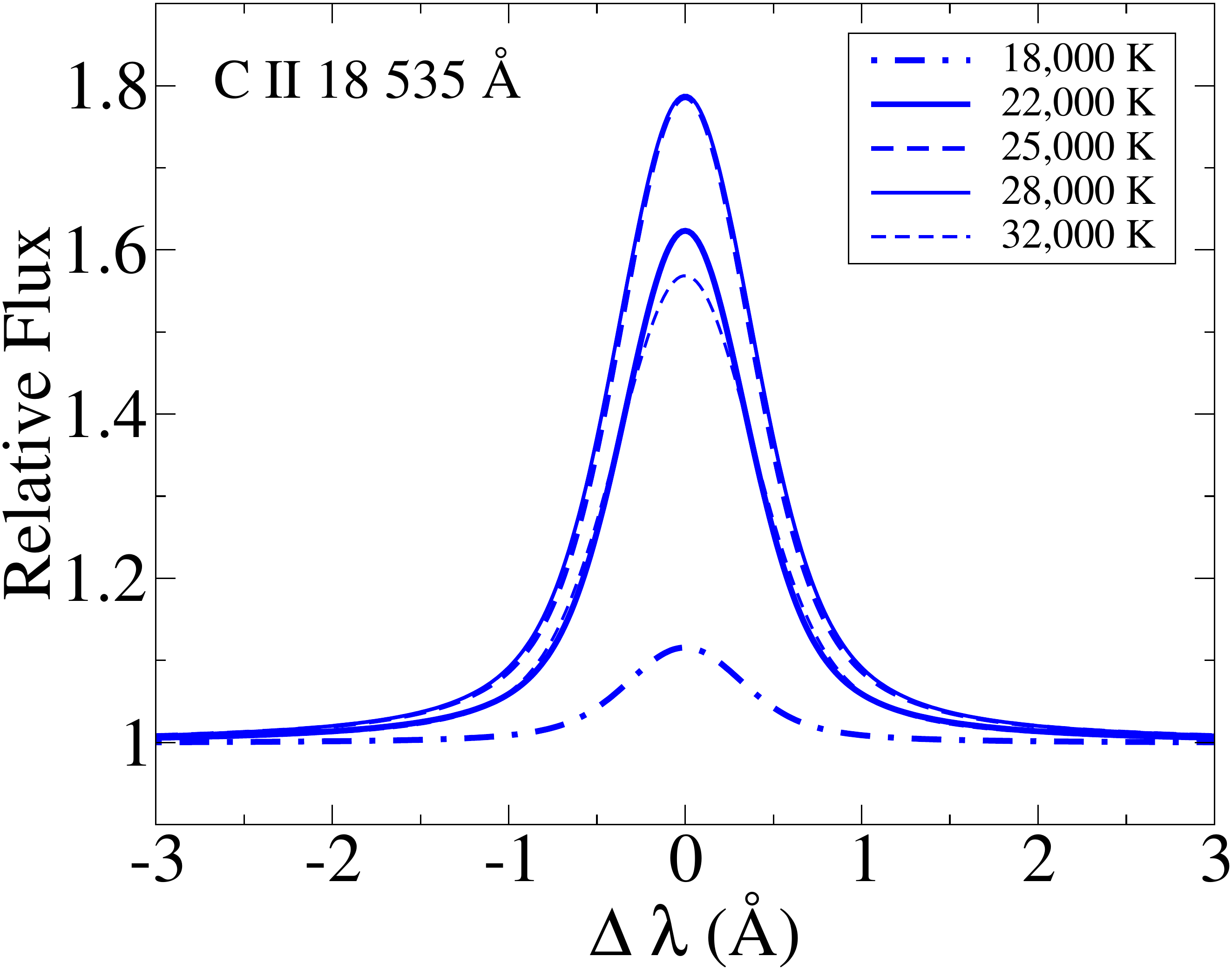}\\
 \centering}
 \hspace{1\linewidth}
 \hfill
 \\[0ex]
  \parbox{0.45\linewidth}{\includegraphics[scale=0.25]{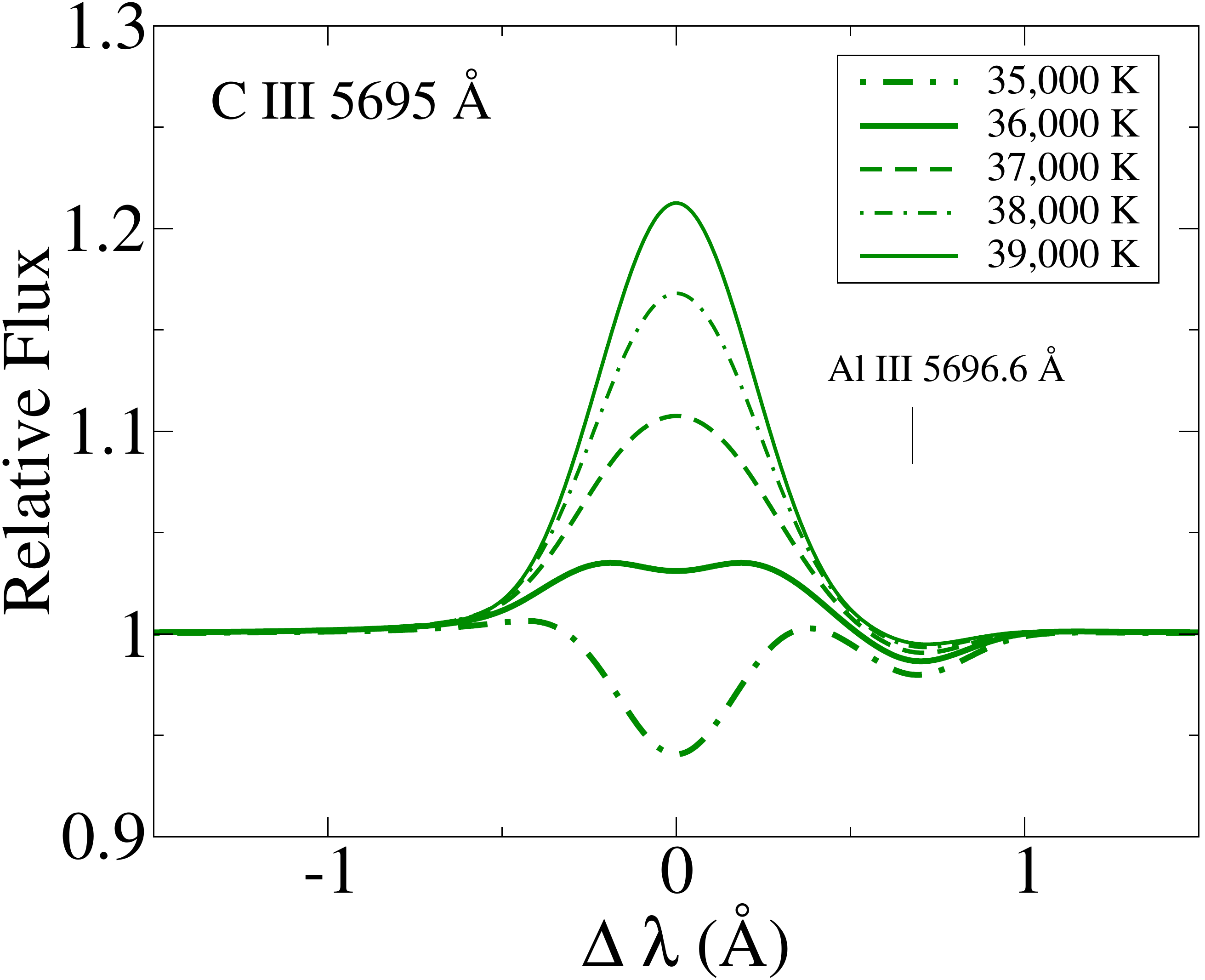}\\
 \centering}
 \parbox{0.45\linewidth}{\includegraphics[scale=0.25]{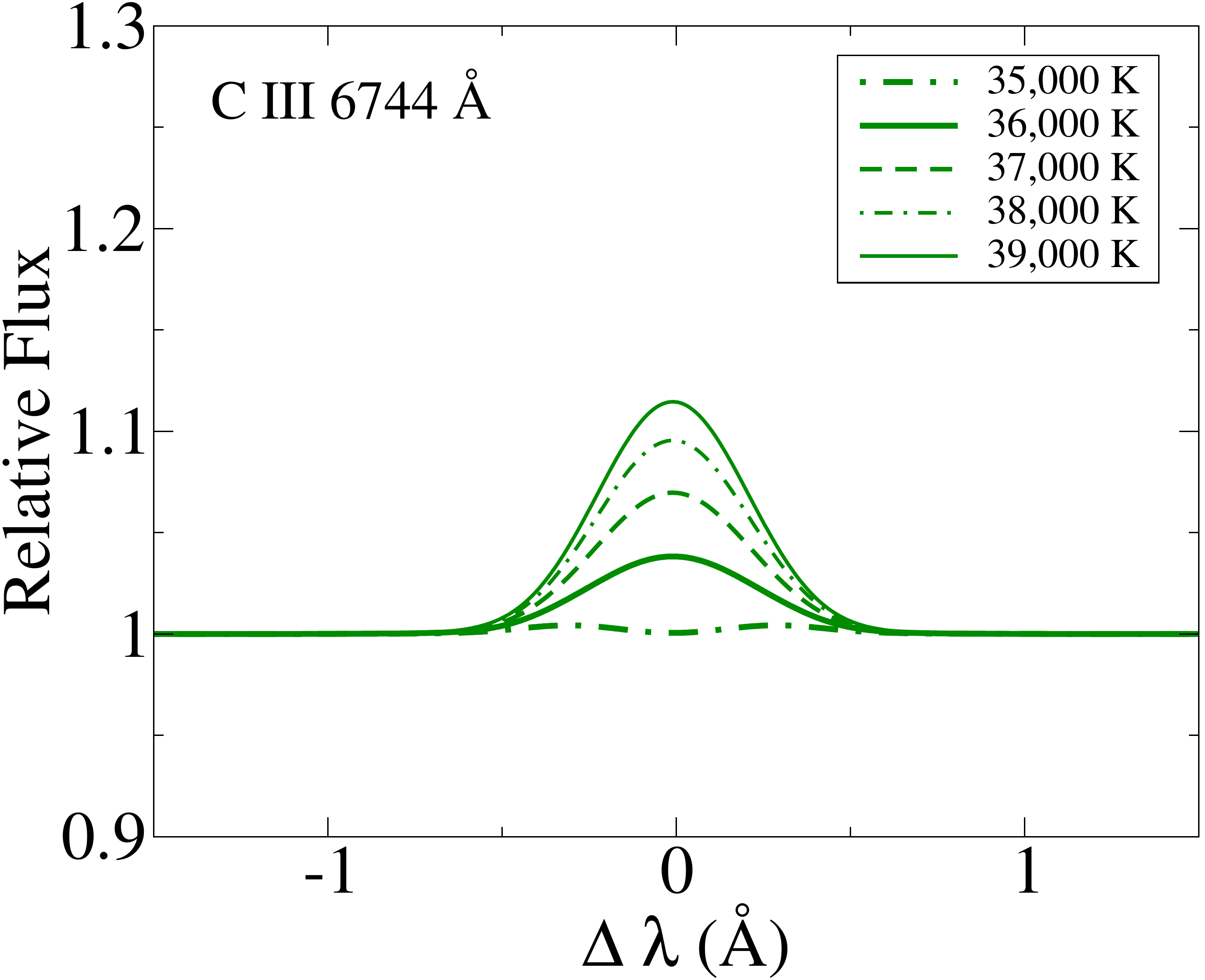}\\
 \centering}
 \hspace{1\linewidth}
 \hfill
 \\[0ex]
  \parbox{0.45\linewidth}{\includegraphics[scale=0.25]{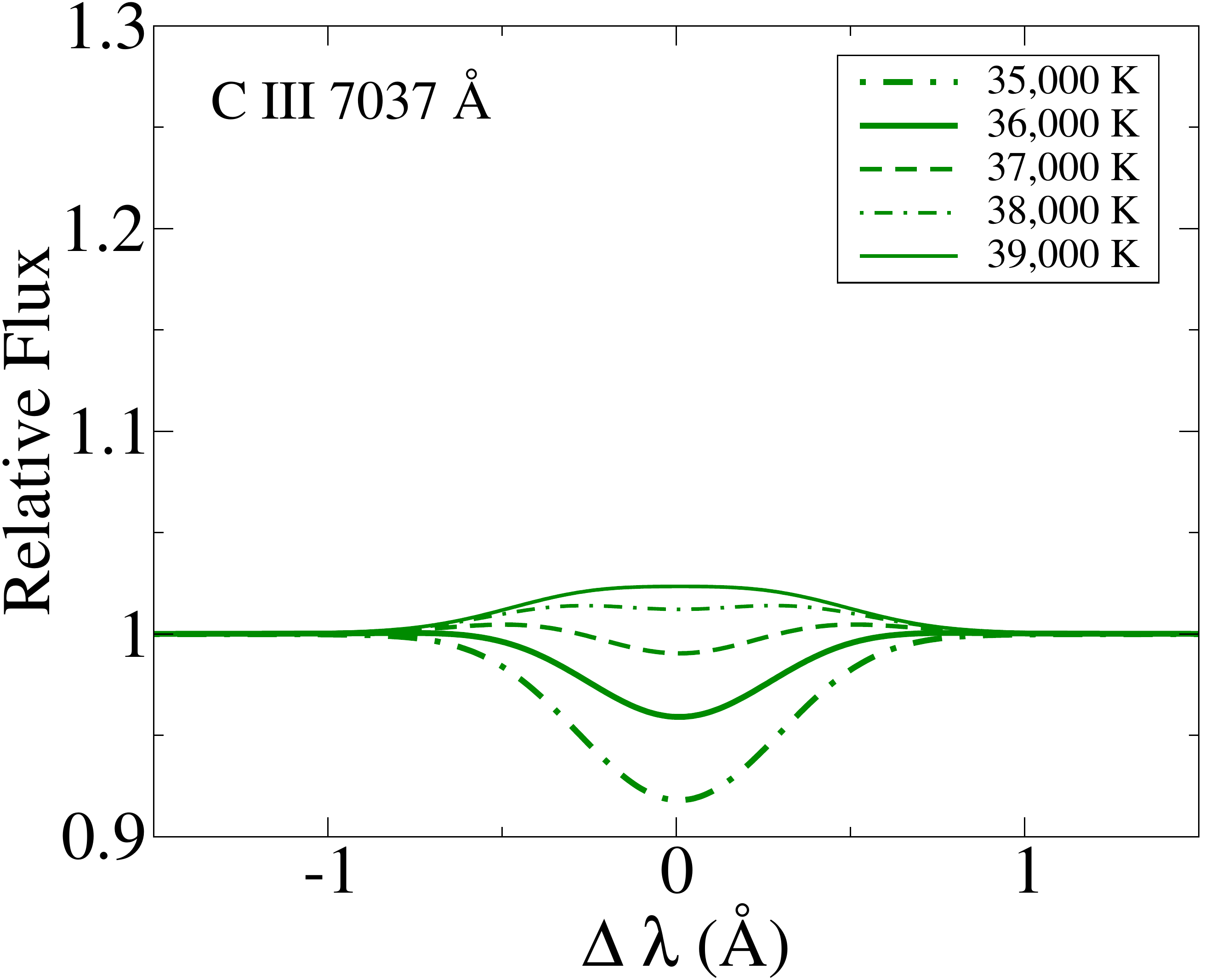}\\
 \centering}
  \parbox{0.45\linewidth}{\includegraphics[scale=0.25]{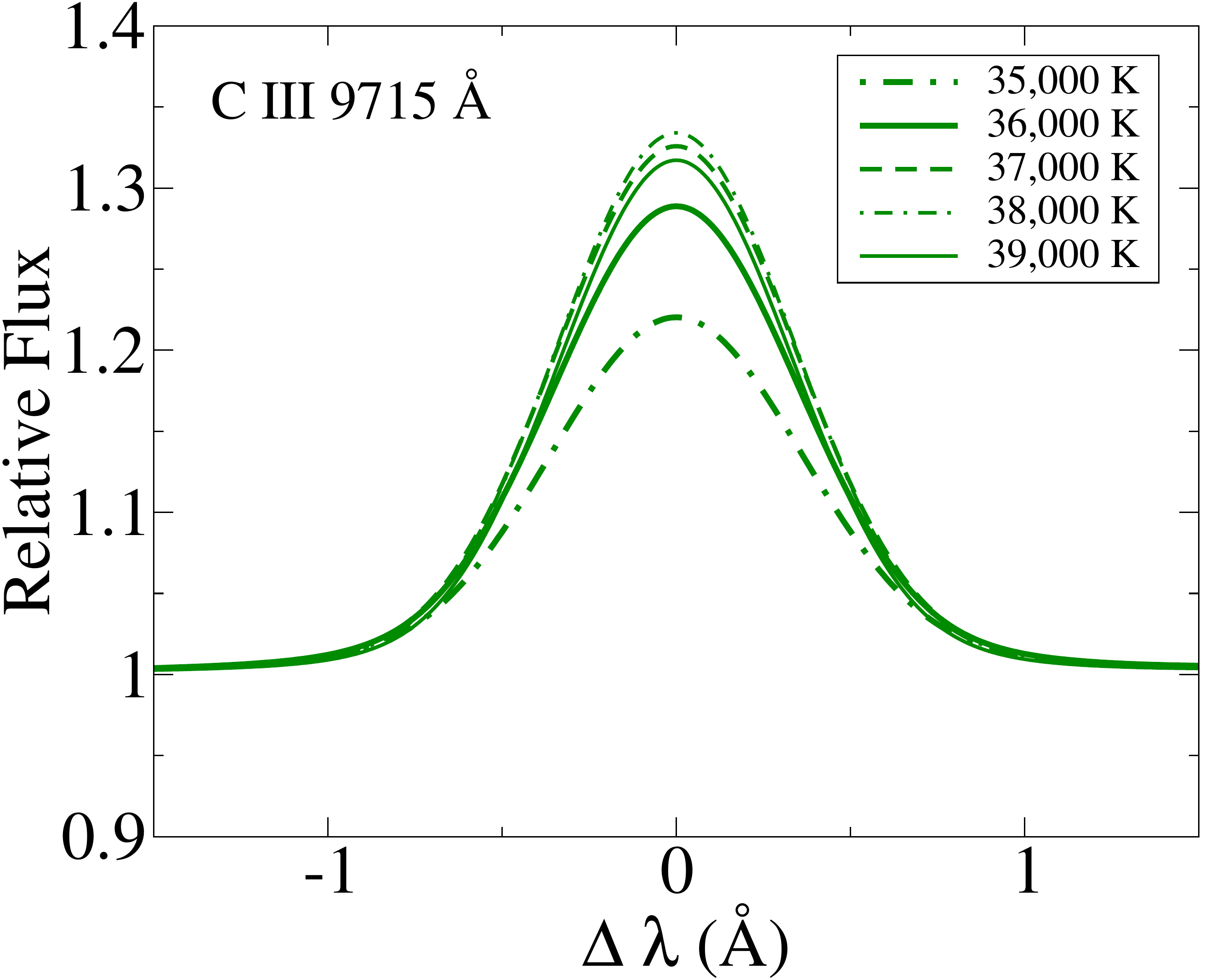}\\
 \centering}
 \hspace{1\linewidth}
 \hfill
 \\[0ex]
 \caption{Evolutions of the C\ii\ 6151, 6461, 9903, and 18\,535~\AA, and C\iii\ 5695, 6744, 7037 and 9715~\AA, line profiles with the effective temperature.
 In every cases, we assume A(C)=8.43, $v$~sin~$i$ = 10~\kms, and $\xi_t$ = 2~\kms. The theoretical spectra are convolved with an instrumental profile of R~=~50\,000.} 
 \label{param34}
 \end{center}
 \end{minipage}
 \end{figure*}

 Our NLTE calculations predict that the emission lines of C\iii\ at 5695~\AA\ (3p $^1$P$^o$ -- 3d $^1$D) and those of 6727--6762~\AA\ (3s $^3$P$^o$ -- 3p $^3$D) 
and 9705--9717~\AA\ (3p $^3$P$^o$  -- 3d $^3$D) to 
appear at 
 effective temperature \Teff $>$ 35\,000 K (log~$g$ = 4). 
The emission line of C\iii\ 7037~\AA\ (3s $^1$P$^o$ -- 4d $^1$D) appears
 at effective temperature \Teff $>$ 37\,000 K (log~$g$ = 4) (Fig. \ref{param34}).
 The analysis of NET values and test calculations show that the photoionization-recombination mechanisms play a crucial role in emission line formation. 
 The lower levels of investigated transitions, 3p $^1$P$^o$, 3s $^3$P$^o$, and 3p $^3$P$^o$, are depopulated due to ionization, while the upper levels, 3d $^1$D, 3p $^3$D, and 3d $^3$D are populated due to recombination
 from C\iv\ reservoir. 
 Extra depopulation of the levels 3p $^1$P$^o$ and 3s $^3$P$^o$ is caused by photon losses in UV transitions 2p$^2$ $^1$D -- 3p $^1$P$^o$ (885~\AA), 2p$^2$ $^1$S -- 3p $^1$P$^o$ (1308~\AA), 
 3s $^3$S -- 3s $^3$P$^o$ (1426 -- 1428~\AA). The upper level 3d $^3$D is effectively populated due to transitions 3d $^3$D -- 3d $^3$F$^o$ (1577~\AA) and 3d $^3$D -- 4f $^3$F$^o$ (1923~\AA).   
 
 On the base of the structure of terms (Fig. \ref{Grot_C3}), the C\iii\ 9705--9717~\AA\ triplet lines are similar to the 5695~\AA\ singlet line. 
 The intensity of emission line at 9705--9717~\AA\ is higher than 5695~\AA. There are two reasons for the difference: its position in the near-IR region, where the continuum flux is lower, and the line formation region of 9705--9717~\AA\ is located closer to outer layers (Fig. \ref{DC}), where the deviations from LTE are stronger.

\section{Carbon lines in the selected stars} \label{sec:stellar}
  
\subsection{Stellar sample, Observations, and stellar parameters }  

  We analyse high resolution spectral data of 22 stars listed in Table \ref{tab_param}.
  The sample contains 20 B-type stars (ranging in spectral types from B3 to B0)
  and two O-type stars. All of these stars show relatively
  sharp spectral lines (slowly rotating). Atmospheric parameters of 
  20 B-type stars are taken from \citet{2012AA...539A.143N} or 
  from \citet{2011AA...532A...2N}. Those of 15 Mon (HD 47839) and HD 42088 are
  taken from \citet{2004AA...413..693M} and from \citet{2015AA...575A..34M}, respectively.
  
  Spectral data from the visible to near IR spectral ranges were obtained 
  with the Echelle Spectro Polarimetric Device for the Observation of Stars 
  (ESPaDOnS) \citep{2006ASPC..358..362D} attached to the 3.6 m telescope 
  of the Canada-France-Hawaii Telescope (CFHT) observatory located on the summit of Mauna Kea,
   Hawaii\footnote{http://www.cfht.hawaii.edu/Instruments/Spectroscopy/Espadons/}. 
    Observations with this spectrograph cover the region from 3690 to 10480 \AA, and we use data from 3855 to 9980 \AA\ in the present study. 
  The resolving power is  $\it R$ = 65\,000. Calibrated intensity spectral data were extracted 
  from the ESPaDOnS archive through Canadian Astronomical Data Centre (CADC).

  After averaging downloaded individual spectral data of each star, we converted the wavelength scale of spectral data of each star into the laboratory scale using measured wavelengths
  of five He~{\sc i} lines (4471.48, 4713.15, 4921.93, 5015.68, and 5875.62 \AA). In the case of O-type stars, we add measured wavelengths of two He~{\sc ii} lines (4685.68 \AA\ and 5411.52 \AA).
  Errors in the wavelength measurements are around $\pm$ 3 km s${}^{-1}$ or smaller. Re-fittings of the continuum level of each spectral order were carried out using  polynomial functions. The signal-to-noise ratios (SN) measured at the continuum near 5550 \AA\ range from 380 to 1800. Five stars among the sample show very high qualities (SN ratio higher than 1000).

 \begin{deluxetable*}{clccccccccc}
\tablecaption{Atmospheric parameters of the selected stars, SN ratio, and sources of the data. \label{tab_param}}
\tablewidth{0pt}
\tablehead{  
\colhead{Number} &\colhead{Star} & \colhead{Name} & \colhead{Sp. T.} & \colhead{ \Teff } & \colhead{log~$g$} & \colhead{$\xi_t$ }  & \colhead{ $V_{mac}$ }  & \colhead{$v$~sin~$i$} & \colhead{SN} & \colhead{Ref.} \\
\colhead{} &\colhead{} & \colhead{} & \colhead{} & \colhead{ K} & \colhead{CGS} & \colhead{\kms } & \colhead{ \kms } & \colhead{ \kms }   & \colhead{} & \colhead{} 
}
\decimalcolnumbers
\startdata
1  & HD~209008 & 18~Peg              & B3 III    &  15\,800  &  3.75   & 4   &  10 & 15       &   710  &  1   \\      
2  & HD~160762 & $\iota$~Her          & B3 IV     &  17\,500  &  3.80   & 1   &     & 6        &   1820 &  1   \\    
3  & HD~35912  & HR~1820             & B2-3 V    &  19\,000  &  4.00   & 2   &  8  & 15       &   650  &  2   \\    
4  & HD~36629  & HIP~26000           & B2 V      &  20\,300  &  4.15   & 2   &  5  & 10       &   630  &  2   \\   
5  & HD~35708  & $o$~Tau             & B2.5 IV   &  20\,700  &  4.15   & 2   &  17 & 25       &   820  &  1    \\   
6  & HD~3360   & $\zeta$ Cas         & B2 IV     &  20\,750  &  3.80   & 2   &  12 & 20       &   1570 &  1   \\    
7  & HD~122980 & $\chi$ Cen          & B2 V      &  20\,800  &  4.22   & 3   &     & 18       &   520  &  1    \\   
8  & HD~16582  & $\delta$ Cet        & B2 IV     &  21\,250  &  3.80   & 2   &  10 & 15       &   1050 &  1    \\   
9  & HD~29248  & $\nu$ Eri           & B2 II     &  22\,000  &  3.85   & 6   &  15 & 26       &   550  &  1    \\                     
10 & HD~886    & $\gamma$ Peg        & B2 IV     &  22\,000  &  3.95   & 2   &  8  & 9        &   1110 &  1   \\       
11 & HD~74575  & $\alpha$ Pix        & B1.5 III  &  22\,900  &  3.60   & 5   &  20 & 11       &   800  &  1   \\      
12 & HD~35299  & HR 1781             & B2 III    &  23\,500  &  4.20   & 0   &     & 8        &   380  &  1   \\      
13 & HD~36959  & HR 1886             & B1 V  Ic  &  26\,100  &  4.25   & 0   &  5  & 12       &   860  &  2   \\                  
14 & HD~61068  & HR 2928             & B2 II     &  26\,300  &  4.15   & 3   &  20 & 14       &   560  &  1   \\      
15 & HD~36591  & HR 1861             & B2 III    &  27\,000  &  4.12   & 3   &     & 12       &   720  &  1  \\                
16 & HD~37042  & $\theta^2$ Ori B    & B0.5 V    &  29\,300  &  4.30   & 2   &  10 & 30       &   690  &  2  \\    
17 & HD~36822  & $\phi^1$ Ori        & B0.5 III  &  30\,000  &  4.05   & 8   &  18 & 28       &   570  &  1  \\    
18 & HD~34816  & $\lambda$ Lep       & B0.5 V    &  30\,400  &  4.30   & 4   &  20 & 30       &   550  &  1   \\   
19 & HD~149438 & $\tau$ Sco          & B0.2 V    &  32\,000  &  4.30   & 5   &  4  & 4        &   1730 &  1    \\   
20 & HD~36512  & $\upsilon$ Ori      & B0 V      &  33\,400  &  4.30   & 4   &  10 & 20       &   700  &  1   \\   
21 & HD~47839  & 15~Mon              & O7 V      &  37\,500  &  3.84   & 12  &  65$^*$ & 62   &   560  &  3  \\   
22 & HD~42088  &                     & O6 V      &  38\,000  &  4.00   & 12$^{**}$  &  37& 41 &   820  &  4  \\   
\enddata                                                
\tablecomments{{\bf Note.} References: 
  1 \citet{2012AA...539A.143N}; 2 \citet{2011AA...532A...2N}; 3 \citet{2004AA...413..693M}; 4 \citet{2015AA...575A..34M}.
  $^*$ \citet{2017AA...597A..22S} ; $^{**}$ \citet{2018AA...613A..65H}.
 Spectral types are extracted from the SIMBAD database. (10) Measured SN ratio near 5550 \AA. }
\end{deluxetable*}

  \begin{figure*}
 \begin{center}
 \includegraphics[scale=0.5]{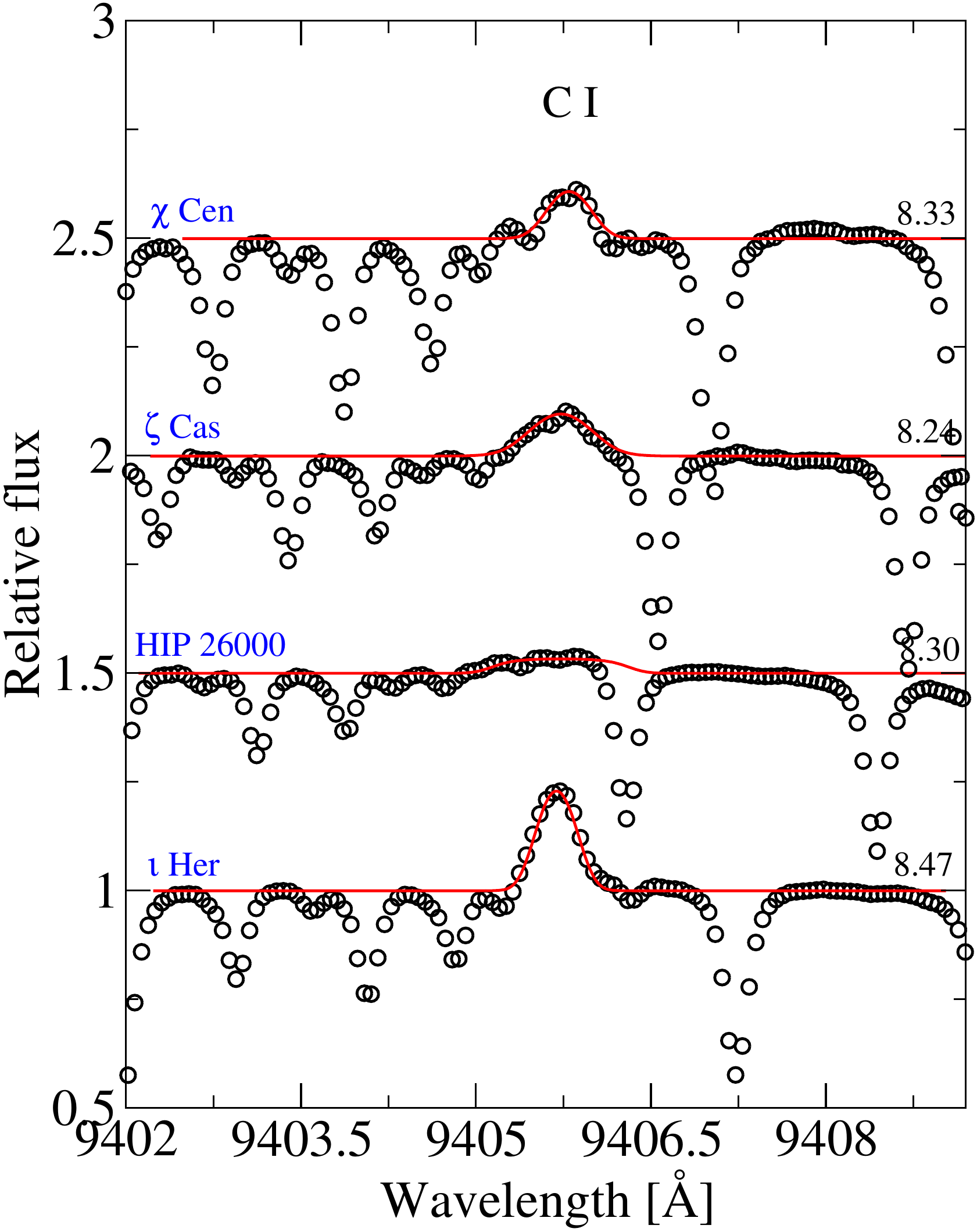}
 \caption{Best NLTE fits (solid curves) of C\ione\ 9405~\AA\ in $\iota$~Her, HIP 26000, $\zeta$ Cas, and $\chi$ Cen. The observed spectra are shown by open circles. }
 \label{9405}
 \end{center}
 \end{figure*}

  \begin{figure*}
  \begin{minipage}{170mm}
  \parbox{0.24\linewidth}{\includegraphics[scale=0.275]{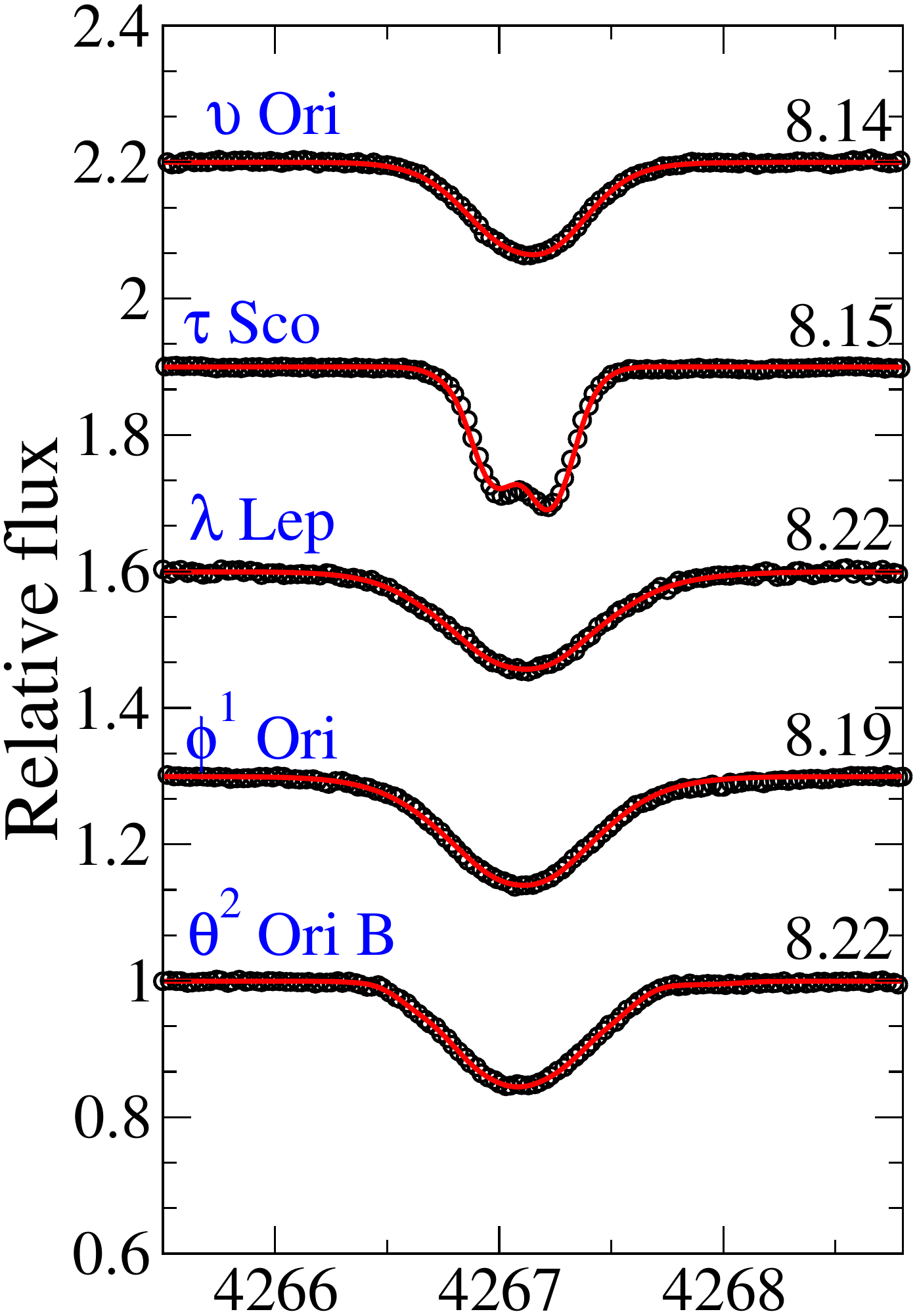}\\
  \centering}
  \parbox{0.24\linewidth}{\includegraphics[scale=0.275]{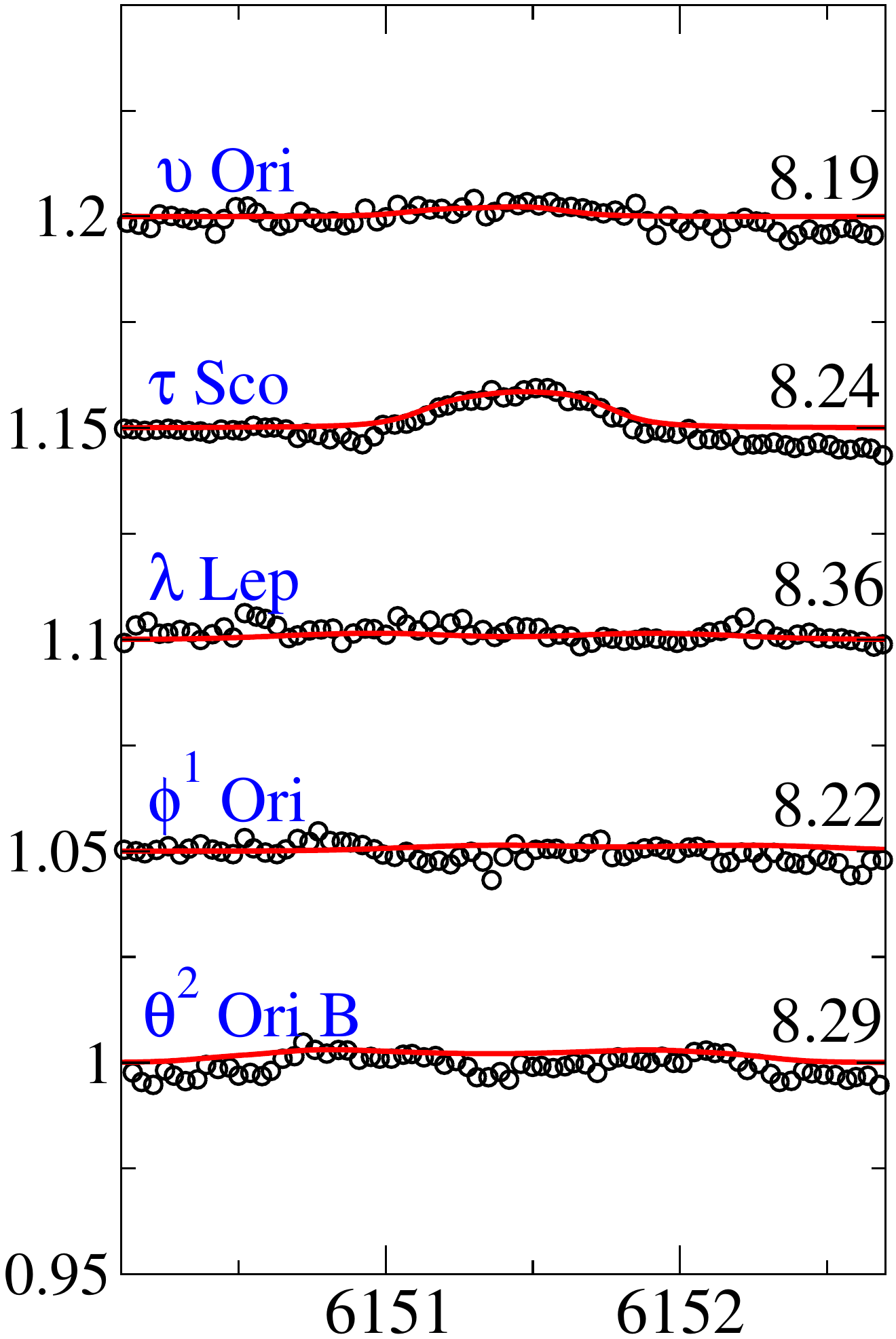}\\
  \centering}
  \parbox{0.24\linewidth}{\includegraphics[scale=0.275]{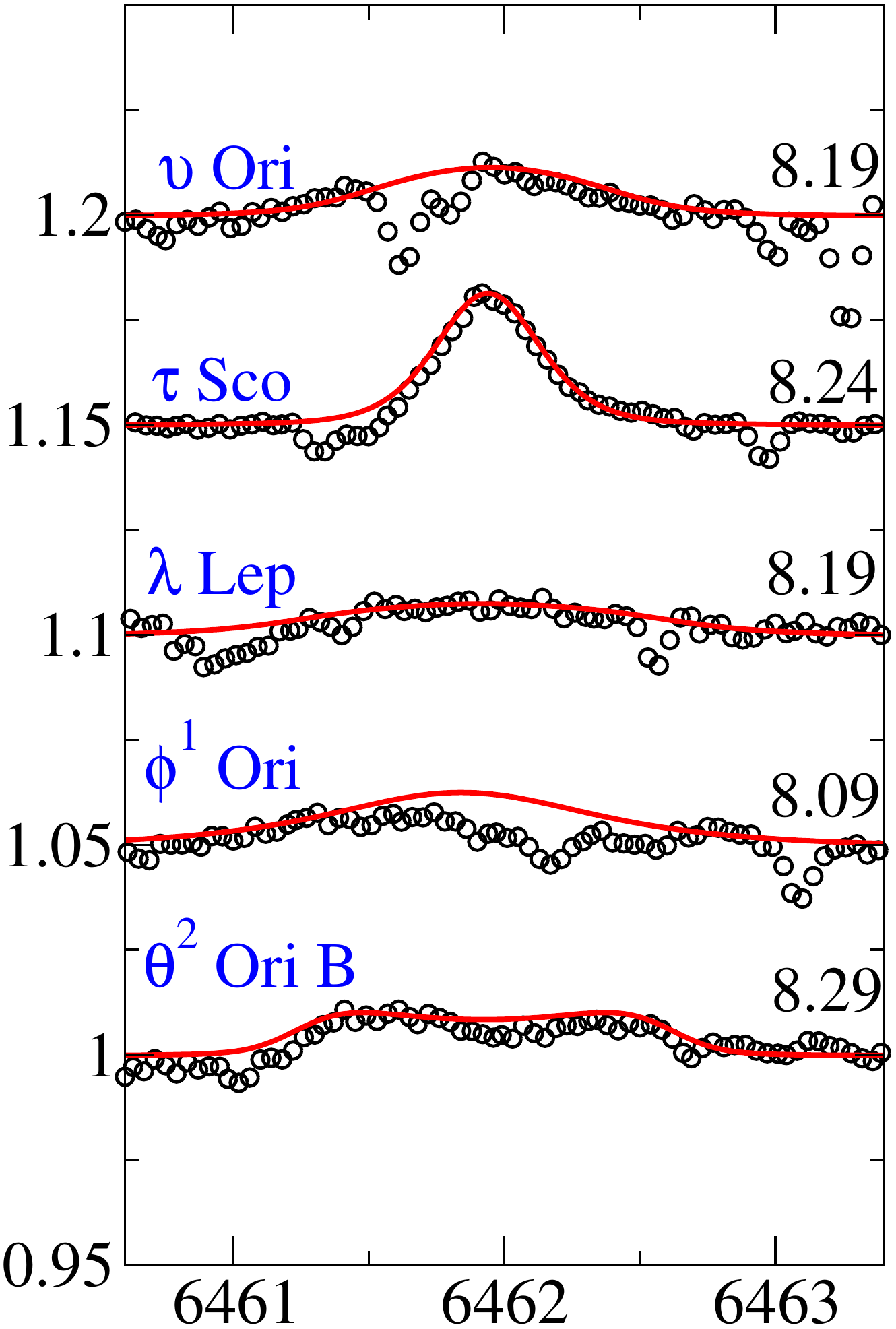}\\
  \centering}
  \parbox{0.24\linewidth}{\includegraphics[scale=0.275]{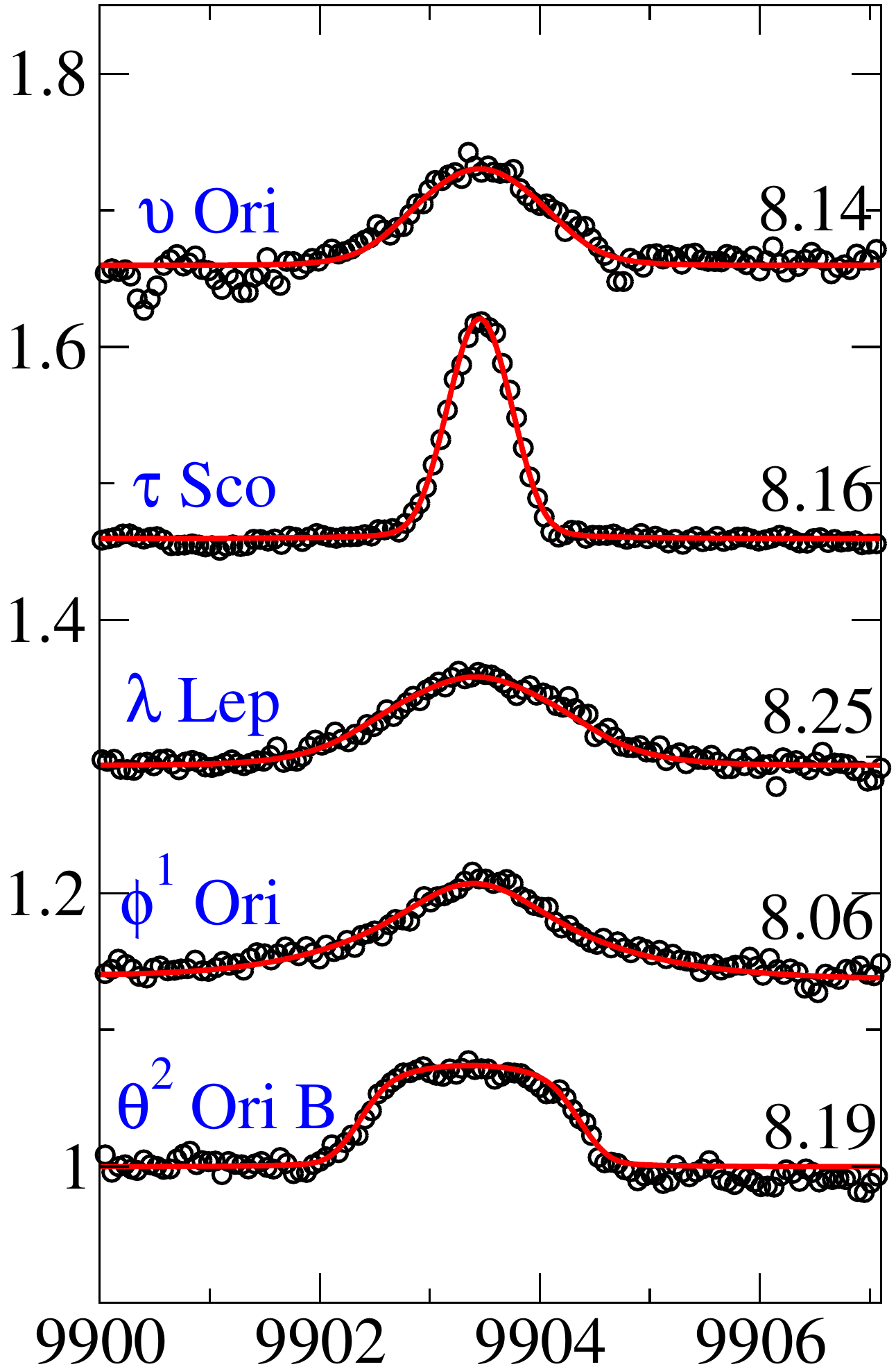}\\
  \centering}
  \hfill
   \\[0ex]
  \parbox{0.24\linewidth}{\includegraphics[scale=0.275]{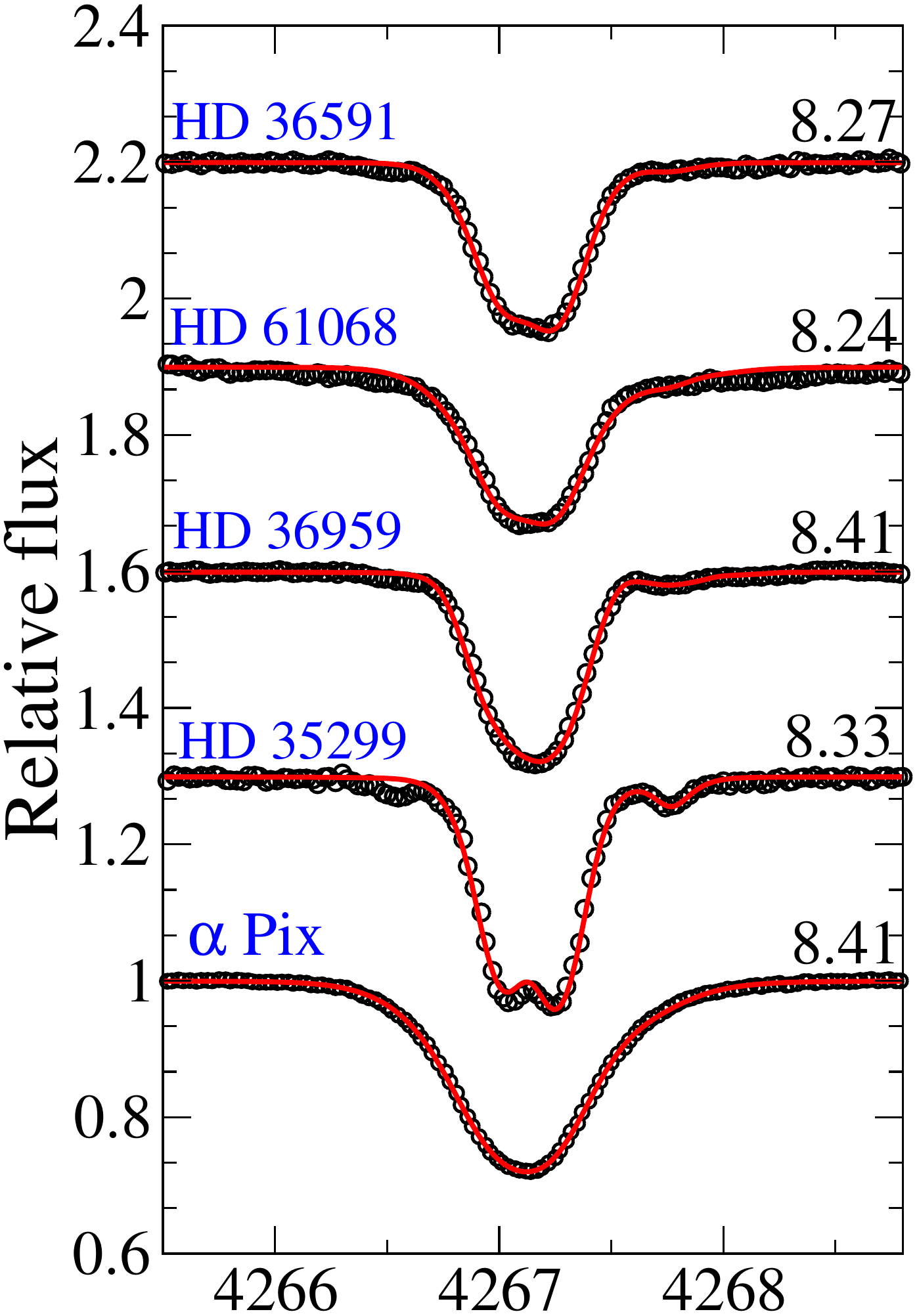}\\
  \centering}
  \parbox{0.24\linewidth}{\includegraphics[scale=0.275]{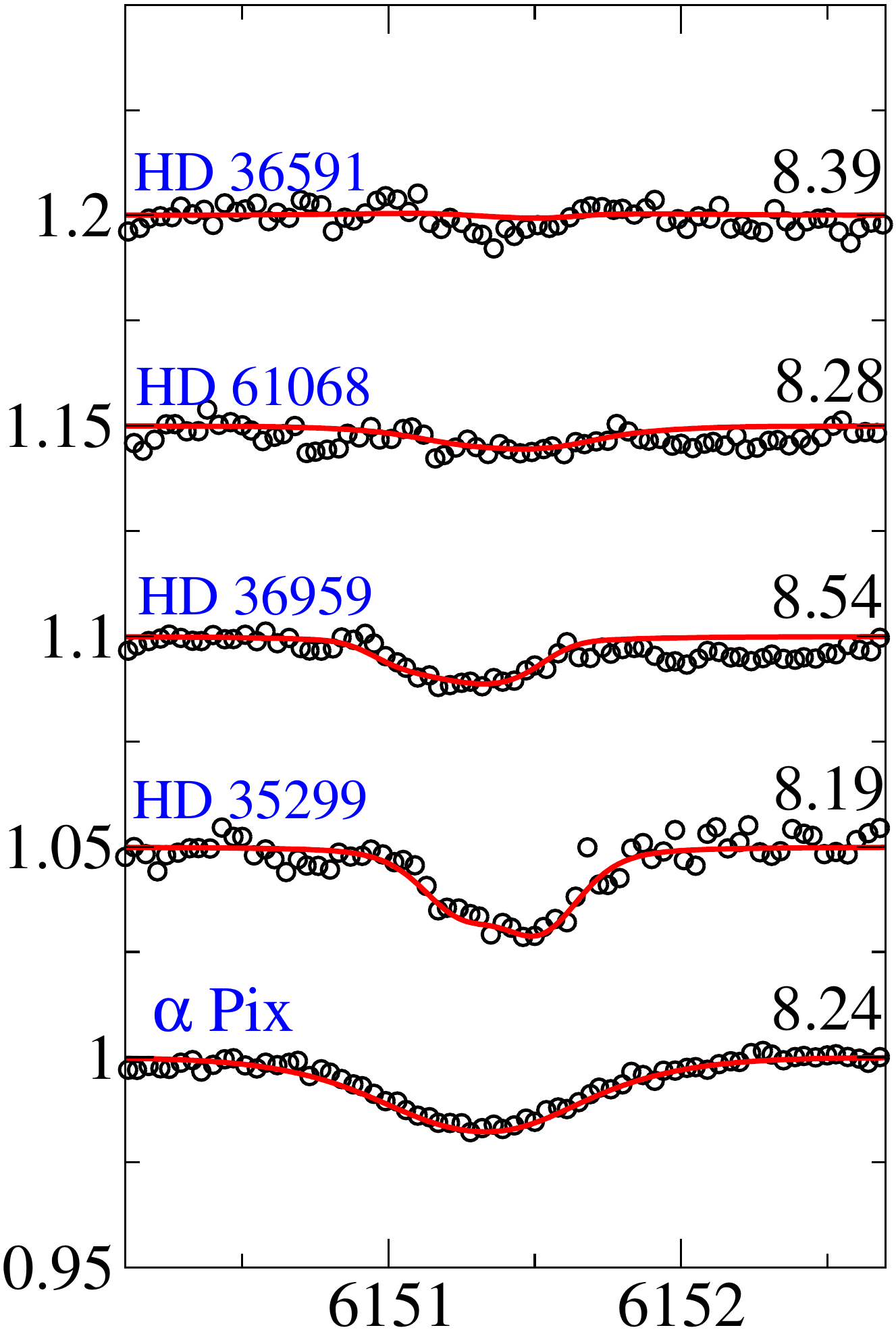}\\
  \centering}
  \parbox{0.24\linewidth}{\includegraphics[scale=0.275]{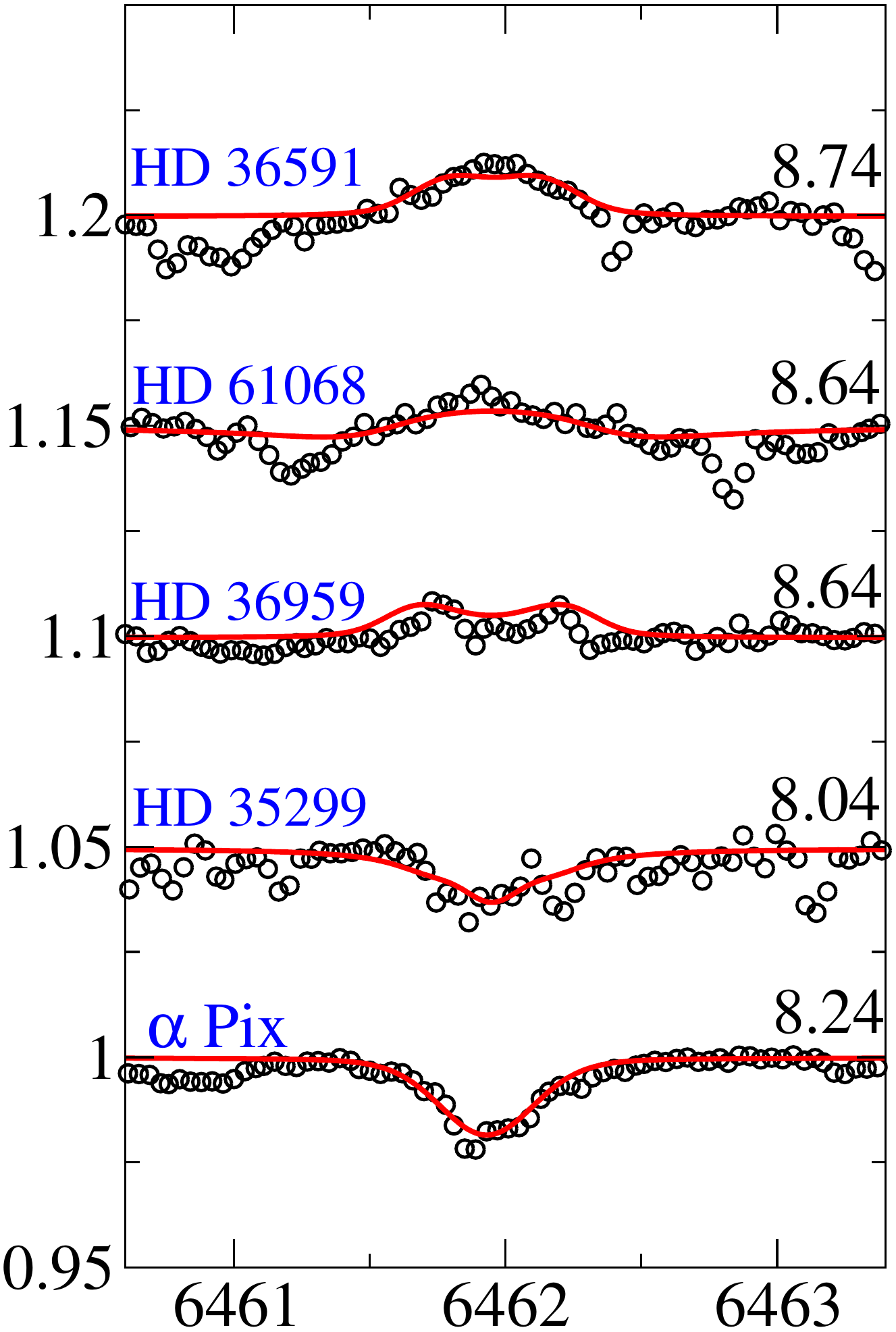}\\
  \centering}
  \parbox{0.24\linewidth}{\includegraphics[scale=0.275]{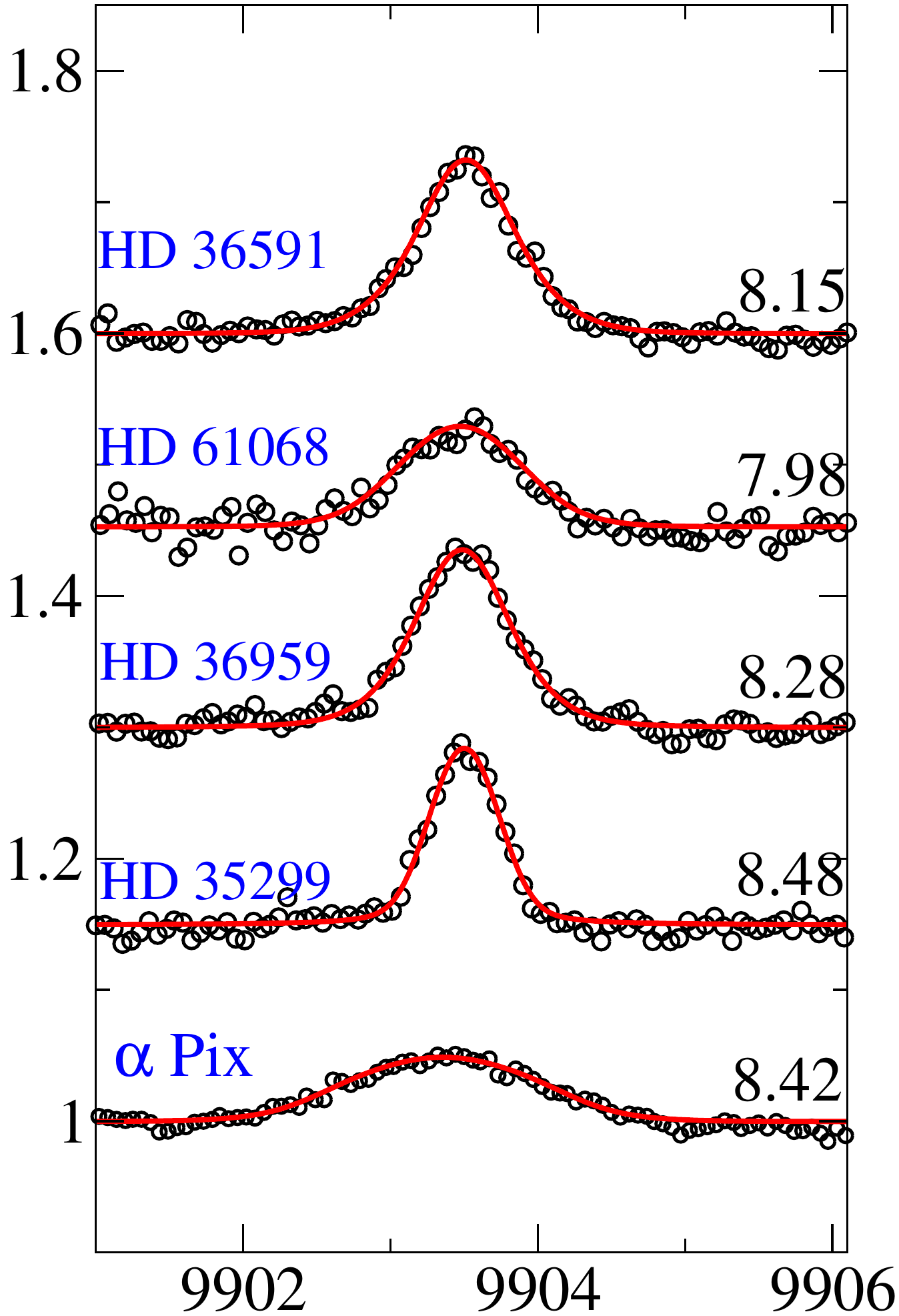}\\
  \centering}
  \hfill
  \\[0ex]
   \parbox{0.24\linewidth}{\includegraphics[scale=0.275]{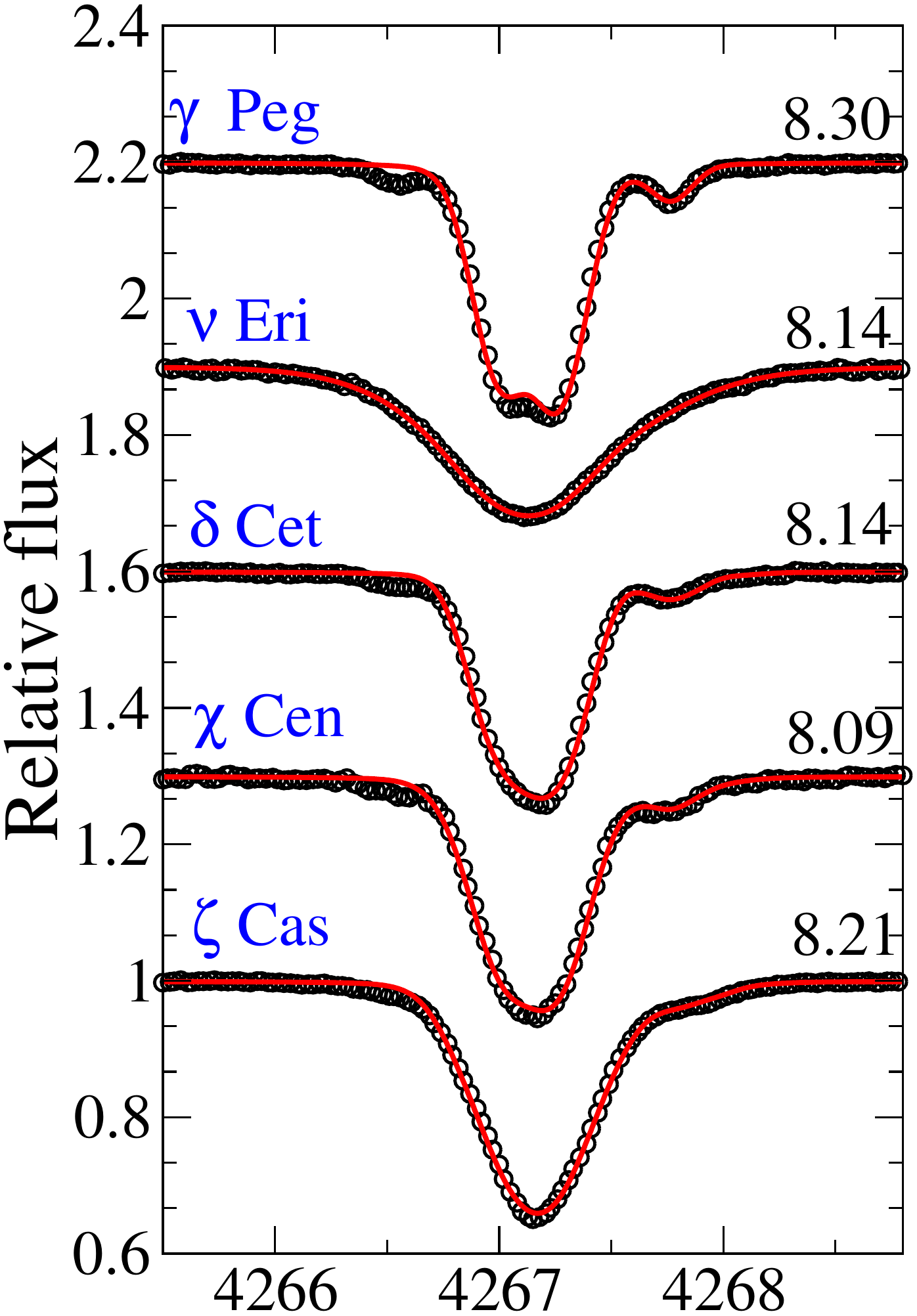}\\
  \centering}
  \parbox{0.24\linewidth}{\includegraphics[scale=0.275]{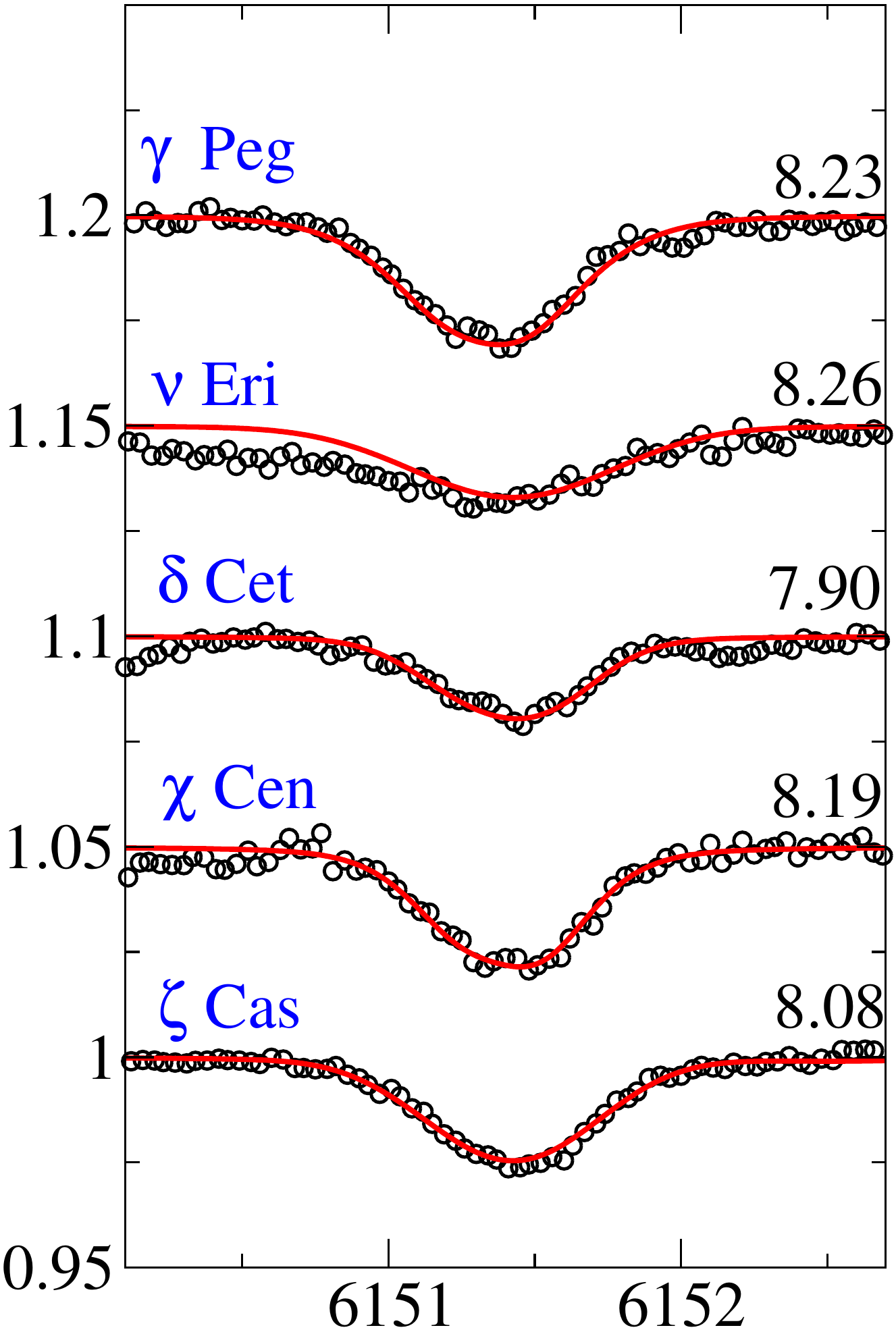}\\
  \centering}
  \parbox{0.24\linewidth}{\includegraphics[scale=0.275]{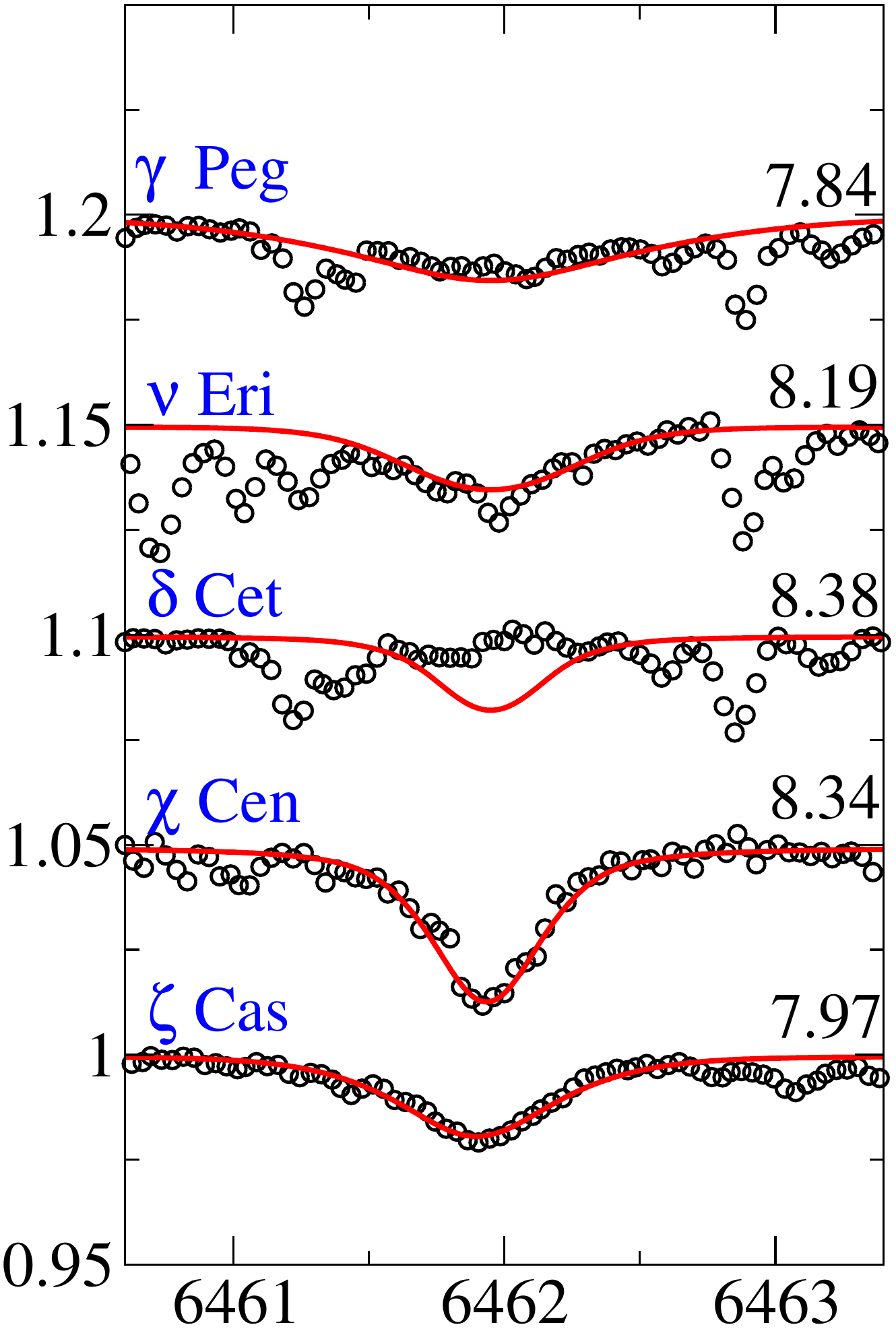}\\
  \centering}
  \parbox{0.24\linewidth}{\includegraphics[scale=0.275]{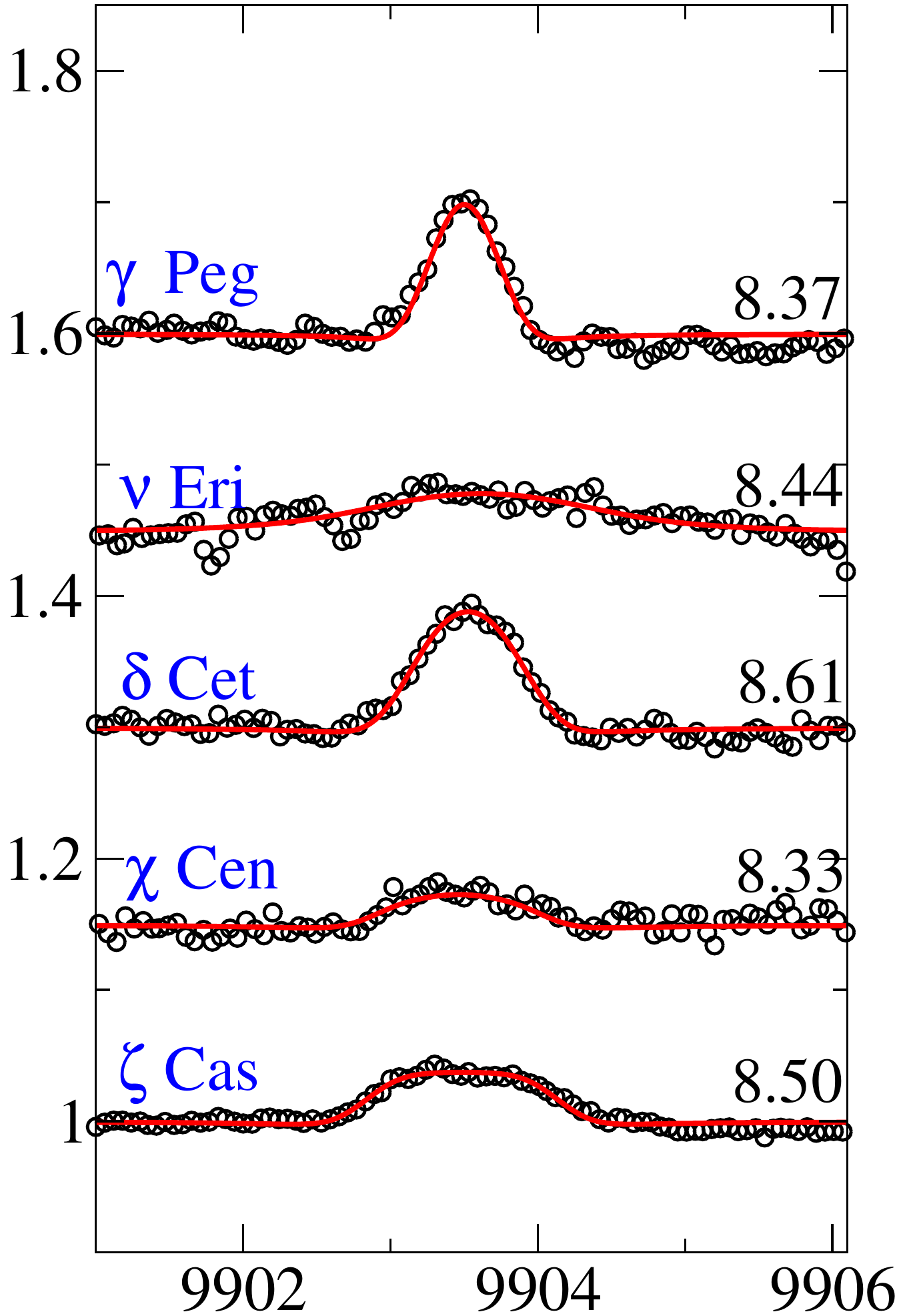}\\
  \centering}
  \hfill
  \\[0ex]
  \parbox{0.24\linewidth}{\includegraphics[scale=0.275]{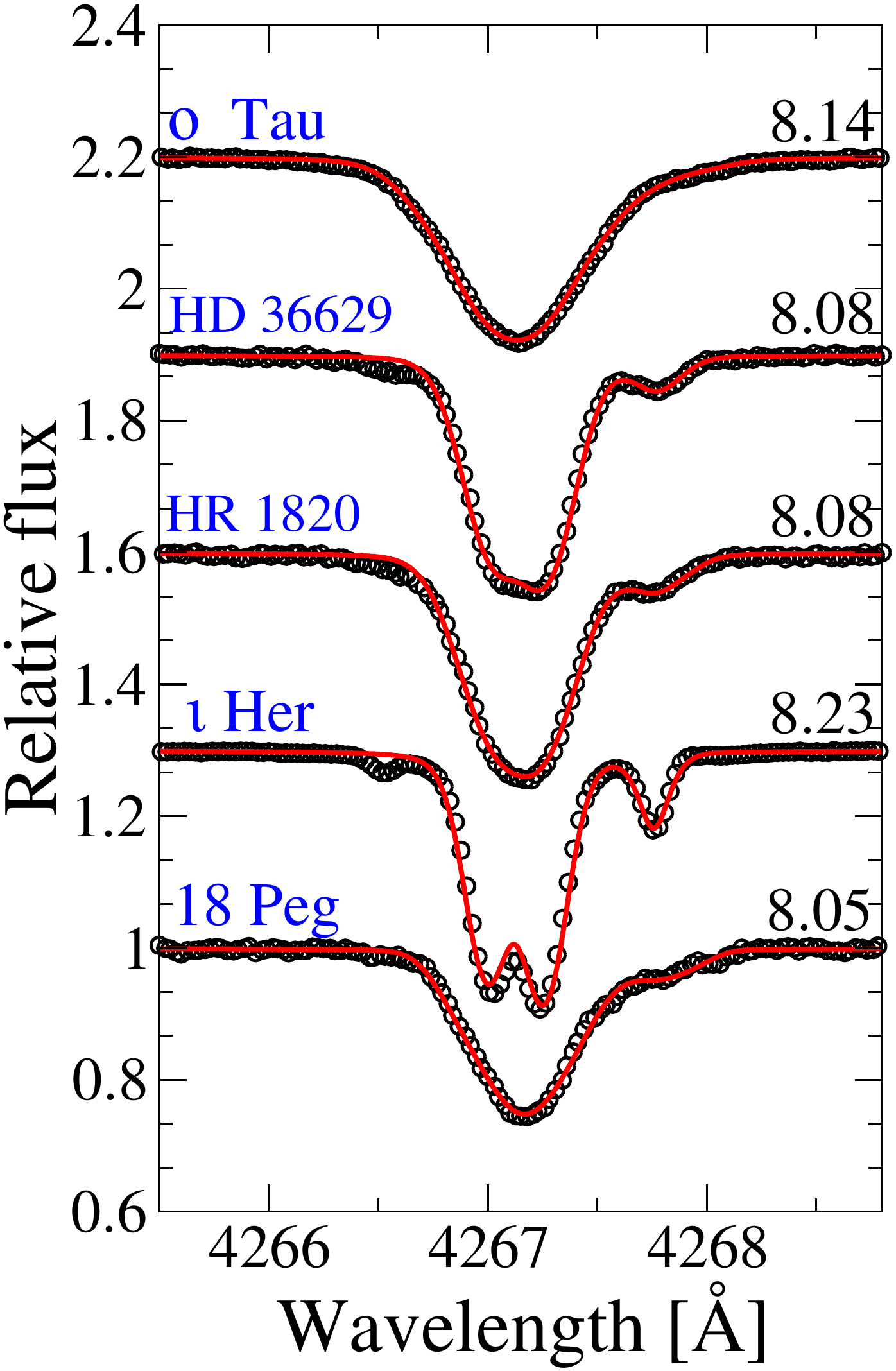}\\
  \centering}
  \parbox{0.24\linewidth}{\includegraphics[scale=0.275]{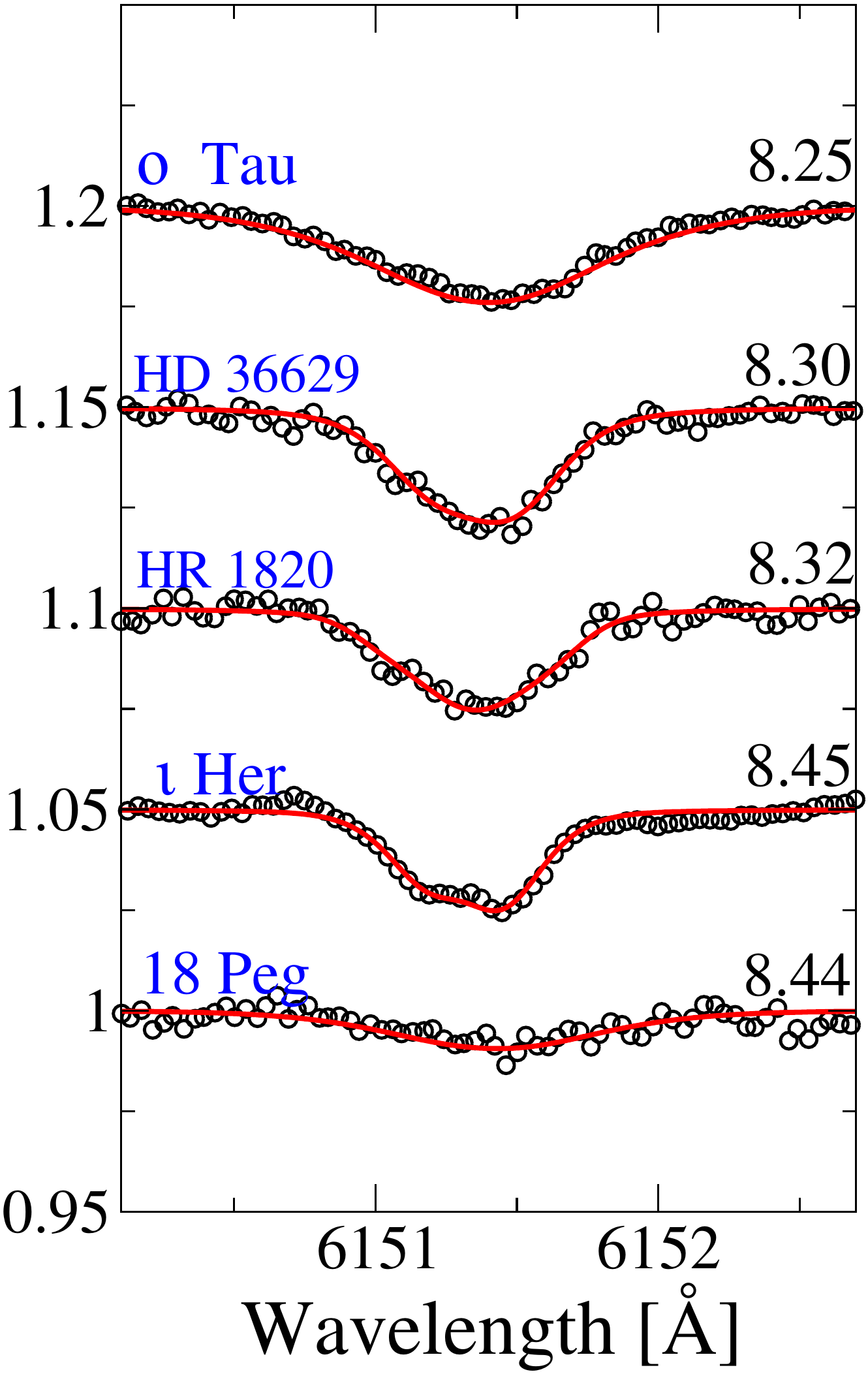}\\
  \centering}
  \parbox{0.24\linewidth}{\includegraphics[scale=0.275]{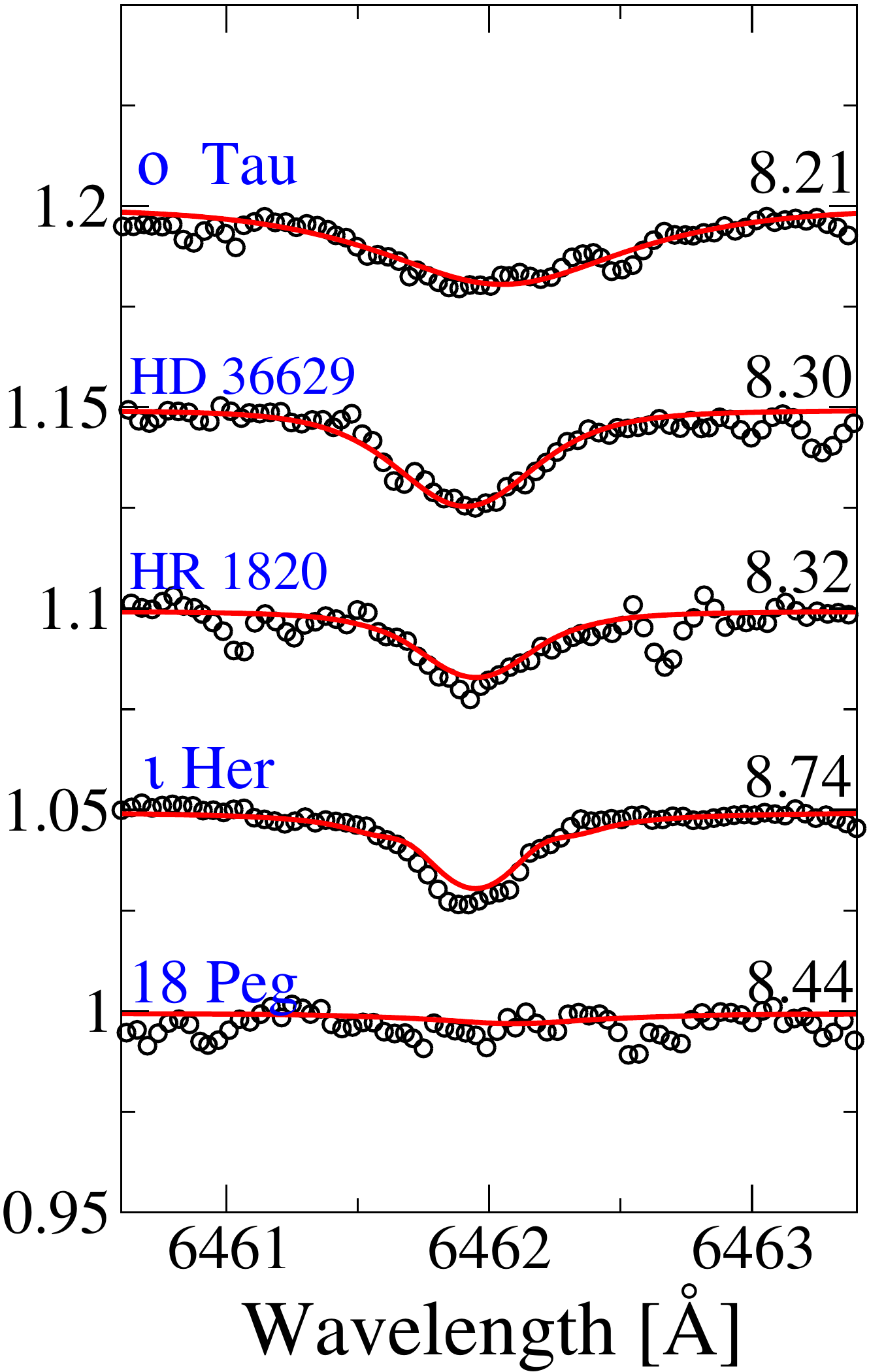}\\
  \centering}
  \parbox{0.24\linewidth}{\includegraphics[scale=0.275]{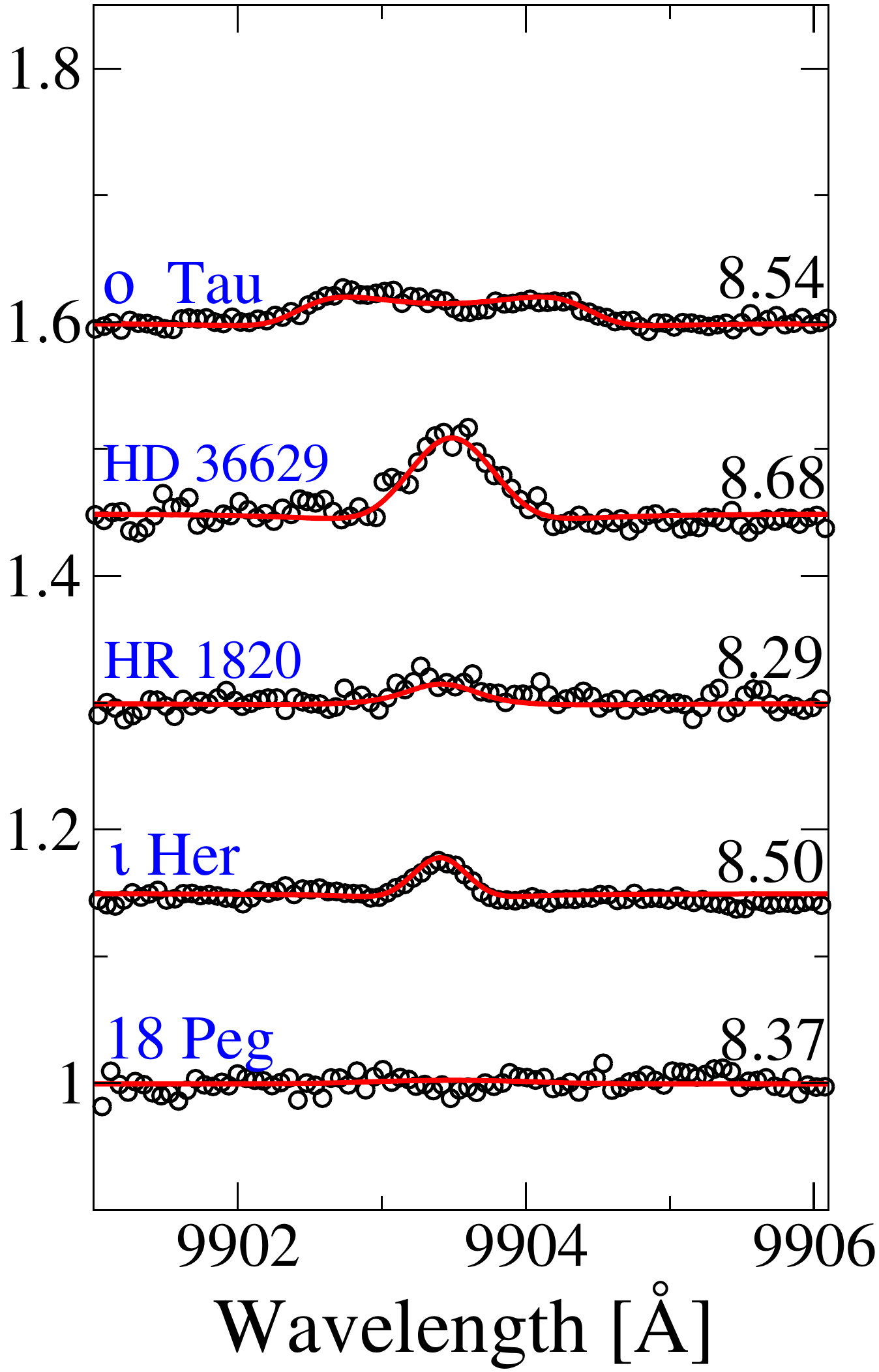}\\
  \centering}
  \hfill
  \\[0ex]
  \caption{Best NLTE fits (solid curves) of C\ii\ 4267, 6151, 6461, and 9903 \AA\ in the program stars (from left to right). The observed spectra are shown by open circles. }
  \label{pics}
  \end{minipage}
  \end{figure*}

\subsection{Determination of carbon abundances in B-stars}\label{Sect:Stars}

  In this section, we derive the carbon abundances of the selected 
  B-stars from emission C\ione, emission and absorption C\ii\, and absorption C\iii\ lines using the atomic data from Table~\ref{tab1}.
  In Table~\ref{tabhot}, we show the NLTE and LTE abundances derived from absorption and emission lines by the spectrum fitting method. 
  Emission lines are marked by $e$ in the corresponding LTE columns.
  The averaged NLTE abundances obtained from emission and absorption lines of four ionization stages are shown in Table~\ref{Final}.
  
  Emission lines of C\ione\ were detected in the observed spectra of the five stars: $\iota$~Her, HR~1820, HIP~26000, $\zeta$~Cas, and $\chi$~Cen.
  In the spectra of 18~Peg and $o$ Tau, the C\ione\ lines were not used, because they are blended with telluric lines located in near-IR spectral region. 
  All the C\ione\ lines almost disappear in the spectra of stars with \Teff\ higher than 22\,000~K, because carbon become predominantly singly- or doubly ionized (Fig.~\ref{balance}). 
  For each star, our NLTE calculations reproduce well the observed C\ione\ emission lines, using our model atom and adopted atmospheric parameters. The best NLTE fits of the C\ione\ 9405~\AA\ emission lines in $\iota$~Her, HIP 26000, $\zeta$ Cas, and $\chi$ Cen are presented in Fig.~\ref{9405}.
  
  The C\ii\ emission line at 9903~\AA\ was detected in the observed spectra of all B-type stars in our sample. The hotter stars, HR 1886, HR 2928, HR 1861, $\theta^2$ Ori B, $\phi^1$ Ori, $\lambda$ Lep, $\tau$ Sco, and $\upsilon$ Ori, also have the emission lines at 6461~\AA\ in their spectra, and the five B-type stars show the emission lines at 6151~\AA, too.
  The best fits of C\ii\ 4267, 6151, 6461, and 9903 \AA, are presented in Figure~\ref{pics}. For each star, our NLTE calculations reproduce satisfactorily the observed C\ii\ 6151, 6461 \AA\ absorption and emission lines. In four stars, 
  $\iota$~Her, $\delta$~Cet, $\gamma$~Peg, and HR~1861, we excluded the abundances obtained from C\ii\ 6151 and 6461~\AA\ lines in the mean value, because these two lines give apparently discrepant
abundances among other lines.
  It should be noticed that these two lines are sensitive to the choice of threshold photoionization cross sections, that were corrected one order of magnitude higher
  for the upper levels, 6f $^2$F$^o$ and 6g $^2$G, of investigated transitions.   
  
   The C\iii\ 8500~\AA\ line should be considered with caution, because it is lying almost in the core of the broad Paschen P16 line at 8502.5~\AA, which also was included in spectral synthesis. We analysed this line in seven stars and the derived abundances were found to be consistent with other C\iii\ lines. 
  
  For each star, NLTE provides smaller line-to-line scatter that is illustrated in Fig.~\ref{CII_CIII_Abn} (left panel).
  The ionization equilibrium C\ione/C\ii/C\iii\ was achieved for the four stars, and C\ii/C\iii\ for eleven stars of our sample.   
  For two stars, $\tau$ Sco and $\upsilon$ Ori, we obtained abundances from three ionization stages, C\ii/C\iii/C\iv. 
   C\ii/C\iii\ ionization equilibrium is achieved in $\tau$ Sco, however the abundance from C\iv\ lines is lower, although still consistent within 2$\sigma$. 
  C\ii/C\iv\ ionization equilibrium is achieved in $\upsilon$ Ori, however the abundance from C\iii\ lines gives a higher abundance than that from the C\ii\ and C\iv\ lines, by 0.23 and 0.22~dex, respectively. 
  In both stars, the abundances from C\iii\ lines are higher compared to C\ii\ and C\iv. 
  For $\phi^1$ Ori, $\lambda$ Lep, $\tau$ Sco the NLTE abundances derived from the C\ii\ and C\iii\ lines agree within the error bars, 
  while in LTE, the abundance differences can reach up to 0.72 dex.

   We obtained an averaged abundance of carbon log~$\epsilon_{\rm C}$=8.36$\pm$0.08 from twenty B-type stars from all lines listed in Table \ref{tabhot}. 
  The resulting value is lower by $\sim$0.07 dex compared to the solar carbon NLTE abundance, log~$\epsilon_{\rm C}$=8.43, which are obtained with this model atom from C\ione\ atomic lines in \citet{2015MNRAS.453.1619A}.
  Our result is in line with the present-day Cosmic Abundance Standard \citep{2012AA...539A.143N}, where log~$\epsilon_{\rm C}$=8.33$\pm$0.04 is presented.
  
  \subsection{Determination of carbon abundances in O-stars}\label{Sect:StarsO} 
  
\begin{figure*}                                                                                    
   \begin{minipage}{190mm}
 \begin{center}
 \parbox{0.38\linewidth}{\includegraphics[scale=0.25]{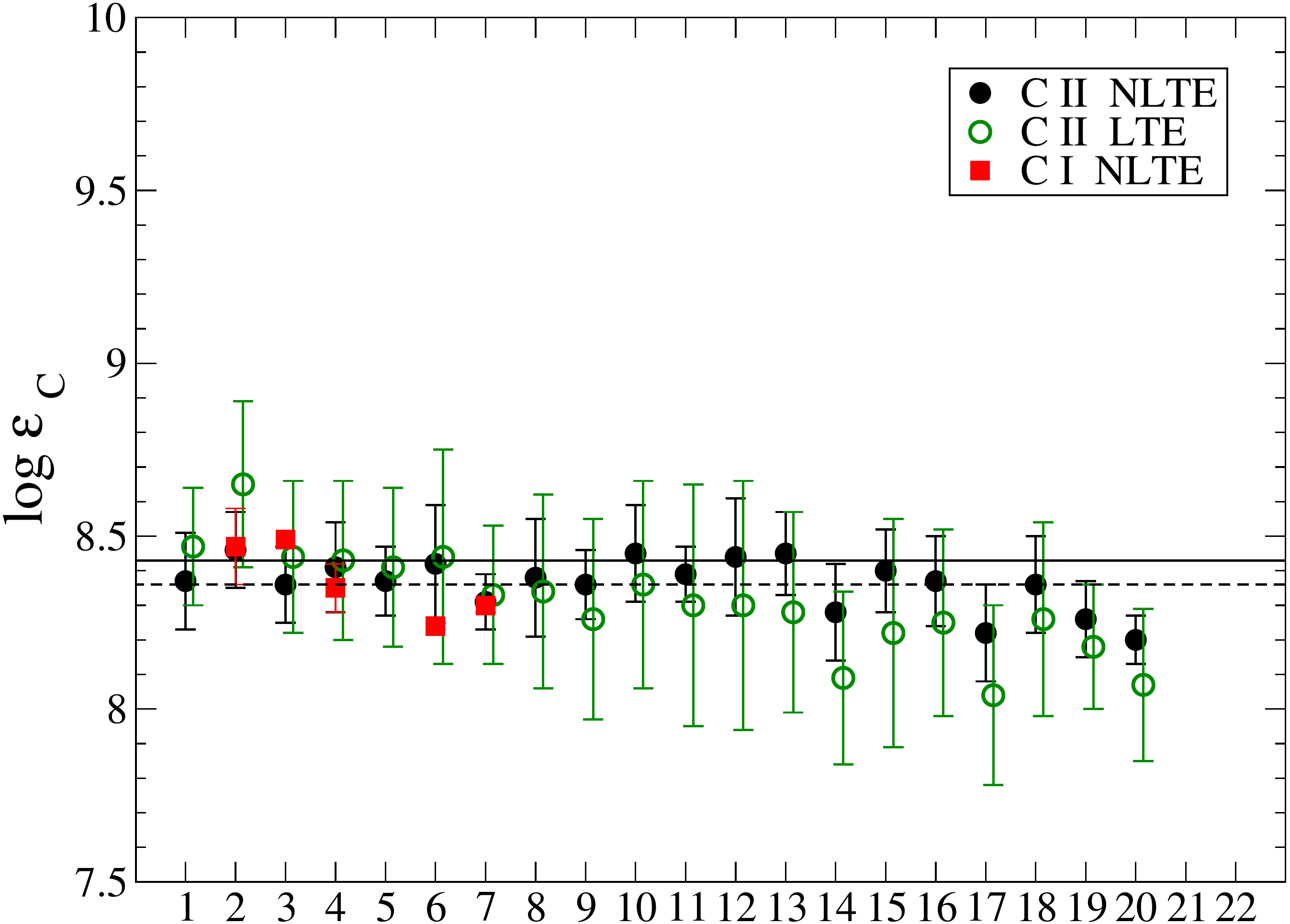}\\
 \centering}
 \parbox{0.38\linewidth}{\includegraphics[scale=0.25]{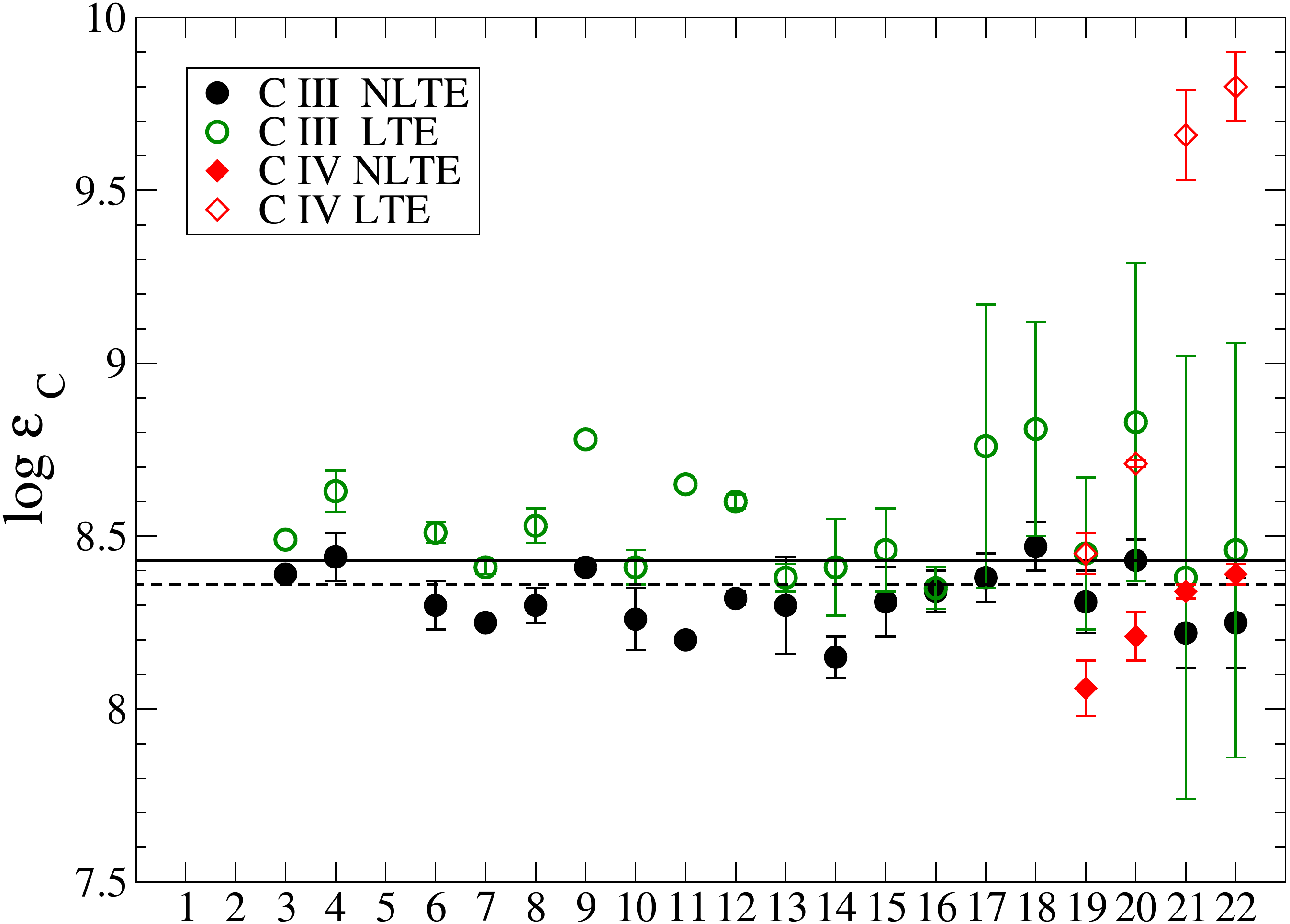}\\
 \centering}
 \hspace{1\linewidth}
 \hfill
 \\[0ex]
 \caption{Carbon LTE and NLTE abundances of
program stars derived from lines of C\ione, C\ii\ (left panel) and C\iii, C\iv\ (right panel). The horizontal axis corresponds to the serial number of star, according to Table~\ref{tab_param}.
The error bars correspond to the dispersion in the single-line measurements about the mean. In each panel, the dashed line represents the
NLTE abundance averaged over all  C\ione, C\ii, C\iii, and C\iv\ lines, log~$\epsilon_{\rm C}$=8.36. The solid line represents the solar value, log~$\epsilon_{\rm C}$=8.43. } 
 \label{CII_CIII_Abn}
 \end{center}
 \end{minipage}
 \end{figure*}

\begin{figure*}
   \begin{minipage}{190mm}
 \begin{center}
 \parbox{0.24\linewidth}{\includegraphics[scale=0.19]{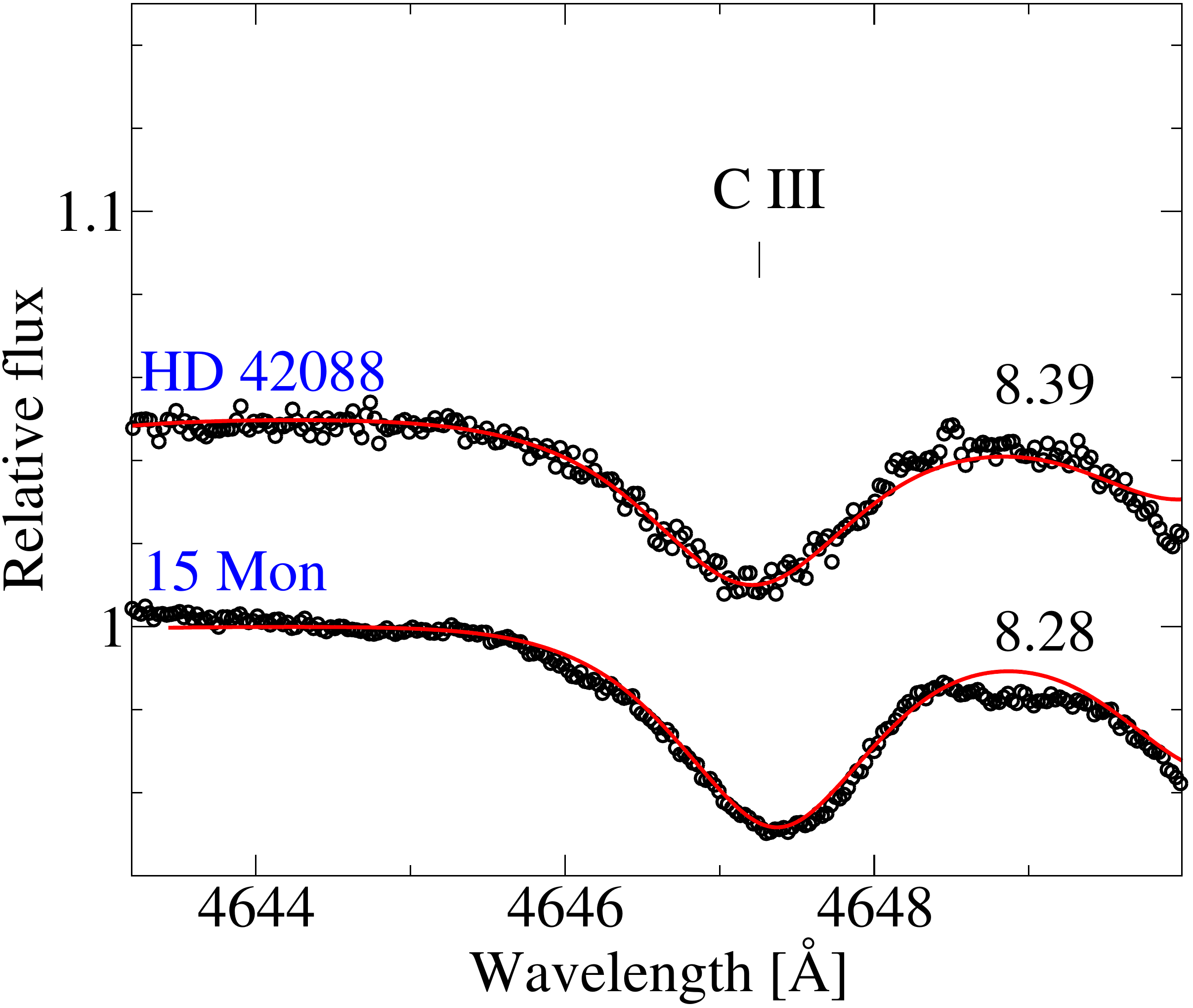}\\
 \centering}
 \parbox{0.24\linewidth}{\includegraphics[scale=0.19]{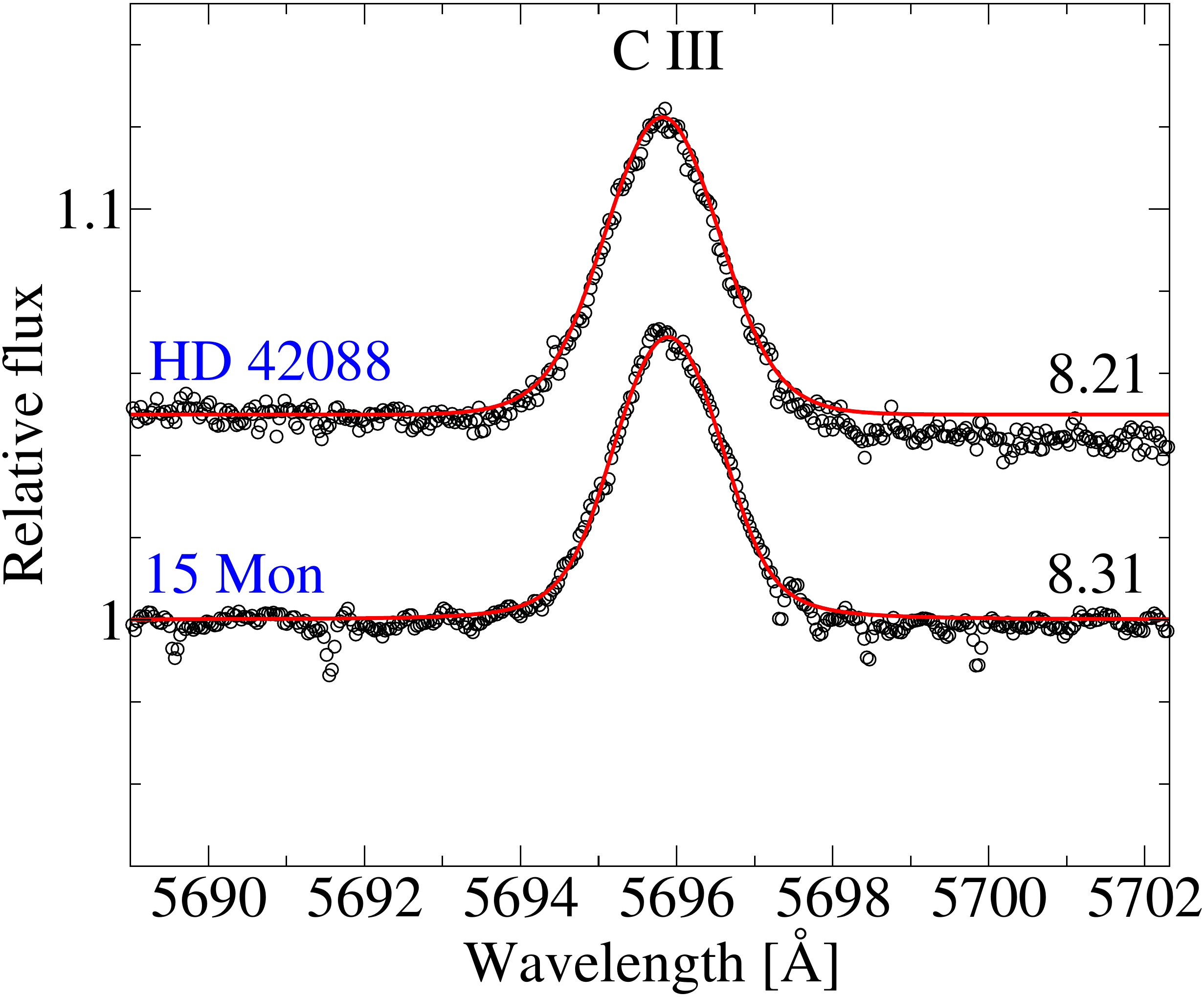}\\
 \centering}
 \parbox{0.24\linewidth}{\includegraphics[scale=0.19]{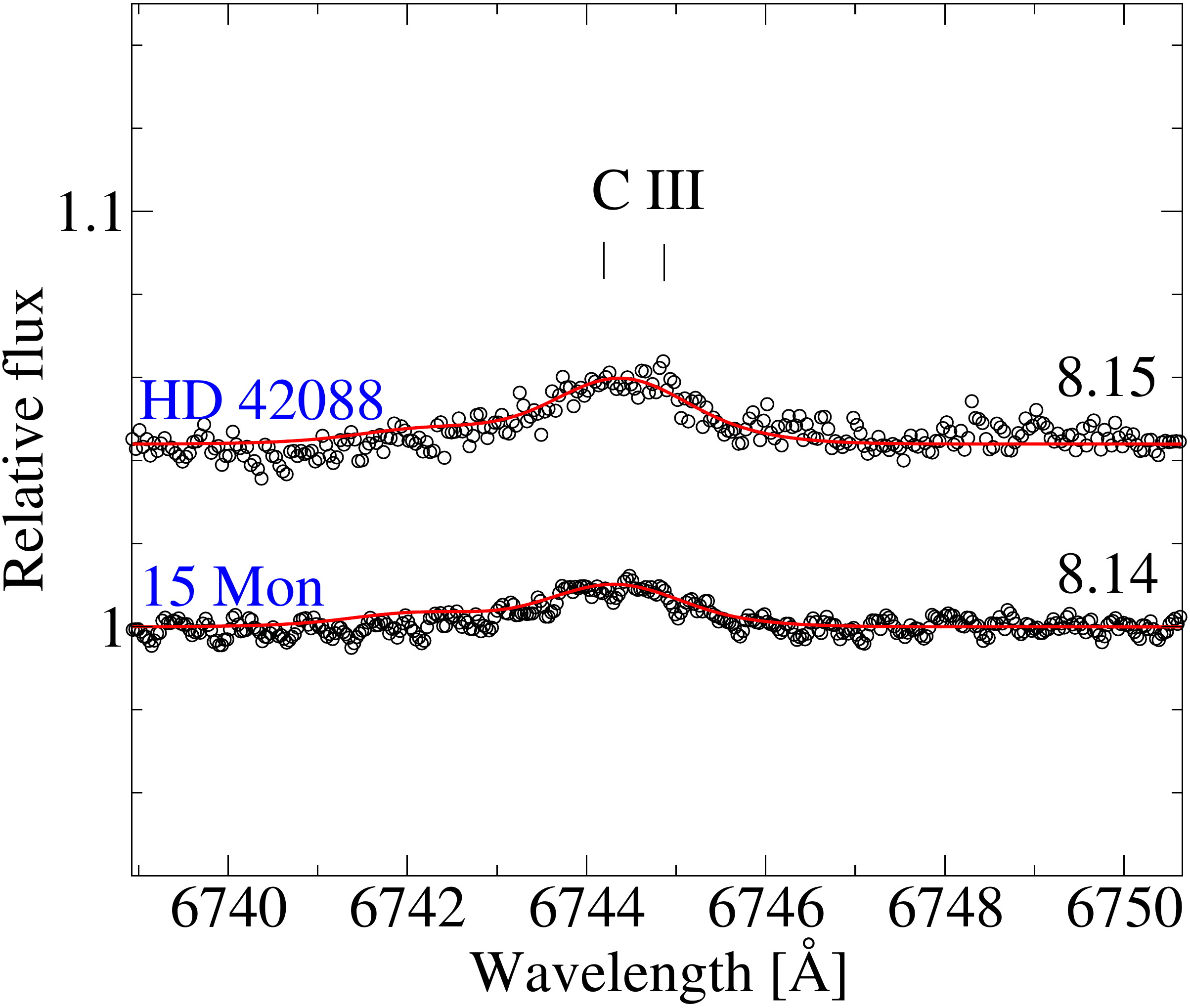}\\
 \centering}
 \parbox{0.24\linewidth}{\includegraphics[scale=0.19]{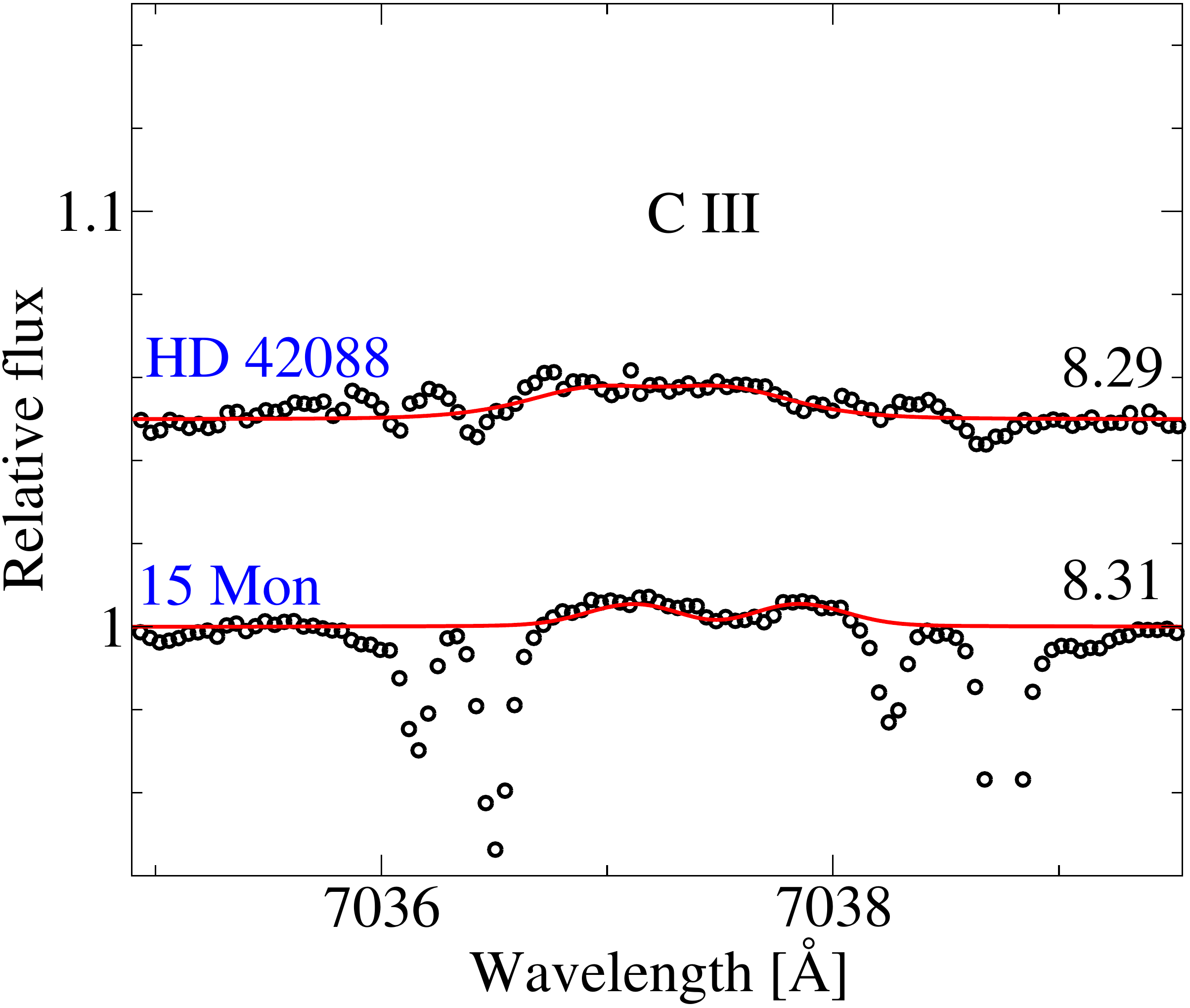}\\
 \centering}
 \hspace{1\linewidth}
 \hfill
 \\[0ex]
 \parbox{0.25\linewidth}{\includegraphics[scale=0.19]{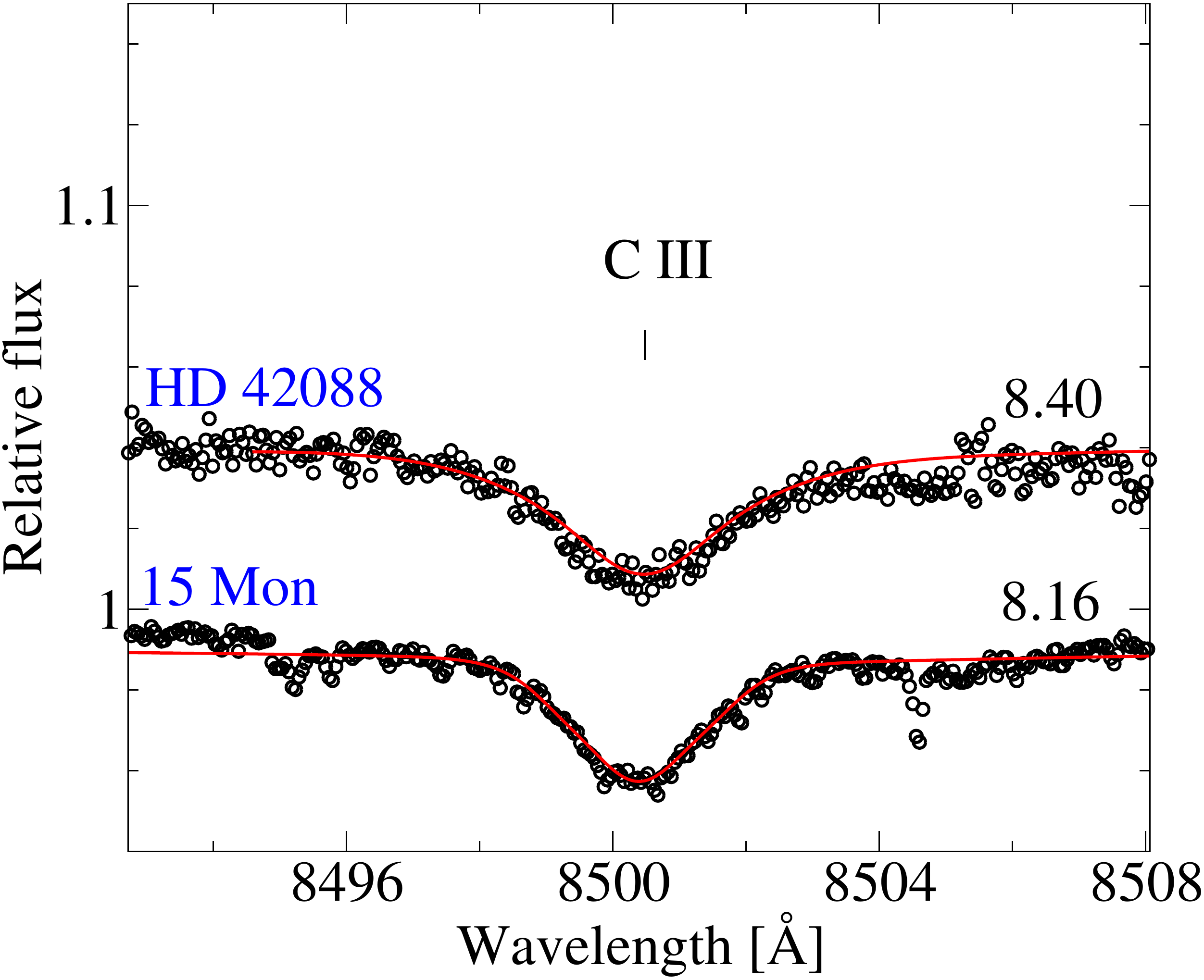}\\
 \centering}
 \parbox{0.25\linewidth}{\includegraphics[scale=0.19]{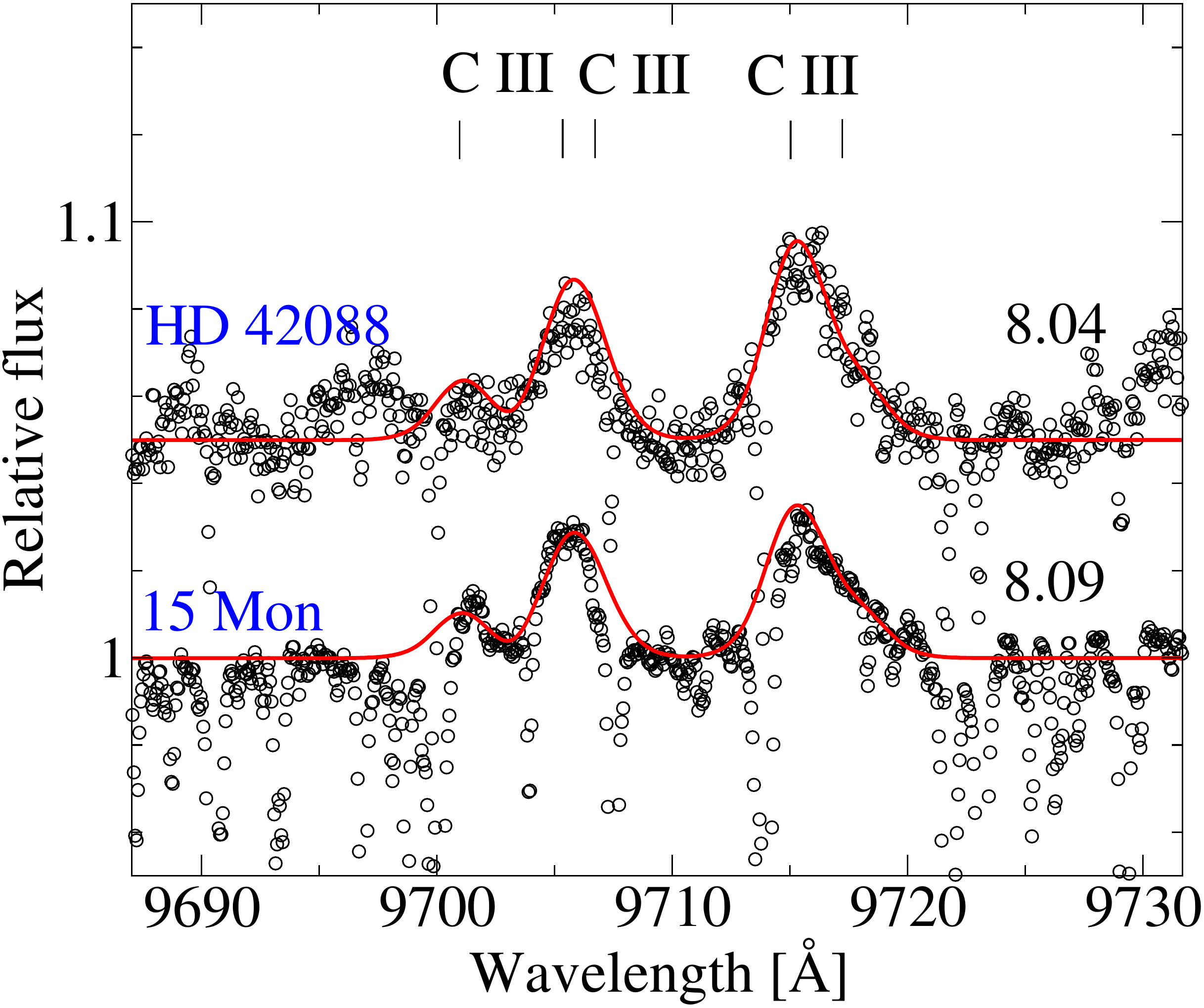}\\
 \centering}
  \parbox{0.25\linewidth}{\includegraphics[scale=0.19]{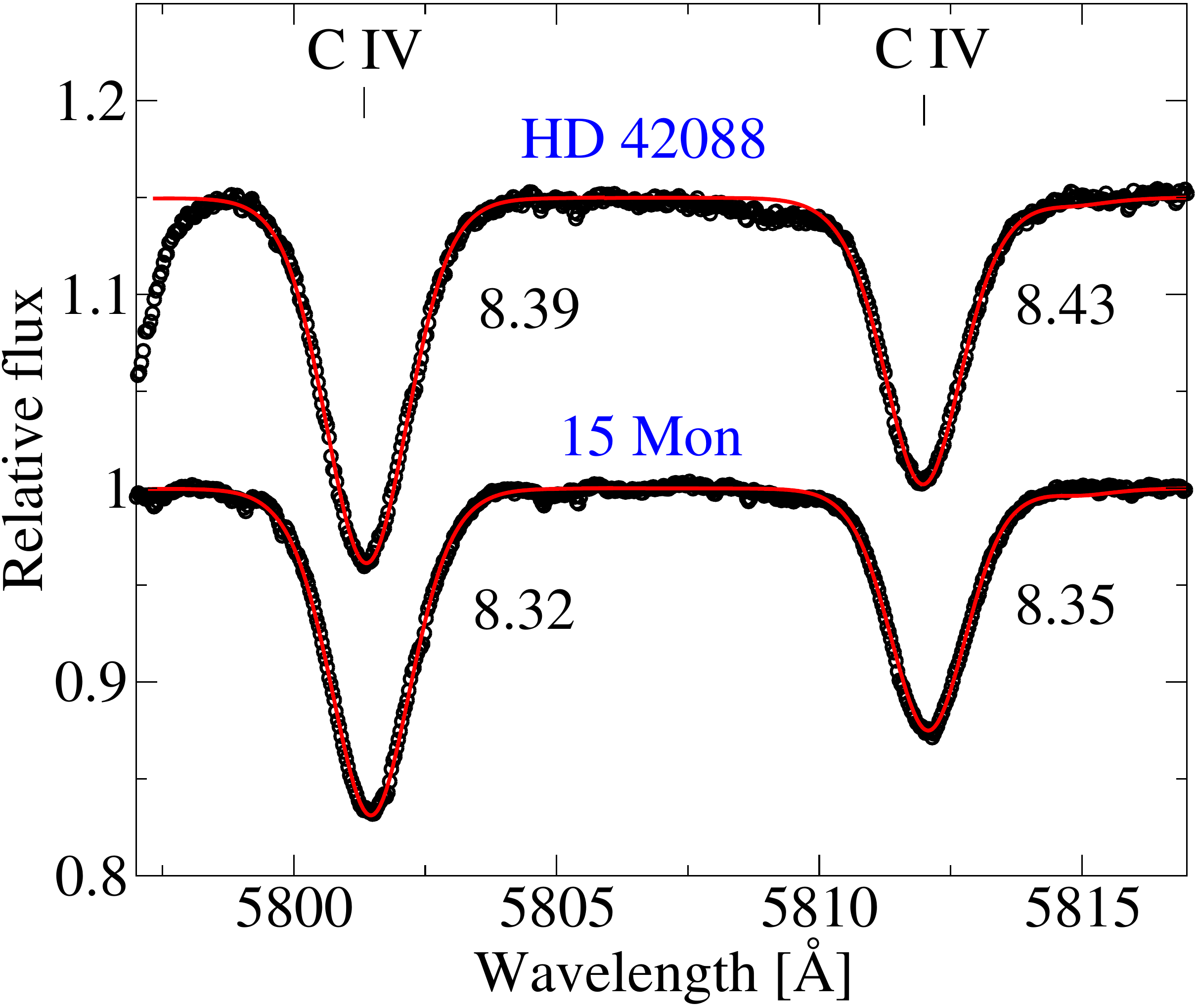}\\
 \centering}
 \hspace{1\linewidth}
 \hfill
 \\[0ex]
 \caption{Best NLTE fits (solid curves) of C\iii\ 4647, 5695, 6744, 7037, 8500, and 9701--17~\AA\ and C\iv\ 5801, 5811~\AA\ in HD~42088 and 15~Mon. The observed spectra are shown by
open circles.} 
 \label{C3profile}
 \end{center}
 \end{minipage}
 \end{figure*}

  Here, we derive the carbon abundances of two O-type stars from emission and absorption C\iii\ and absorption C\iv\ lines by spectrum fitting. 
  In Table~\ref{C3abn}, we present the abundances derived from absorption and emission lines in two O-type stars.
  Best NLTE fits of C\iii\ 5695, 6744, 8500, and 9701--17~\AA\ and C\iv\ 5801, 5811~\AA\ in HD~42088 and 15~Mon are presented on Figure~\ref{C3profile}.
  Everywhere NLTE provides smaller line-to-line scatter that is illustrated by Fig.~\ref{CII_CIII_Abn} (right panel).
  The abundances derived from emission and absorption C\iii\ are slightly lower (by $\sim$0.14~dex) compared to C\iv, although still consistent within the error bars.
  In both stars, NLTE provides consistent values within 0.14~dex abundances from the two ionization stages, C\iii\ and C\iv, while the LTE abundance
  difference can be up to 1.34~dex in the absolute value. 
  It should be noticed that emission lines of C\iii\ 9705--17~\AA\ tend to be stronger in our modeling, that leads to lower abundances from this line in both stars. 
  After excluding C\iii\ 9705--17~\AA\ lines, we obtained log~$\epsilon_{\rm C}$=8.31$\pm$0.11 in HD~42088 and log~$\epsilon_{\rm C}$=8.27$\pm$0.08 in 15~Mon from the lines of both ionization stages.

 \begin{deluxetable*}{cccccccccccccccccccccccccccc}
 \tabletypesize{\scriptsize}
 \tablecaption{NLTE and LTE Carbon abundances in 20 B-type stars. \label{tabhot}}
\def\arraystretch{0.7}
\setlength{\tabcolsep}{2pt} 
\tabletypesize{\scriptsize }
\tablewidth{20pt}
\tablehead{
\colhead{$\lambda$ (\AA\,)} & \colhead{ \tiny NLTE} & \colhead{ \tiny LTE} & \colhead{ \tiny NLTE} & \colhead{ \tiny LTE}  & \colhead{\tiny NLTE} & \colhead{ \tiny LTE} & \colhead{ \tiny NLTE} & \colhead{ \tiny LTE}  & \colhead{ \tiny NLTE} & \colhead{ \tiny LTE}  & \colhead{ \tiny NLTE} & \colhead{ \tiny LTE} & \colhead{\tiny NLTE} & \colhead{\tiny LTE} & \colhead{\tiny NLTE} & \colhead{ \tiny LTE}  & \colhead{ \tiny NLTE} & \colhead{ \tiny LTE} & \colhead{ \tiny NLTE} & \colhead{ \tiny LTE} 
}
\startdata
 \multicolumn{21}c{ }      \\
 &\multicolumn2c{18 Peg} &\multicolumn2c{$\iota$~Her} &\multicolumn2c{HR 1820} &\multicolumn2c{HIP 26000} &\multicolumn2c{ $o$ Tau} &\multicolumn2c{$\zeta$ Cas} &\multicolumn2c{$\chi$ Cen} &\multicolumn2c{$\delta$ Cet } &\multicolumn2c{$\nu$ Eri } &\multicolumn2c{$\gamma$ Peg  } \\ 
  \multicolumn{21}c{ }      \\\hline
 C\i\ & \multicolumn{20}c{ }      \\ 
 9078    &\nodata &\nodata & ~8.62 & $e$    &\nodata &\nodata&\nodata &\nodata&\nodata &\nodata& \nodata  &\nodata & \nodata&\nodata&\nodata &\nodata&\nodata &\nodata&\nodata &\nodata  \\
 9088    &\nodata &\nodata & ~8.51 & $e$    &\nodata &\nodata& ~8.40  &$e$    &\nodata &\nodata& \nodata  &\nodata & \nodata&\nodata&\nodata &\nodata&\nodata &\nodata&\nodata &\nodata  \\
 9094    &\nodata &\nodata & ~8.39 & $e$    & ~8.49  & $e$   &\nodata &\nodata&\nodata &\nodata& \nodata  &\nodata &  ~8.29  &$e$    &\nodata &\nodata&\nodata &\nodata&\nodata &\nodata  \\
 9111    &\nodata &\nodata & ~8.47 & $e$    &\nodata &\nodata&\nodata &\nodata&\nodata &\nodata& ~8.23    &$e$     &  ~8.29  &$e$    &\nodata &\nodata&\nodata &\nodata&\nodata &\nodata  \\
 9405    &\nodata &\nodata & ~8.47 & $e$    &\nodata &\nodata& ~8.30  &$e$    &\nodata &\nodata& ~8.24    &$e$     &  ~8.33  &$e$    &\nodata &\nodata&\nodata &\nodata&\nodata &\nodata  \\
 9658    &\nodata &\nodata & ~8.29 & $e$    &\nodata &\nodata&\nodata &\nodata&\nodata &\nodata& \nodata  &\nodata & \nodata&\nodata&\nodata &\nodata&\nodata &\nodata&\nodata &\nodata  \\\hline
 Mean    &\nodata &\nodata & ~8.47 &\nodata & ~8.49  &\nodata& ~8.35  &\nodata&\nodata &\nodata& ~8.24    &\nodata &  ~8.30  &\nodata&\nodata &\nodata&\nodata &\nodata&\nodata &\nodata  \\
 $\sigma$&\nodata &\nodata & ~0.11 &\nodata &\nodata &\nodata& ~0.07  &\nodata&\nodata &\nodata& ~0.01    &\nodata &  ~0.02  &\nodata&\nodata &\nodata&\nodata &\nodata&\nodata &\nodata  \\ \hline
 C\ii\ & \multicolumn{20}c{ }      \\ 
 3918    & ~8.30	& ~8.35	& ~8.33	  & ~8.50	& ~8.24	& ~8.24	& ~8.24	& ~8.24	&\nodata	&\nodata	  & ~8.30	& ~8.32	& ~8.21	& ~8.16	& ~8.11	& ~8.10	& ~8.43	& ~8.09	& ~8.26	& ~8.15   \\
 3920    & ~8.30	& ~8.24	& ~8.28	  & ~8.51	& ~8.23	& ~8.30	& ~8.24	& ~8.28	&\nodata	&\nodata	  & ~8.25	& ~8.34	& ~8.20	& ~8.23	& ~8.09	& ~8.09	&\nodata	&\nodata	& ~8.26	& ~8.21   \\
 4267    & ~8.05	& ~8.24	& ~8.23	  & ~8.42	& ~8.08	& ~8.12	& ~8.08	& ~8.02	& ~8.14	& ~8.03	& ~8.21	  & ~7.96	& ~8.09	&~7.90	& ~8.14	&~7.72	& ~8.14	&~7.63	& ~8.30 &~7.83   \\
 5132    & ~8.43	& ~8.46	& ~8.52	  & ~8.54	& ~8.42	& ~8.41	& ~8.43	& ~8.44	& ~8.34	& ~8.34	& ~8.41	  & ~8.39	& ~8.32	& ~8.30	& ~8.30	& ~8.24	& ~8.42	& ~8.35	& ~8.46	& ~8.38   \\
 5137    &\nodata	&\nodata& ~8.54	  & ~8.53	& ~8.44	& ~8.46	& ~8.40	& ~8.41	& ~8.35	& ~8.31	& ~8.48	  & ~8.44	& ~8.41	& ~8.41	& ~8.31	& ~8.25	& ~8.44	& ~8.19	& ~8.47	& ~8.39   \\
 5139    &\nodata	&\nodata& ~8.47	  & ~8.47	& ~8.44	& ~8.40	& ~8.34	& ~8.35	& ~8.35	& ~8.35	& ~8.43	  & ~8.35	& ~8.28	& ~8.23	& ~8.29	& ~8.23	& ~8.42	& ~8.22	& ~8.46	& ~8.37   \\
 5143    & ~8.48	& ~8.58	& ~8.53	  & ~8.57	& ~8.44	& ~8.45	& ~8.43	& ~8.44	& ~8.36	& ~8.39	& ~8.44	  & ~8.42	& ~8.34	& ~8.33	& ~8.37	& ~8.31	& ~8.42	& ~8.36	& ~8.47	& ~8.45   \\
 5145    & ~8.46	& ~8.49	& ~8.55	  & ~8.63	& ~8.43	& ~8.49	& ~8.49	& ~8.51	& ~8.36	& ~8.43	& ~8.49	  & ~8.45	& ~8.36	& ~8.37	& ~8.39	& ~8.33	& ~8.44	& ~8.36	& ~8.47	& ~8.47   \\
 5151    & ~8.48	& ~8.50	& ~8.53	  & ~8.56	& ~8.41	& ~8.41	& ~8.44	& ~8.46	& ~8.35	& ~8.36	& ~8.44	  & ~8.43	& ~8.35	& ~8.36	& ~8.35	& ~8.29	& ~8.45	& ~8.38	& ~8.46	& ~8.45   \\
 5648    &\nodata	&\nodata& ~8.46	  & ~8.59	& ~8.46	& ~8.55	& ~8.40	& ~8.49	& ~8.39	& ~8.48	& ~8.37	  & ~8.47	& ~8.38	& ~8.44	& ~8.26	& ~8.36	& ~8.31	& ~8.45	& ~8.35	& ~8.44   \\
 5662    &\nodata	&\nodata& ~8.43	  & ~8.52	& ~8.37	& ~8.45	& ~8.32	& ~8.41	& ~8.26	& ~8.34	& ~8.24	  & ~8.35	& ~8.21	& ~8.29	& ~8.15	& ~8.27	& ~8.25	& ~8.42	& ~8.30	& ~8.40   \\
 6151    & ~8.44	& ~8.51	& ~8.45	  & ~8.28	& ~8.32	& ~8.09	& ~8.30	& ~8.03	& ~8.25	& ~8.04	& ~8.08	  & ~7.80	& ~8.19	& ~7.98	& ~7.90$^*$	& ~7.58$^*$	& ~8.26	& ~7.80	& ~8.23	&~7.77   \\
 6461    & ~8.44	& ~8.14	&~8.74$^*$&~8.17$^*$& ~8.32	&~7.86	& ~8.30	&~7.85	& ~8.21	&~7.84	& ~7.97   & ~7.59   & ~8.34	&~7.92&\nodata&\nodata	& ~8.19	&~7.50&~7.84$^*$&~7.44$^*$   \\
 6578    & ~8.23	& ~8.66	& ~8.36	  &~9.13	& ~8.18	& ~8.94	& ~8.31	& ~8.79	& ~8.39	& ~8.81	& ~8.59	  & ~8.93	& ~8.26	& ~8.70	& ~8.62	& ~8.82	&\nodata	&\nodata	& ~8.67	& ~8.65   \\
 6582    & ~8.26	& ~8.66	& ~8.42	  &~9.12	& ~8.22	& ~8.79	& ~8.37	& ~8.73	& ~8.45	& ~8.72	& ~8.67	  & ~8.83	& ~8.33	& ~8.59	& ~8.72	& ~8.65	&\nodata	&\nodata	& ~8.77	& ~8.62   \\
 6779    &\nodata	&\nodata&\nodata  &\nodata  & ~8.40	& ~8.35	& ~8.46	& ~8.42	& ~8.41	& ~8.36	& ~8.45	  & ~8.39	& ~8.36	& ~8.29	& ~8.33	& ~8.23	& ~8.28	& ~8.30	& ~8.47	& ~8.36   \\
 6780    &\nodata	&\nodata&\nodata  &\nodata  & ~8.40	& ~8.35	& ~8.46	& ~8.44	& ~8.41	& ~8.36	& ~8.45	  & ~8.39	& ~8.36	& ~8.29	& ~8.33	& ~8.23	& ~8.28	& ~8.30	& ~8.47	& ~8.36   \\
 6783    &\nodata	&\nodata& ~8.61	  & ~8.63	& ~8.53	& ~8.47	& ~8.59	& ~8.52	& ~8.49	& ~8.44	& ~8.62	  & ~8.51	& ~8.40	& ~8.35	& ~8.47	& ~8.35	& ~8.42	& ~8.40	& ~8.59	& ~8.46   \\
 6800    &\nodata	&\nodata&\nodata  &\nodata  & ~8.40	& ~8.36	& ~8.46	& ~8.41	& ~8.41	& ~8.36	& ~8.47	  & ~8.39	& ~8.35	& ~8.31	& ~8.32	& ~8.23	& ~8.36	& ~8.38	& ~8.42	& ~8.32   \\
 9903    &\nodata	&\nodata& ~8.50   &	$e$	    & ~8.29	&$e$	& ~8.68	&$e$    & ~8.54	&$e$	& ~8.50	  &$e$      & ~8.33	&$e$    & ~8.61	&$e$	& ~8.44 &$e$    & ~8.37	&$e$   \\ \hline
 Mean    & ~8.37	& ~8.47	& ~8.46	  & ~8.65	& ~8.36	& ~8.44	& ~8.41	&~8.43	& ~8.37	& ~8.41	& ~8.42	  & ~8.44	& ~8.31	& ~8.33	& ~8.38	& ~8.34	& ~8.36	& ~8.26	& ~8.45	& ~8.36   \\
 $\sigma$&~0.14	    &~0.17	&~0.11	  &~0.24	& ~0.11	&~0.22	& ~0.13	&~0.23	&~0.10	&~0.23	& ~0.17	  & ~0.31	& ~0.08	& ~0.20	& ~0.17	& ~0.28	& ~0.10	& ~0.29	& ~0.14	&~0.30   \\ \hline
 C\iii\  & \multicolumn{20}c{ }      \\ 
 4647    & \nodata  &  \nodata &\nodata& \nodata & ~8.39   &~8.49  & ~8.39   & ~8.58    &\nodata & \nodata & ~8.23   & ~8.47	&~8.24	 &~8.39  &~8.24 &~8.47  &~8.41   &~8.78  & \nodata& \nodata \\
 4650    & \nodata  &  \nodata &\nodata& \nodata & \nodata &\nodata& ~8.49   & ~8.67    &\nodata & \nodata & ~8.30   & ~8.53	&~8.26	 &~8.42  &~8.31 &~8.55  &\nodata &\nodata&~8.17   &~8.43 \\
 4651    & \nodata  &  \nodata &\nodata& \nodata & \nodata &\nodata& \nodata & \nodata  &\nodata & \nodata & ~8.36   & ~8.53	&\nodata &\nodata&~8.34	&~8.57  &\nodata &\nodata&~8.25   &~8.44  \\
 5696    & \nodata  &  \nodata &\nodata& \nodata & \nodata &\nodata& \nodata & \nodata  &\nodata & \nodata & \nodata & \nodata  &\nodata &\nodata&\nodata&\nodata&\nodata&\nodata&~8.34   &~8.35  \\ \hline
 Mean    & \nodata  &  \nodata &\nodata& \nodata & ~8.39   &~8.49  & ~8.44   & ~8.63    &\nodata & \nodata & ~8.30   &~8.51     &~8.25	 &~8.41	&~8.30	&~8.53	&~8.41	 &~8.78  &~8.26	  &~8.41  \\
 $\sigma$& \nodata  &  \nodata &\nodata& \nodata & \nodata &\nodata& ~0.07   & ~0.06    &\nodata & \nodata & ~0.07   &~0.03     &~0.01	 &~0.02	&~0.05	&~0.05	&\nodata &\nodata&~0.09	  &~0.05  \\\hline
 \multicolumn{21}c{ }      \\
 &\multicolumn2c{$\alpha$ Pix } &\multicolumn2c{HR 1781 } &\multicolumn2c{HR 1886 } &\multicolumn2c{HR 2928}
&\multicolumn2c{HR 1861} &\multicolumn2c{$\theta^2$ Ori B} &\multicolumn2c{$\phi^1$ Ori} &\multicolumn2c{$\lambda$ Lep } &\multicolumn2c{ $\tau$ Sco } &\multicolumn2c{ $\upsilon$ Ori} \\
\multicolumn{21}c{ }      \\\hline
 C\ii\ & \multicolumn{20}c{ }      \\
 3918     &~8.36	&~8.26	 &~8.19	&~8.09	&~8.29	&~8.13	&~\nodata&\nodata&\nodata&\nodata &\nodata	&\nodata	&~8.26	    &~8.18	    &\nodata	&\nodata	&~8.25	&~8.25	&8.31	 &8.26    \\
 3920     &\nodata  &\nodata &~8.19	&~8.13	&~8.20	&~8.06	&~8.15	 &~7.93	 &\nodata&\nodata &\nodata	&\nodata	&\nodata	&\nodata	&\nodata	&\nodata	&\nodata&\nodata&\nodata &\nodata   \\
 4267     &~8.41	&~7.70	 &~8.33	&~7.72	&~8.41	&~7.64	&~8.24	 &~7.54	 &~8.27	 &~7.55	  &~8.22	&~7.58	&~8.19	 &~7.75	    &~8.22	&~7.71  &~8.15  &~7.82  &~8.14    &~8.03    \\
 5132     &~8.42	&~8.33	 &~8.49	&~8.38	&~8.54	&~8.35	&~8.33	 &~8.14	 &~8.51	 &~8.32	  &~8.51	&~8.30	&~8.34	 &~8.17	    &~8.48	&~8.31  &~8.39  &~8.24  & \nodata &\nodata    \\
 5137     &~8.43	&~8.42	 &~8.52	&~8.40	&~8.54	&~8.34	&~8.31	 &~8.14	 &~8.47	 &~8.29	  &~8.47	&~8.27	&\nodata &\nodata	&\nodata&\nodata&\nodata&\nodata&\nodata &\nodata    \\
 5139     &~8.42	&~8.34	 &~8.45	&~8.34	&~8.55	&~8.34	&~8.27	 &~8.09	 &~8.45	 &~8.27	  &~8.53	&~8.33	&\nodata &\nodata	&\nodata&\nodata&\nodata&\nodata&\nodata &\nodata    \\
 5143     &~8.44	&~8.33	 &~8.55	&~8.44	&~8.55	&~8.37	&~8.36	 &~8.19	 &~8.55	 &~8.37	  &~8.53	&~8.34	&~8.39	 &~8.16	    &~8.49	&~8.36  &~8.37  &~8.29  &\nodata &\nodata    \\
 5145     &~8.44	&~8.42	 &~8.54	&~8.43	&~8.54	&~8.40	&~8.36	 &~8.19	 &~8.53	 &~8.33	  &~8.54	&~8.34	&~8.37	 &~8.19	    &~8.49	&~8.36  &~8.41  &~8.29  &~8.25   &~8.18    \\
 5151     &~8.42	&~8.32	 &~8.47	&~8.35	&~8.54	&~8.43	&~8.29	 &~8.11	 &~8.47	 &~8.28	  &~8.52	&~8.32	&\nodata &\nodata	&\nodata&\nodata&~8.41	&~8.28	&~8.24	 &~8.18    \\
 5648     &~8.31	&~8.38	 &~8.35	&~8.42	&~8.42	&~8.40	&~8.29	 &~8.34	 &~8.33	 &~8.33	  &~8.33	&~8.37	&\nodata &\nodata	&\nodata&\nodata&~8.17	&~8.24	&\nodata &\nodata   \\
 5662     &~8.21	&~8.35	 &~8.32	&~8.36	&~8.31	&~8.32	&~8.08	 &~8.12	 &~8.21	 &~8.27	  &~8.27	&~8.34	&\nodata &\nodata	&~8.28	&~8.31	&~8.12	&~8.22	&\nodata &\nodata   \\
 6151     &~8.24	&~7.64	 &~8.19	&~7.51	&~8.54	&~7.35	&~8.28	 &~7.31	 &~8.39	 &~7.16	  &~8.29	&~$e$	&~8.22	 &$e$	    &~8.36	&$e$	&~8.24	&$e$	&~8.19	 & $e$      \\
 6461     &~8.24	&~7.23	 &~8.04	&~7.14	&~8.64	&~$e$	&~8.64	 &~$e$	 &~8.74$^*$	&~$e$ &~8.29	&~$e$	&~8.09	 &$e$	    &~8.19	&$e$	&~8.24	&$e$	&~8.19	 & $e$  \\
 6578     &~8.32	&~8.52	 &~8.62	&~8.41	&~8.54	&~8.24	&~8.32	 &~8.09	 &~8.42	&~8.19	  &~8.16	&~7.93	&~8.04	 &~7.69	    &\nodata&\nodata&~8.14	&~7.92	&~8.09	 &~7.69    \\
 6582     &~8.41	&~8.41	 &~8.71	&~8.36	&~8.63	&~8.18	&~8.39	 &~7.99	 &~8.49	&~8.16	  &\nodata  &\nodata&~8.09	 &~7.74	    &\nodata&\nodata&~8.18	&~7.96	&~8.18	 &~7.85    \\
 6779     &~8.40	&~8.40	 &~8.42	&~8.34	&~8.40	&~8.34	&~8.17	 &~8.11	 &~8.29	&~8.22	  &~8.25	&~8.23	&\nodata &\nodata	&\nodata&\nodata&\nodata&\nodata&\nodata&\nodata   \\
 6780     &~8.39	&~8.27	 &~8.37	&~8.35	&~8.40	&~8.34	&~8.17	 &~8.11	 &~8.29	&~8.22	  &~8.25	&~8.23	&\nodata &\nodata	&\nodata&\nodata&\nodata&\nodata&\nodata&\nodata   \\
 6783     &~8.54	&~8.40	 &~8.59	&~8.45	&~8.51	&~8.42	&~8.17	 &~8.15	 &~8.37	&~8.31	  &~8.31	&~8.31	&\nodata &\nodata	&\nodata&\nodata&\nodata&\nodata&\nodata&\nodata   \\
 6800     &~8.40	&~8.29	 &~8.41	&~8.34	&~8.44	&~8.32	&~8.18	 &~8.14	 &~8.28	&~8.22	  &~8.31	&~8.31	&\nodata &\nodata	&\nodata&\nodata&\nodata&\nodata&\nodata&\nodata   \\
 9903     &~8.42	&~$e$	 &~8.48	&~$e$	&~8.28	&~$e$	&~7.98	 &~$e$	 &~8.15	&~$e$	  &~8.19	&~$e$	&~8.06	 &$e$	    &~8.25	&$e$	&~8.16	&$e$	&~8.14	&$e$     \\ \hline
 Mean     &~8.39	&~8.30	 &~8.44	&~8.30	&~8.45	&~8.28	&~8.28	 &~8.09	 &~8.40	&~8.22	  &~8.37	&~8.25	&~8.22	 &~8.04	    &~8.36	&~8.26	&~8.26	&~8.18	&~8.20	&~8.07    \\ 
 $\sigma$ &~0.08	&~0.35	 &~0.17	&~0.36	&~0.12	&~0.29	&~0.14	 &~0.25	 &~0.12	&~0.33	  &~0.13	&~0.27	&~0.14	 &~0.26	    &~0.14	&~0.28	&~0.11	&~0.18	&~0.07	&~0.22   \\ \hline
 C\iii\ & \multicolumn{20}c{ }      \\                                     
 4056& \nodata	&\nodata	&\nodata	&\nodata	&~8.43	&~8.33	 &~8.15   &~8.12   &~8.35	 &~8.33	  &~8.31	&~8.31	 &~8.34	   &~8.34	&~8.43	&~8.44	&~8.32	&~8.32	&~8.34	&~8.37     \\
 4153& \nodata	&\nodata	&\nodata	&\nodata	&~8.42	&~8.36	 &\nodata &\nodata &~8.39	 &~8.37	  &~8.31	&~8.31	 &\nodata  &\nodata &\nodata&\nodata&~8.31	&~8.41	&~8.39	&~8.58     \\
 4163& \nodata	&\nodata	&\nodata	&\nodata	&\nodata&\nodata &\nodata &\nodata &\nodata	 &\nodata & \nodata	&\nodata &~8.34	   &~8.45	&\nodata&\nodata&~8.30	&~8.38	&~8.44	&~8.65  \\
 4647& ~8.20    &~8.65	    &~8.30	    &~8.61	    &\nodata&\nodata &\nodata &\nodata &\nodata	 &\nodata & \nodata	&\nodata &~8.47	   &~9.22	&~8.49	&~9.15	&\nodata&\nodata&~8.42	&~9.18   \\
 4650& \nodata	&\nodata	&~8.31	    &~8.62	    &~8.11  &~8.44   &~8.10	  &~8.46   &~8.17    &~8.63	  & \nodata	&\nodata &\nodata  &\nodata	&\nodata&\nodata&~8.17	&~8.78	&~8.44	&~9.24  \\
 4651& \nodata	&\nodata	&~8.34	    &~8.60	    &~8.19  &~8.39	 &~8.10	  &~8.32   &~8.21    &~8.55	  & \nodata	&\nodata &\nodata  &\nodata	&\nodata&\nodata&~8.30	&~8.65	&~8.46	&~9.27  \\
 4664& \nodata	&\nodata	&\nodata	&\nodata	&\nodata&\nodata &\nodata &\nodata &\nodata  &\nodata & \nodata	&\nodata &~8.27	   &~8.27	&~8.36	&~8.47	&~8.33	&~8.33	&~8.41	&~8.38  \\
 4666& \nodata	&\nodata	&\nodata	&\nodata	&\nodata&\nodata &\nodata &\nodata &~8.29    &~8.33   &~8.39	&~8.43	 &~8.39	   &~8.39	&~8.49	&~8.59	&~8.33	&~8.35	&~8.51	&~8.51  \\
 5696& \nodata	&\nodata	&\nodata	&\nodata	&~8.26	&~8.38	 &~8.22	  &~8.42   &~8.29    &~8.41   &~8.25	&~8.34	 &\nodata  &\nodata &\nodata&\nodata&~8.18	&~8.12	&~8.39	&~8.07  \\
 6731& \nodata	&\nodata	&\nodata	&\nodata	&\nodata&\nodata &\nodata &\nodata &\nodata  &\nodata &\nodata	&\nodata &\nodata  &\nodata &\nodata&\nodata&~8.47	&~8.22	&~8.49	&~8.11  \\
 6744& \nodata	&\nodata	&\nodata	&\nodata	&\nodata&\nodata &\nodata &\nodata &\nodata  &\nodata &\nodata	&\nodata &~8.44	   &~8.96	&\nodata&\nodata&~8.41	&~8.12	&~8.53	&~7.94  \\
 8500& \nodata	&\nodata	&\nodata	&\nodata	&\nodata&\nodata &\nodata &\nodata &~8.44    &~8.51   &\nodata	&\nodata &~8.34	   &~8.34	&~8.54	&~8.93	&~8.23	&~8.67	&~8.32	&~8.92  \\ \hline
 Mean& ~8.20	&~8.65	    &~8.32	    &~8.60      &~8.30	&~8.38	 &~8.15	  &~8.41   &~8.31    &~8.46   &~8.34	&~8.35	 &~8.38	   &~8.76	&~8.47 	&~8.81 	&~8.31	&~8.45	&~8.43	&~8.83  \\
 $\sigma$&\nodata&\nodata	&~0.02	    &~0.01	    &~0.14	&~0.04	 &~0.06	  &~0.14   &~0.10	 &~0.12   &~0.06	&~0.06	 &~0.07	   &~0.41	&~0.07 	&~0.31 	&~0.09	&~0.22	&~0.06	&~0.46  \\ \hline
 C\iv\ & \multicolumn{20}c{ }      \\    
 5801& \nodata	&\nodata	&\nodata	&\nodata	&\nodata&\nodata &\nodata &\nodata &\nodata  &\nodata &\nodata  &\nodata &\nodata  &\nodata  &\nodata &\nodata &~8.00 &~8.41  &~8.16  & ~8.70  \\
 5811& \nodata	&\nodata	&\nodata	&\nodata	&\nodata&\nodata &\nodata &\nodata &\nodata  &\nodata &\nodata  &\nodata &\nodata  &\nodata  &\nodata &\nodata &~8.12 &~8.49  &~8.26  & ~8.71  \\ \hline
 Mean& \nodata	&\nodata	&\nodata	&\nodata	&\nodata&\nodata &\nodata &\nodata &\nodata  &\nodata &\nodata  &\nodata &\nodata  &\nodata  &\nodata &\nodata &~8.06 &~8.45  &~8.21  & ~8.71 \\
$\sigma$&\nodata&\nodata	&\nodata	&\nodata	&\nodata&\nodata &\nodata &\nodata &\nodata  &\nodata &\nodata  &\nodata &\nodata  &\nodata  &\nodata &\nodata &~0.08 &~0.06  &~0.07  & ~0.01 \\\hline
\enddata                                                                                                                                                                  
\tablecomments{{\bf Notes.} Emission lines are marked by symbol $e$. Lines and abundances, which were not used in mean calculations are marked by (*). $\sigma$ means standard deviation.}
\end{deluxetable*}

  \begin{deluxetable*}{ccccc}
  \tablecaption{NLTE and LTE Carbon abundances in two O-type stars \label{C3abn}}
  \tablewidth{0pt}
  \tablehead{ \colhead{$\lambda$ (\AA\,)} & \colhead{ \tiny NLTE} & \colhead{ \tiny LTE} & \colhead{ \tiny NLTE} & \colhead{ \tiny LTE}  }
  \startdata
          &  \multicolumn2c{HD~42088} &\multicolumn2c{15~Mon} \\ \hline
  \decimalcolnumbers
   C\iii\ & \multicolumn{4}c{ }          \\ 
  4163              &   8.17  & 8.60     &  --  & --      \\
  4647              &   8.39  & 7.57     &  8.28    & 7.73     \\
  8500              &   8.40  & 8.63     &  8.16    & 8.63     \\ \hline
  Mean              &   8.33  & 8.46     &  8.22    & 8.38     \\
  $\sigma$          &   0.13  & 0.60     &  0.08    & 0.64    \\ \hline
  5695              &   8.21  & $e$      &  8.31    & $e$      \\
  6744              &   8.15  & $e$      &  8.14    & $e$      \\
  7037              &   8.29  & $e$      &  8.31    & $e$      \\
  9705--17          &   8.04  & $e$      &  8.09    & $e$     \\\hline
  Mean              &   8.18  & --       &  8.22    &  --      \\
  $\sigma$          &   0.11  & --       &  0.11    &  --      \\ \hline
  Mean C\iii\       &   8.25  & 8.46     &  8.22    & 8.38   \\
  $\sigma$          &   0.13  & 0.60     &  0.10    & 0.64    \\ \hline
   C\iv\            &   \multicolumn{4}c{ }                    \\ 
  5801              &   8.37  & 9.86   & 8.32   &   9.74      \\
  5811              &   8.41  & 9.72   & 8.35   &   9.56      \\ \hline
  Mean C\iv\        &   8.39  & 9.80   & 8.34   &   9.66    \\ 
  $\sigma$          &   0.03  & 0.10   & 0.02   &   0.13   \\ \hline 
  \enddata
  \tablecomments{{\bf Notes.} Emission lines are marked by symbol $e$. }
  \end{deluxetable*}

 \begin{deluxetable*}{clcc}
\tablecaption{ The Carbon NLTE abundances obtained from emission and absorption lines of four ionization stages \label{Final}}
\tablewidth{0pt}
\tablehead{  
\colhead{Number} &\colhead{Star} & \colhead{Name} & \colhead{ log~$\epsilon_{\rm C}$ }  \\
}
\decimalcolnumbers
\startdata
1  & HD~209008 & 18 Peg          & 8.37$\pm$0.14  \\                    
2  & HD~160762 & $\iota$~Her      & 8.46$\pm$0.11  \\                  
3  & HD~35912  & HR 1820         & 8.37$\pm$0.11   \\                 
4  & HD~36629  & HIP 26000       & 8.40$\pm$0.12  \\                  
5  & HD~35708  & $o$ Tau         & 8.37$\pm$0.10  \\                  
6  & HD~3360   & $\zeta$ Cas     & 8.40$\pm$0.16  \\                  
7  & HD~122980 & $\chi$ Cen      & 8.30$\pm$0.07  \\                  
8  & HD~16582  & $\delta$ Cet    & 8.37$\pm$0.16   \\                 
9  & HD~29248  & $\nu$ Eri       & 8.36$\pm$0.10   \\                 
10 & HD~886    & $\gamma$ Peg    & 8.40$\pm$0.11  \\                  
11 & HD~74575  & $\alpha$ Pix    & 8.39$\pm$0.08   \\                 
12 & HD~35299  & HR 1781         & 8.43$\pm$0.17   \\                 
13 & HD~36959  & HR 1886         & 8.46$\pm$0.14   \\                 
14 & HD~61068  & HR 2928         & 8.26$\pm$0.14    \\                
15 & HD~36591  & HR 1861         & 8.37$\pm$0.11    \\                
16 & HD~37042  & $\theta^2$ Ori B & 8.36$\pm$0.12    \\                             
17 & HD~36822  & $\phi^1$ Ori    & 8.29$\pm$0.13    \\                             
18 & HD~34816  & $\lambda$ Lep   & 8.40$\pm$0.12   \\                            
19 & HD~149438 & $\tau$ Sco      & 8.28$\pm$0.10   \\                           
20 & HD~36512  & $\upsilon$ Ori  & 8.32$\pm$0.13   \\                            
21 & HD~47839  & 15 Mon AaAb     & 8.27$\pm$0.11   \\               
22 & HD~42088  &                 & 8.31$\pm$0.11   \\\hline         
\enddata                                                           
\end{deluxetable*}

 \subsection{Discussions}\label{sect:Disc} 
 
 We examine the reliability of carbon abundances obtained from C\ii\ 4267, 9903~\AA\ and C\iii\ 5695~\AA\ lines. 
 The C\ii\ 4267, 9903~\AA\ lines reach a maximum strength around \Teff\ $\sim$21\,000 and ∼30\,000~K, respectively. 
 The C\iii\ 5695~\AA\ line reaches a maximum strength in absorption at \Teff\ $=$33\,400~K, and is still in emission at \Teff\ $=$38\,000~K.
 
\begin{figure*}                                                                                    
   \begin{minipage}{190mm}
 \begin{center}
 \parbox{0.38\linewidth}{\includegraphics[scale=0.25]{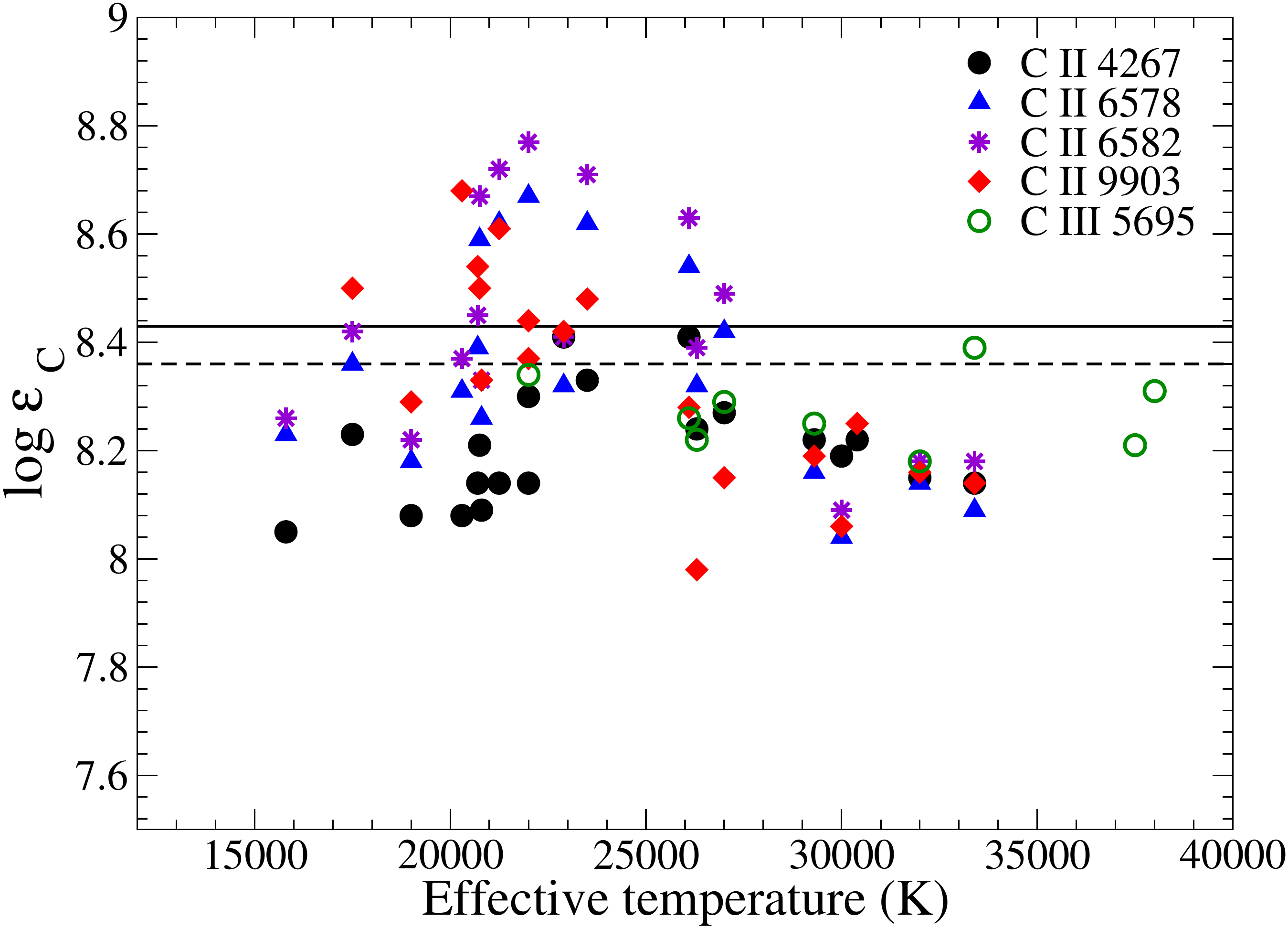}\\
 \centering}
 \hspace{1\linewidth}
 \hfill
 \\[0ex]
 \caption{Carbon NLTE abundances in program stars derived from lines of C\ii\ at 4267, 6578, 6582, and 9903~\AA\ and of C\iii\ at 5695~\AA\ in a wide temperature range.
 The solid line represents solar value, log~$\epsilon_{\rm C}$=8.43. The dashed line represents the NLTE abundance averaged over all C\ione, C\ii\ and C\iii\ lines, log~$\epsilon_{\rm C}$=8.36.  } 
 \label{Teff_Abn}
 \end{center}
 \end{minipage}
 \end{figure*}

 It is known from the observations of the CHANDRA satellite \citep{2011ApJS..194....7N}, that the X-ray fluxes in O-B stars can be higher than theoretical one, that 
 can lead to underestimation of radiation in the far UV and, as a result, to the underestimation of the degree of ionization, that can lead to lower abundances dependent on the temperature.
 Our stars have effective temperatures lying in the range 15\,800--38\,000~K. No trend between the abundances from C\ii\ 4267, 6578, 6582~\AA\ and C\iii\ 5695~\AA\ lines and effective temperature is apparent 
 (Fig.~\ref{Teff_Abn}).
 
  The obtained abundances from C\ii\ 4267~\AA\ line are systematically lower by 0.20~$-$~0.30~dex compared to other C\ii\ lines in the stars with \Teff~$\leq$~22\,000~K.
 The C\ii\ 4267~\AA\ line provides consistent abundances at the temperature range from 22\,900~K to 30\,000~K only, 
 where the agreement between C\ii\ 4267~\AA\ and the other carbon lines is within 0.14~dex. 

 Other two classical strong lines of C\ii\ 6578/82~\AA\ show too high abundances in several of the cooler sample stars, switching to too low abundance in some of 
 the higher temperature stars, that well demonstrated on Figure~\ref{Teff_Abn}. 
 However, for eight stars in our sample, $\iota$~Her, HIP~26000, $o$~Tau, $\chi$~Cen, $\alpha$~Pix, HR~1886, HR~2928, HR~1861, the discrepancies between C\ii\ 6578~\AA\ and other carbon lines do not exceed 0.10~dex. 
 We have noticed that the line C\ii\ 6578~\AA\ shows systematically lower abundance by about 0.07~dex compared to the C\ii\ 6582~\AA\ in all sample stars, however both of them belong to same multiplet. 
 
 The prominent discrepancies between C\ii\ 4267~\AA\ and C\ii\ 6582~\AA\ were found in the most of the stars, except for six of them, HR~1820, $\alpha$~Pix, $\phi^1$~Ori, $\lambda$~Lep, $\tau$~Sco  
 in which the value, log~$\epsilon_{\rm C} (6582)$ $-$ log~$\epsilon_{\rm C} (4267)$, is lower than 0.14~dex.
 For six stars, HR~1820, $\alpha$~Pix, HR~2928, $\theta^2$ Ori B, $\tau$ Sco, and $\upsilon$ Ori, the abundances from C\ii\ 4267~\AA\ and C\ii\ 6578~\AA\ are found to be consistent within 0.10~dex.

 We have found that the mean abundance from emission C\ii\ 9903~\AA\ line in the stars with \Teff $<$25\,000~K is log~$\epsilon_{\rm C}$=8.48$\pm$0.12, while in the stars 
 with \Teff$>$25\,000~K is log~$\epsilon_{\rm C}$=8.16$\pm$0.10 (Fig.~\ref{Teff_Abn}). 
  We find a small (within 3~$\sigma$) 
temperature dependence in the C abundances between hot and cool stars.  
  The most probable reason for this apparent difference is our treatment of the model atom, in which 
 R-matrix calculations for effective collision strengths (dependent on temperature)  have been carried out only for 30 lowest fine-structure levels of C\ii\ available in the literature.

\subsection{Comparison with previous studies}

  We obtained log~$\epsilon_{\rm C}$=8.36$\pm$0.08 from twenty B-type stars, that is in a good agreement with present-day Cosmic abundance standard \citep{2012AA...539A.143N}, where
  log~$\epsilon_{\rm C}$=8.33$\pm$0.04 was obtained in twenty B-stars. It is not surprising, because the sixteen stars in our sample are common with \citet{2012AA...539A.143N} and our methods of calculations are similar. However, there are some differences in the used model atom. 
  \citet{2008A&A...481..199N} performed the NLTE calculations for C\ii/C\iii/C\iv\ on the basis of the ATLAS9 model atmosphere using 
  Maxwellian-averaged collision strengths for electron-impact excitation among the lowest 24 terms of C\iii\ from the R-matrix computations of \citet{2003JPhB...36..717M}, and
  effective collision strengths for the lowest 24 fine-structure levels of C\iv\ from \citet{2004PhyS...69..385A}.
  The departure coefficients computed for the atmosphere with 32\,000 / 4.3, corresponding to $\tau$ Sco (Fig. \ref{DC} for the levels of C\ii\ and C\iii\ and Fig. \ref{DCC4} for C\iv) are similar to departure coefficients 
  in \citet{2008A&A...481..199N}. 
  We briefly comment on some individual stars as follows.
 
  {\bf HD~160762 ($\iota$~Her).} In \citet{2016MNRAS.462.1123A}, the star was found to have log~$\epsilon_{\rm C}$=8.43$\pm$0.10 from emission C\ione\ and absorption C\ii\ lines (totally 20 lines), that is 
  in agreement with this work, log~$\epsilon_{\rm C}$=8.46$\pm$0.11, where we added the abundance from emission C\ii\ 9903~\AA\ line.  
   \citet{2012AA...539A.143N} obtained log~$\epsilon_{\rm C}$=8.40$\pm$0.07 from 13 lines of C\ii, that is consistent with both \citet{2016MNRAS.462.1123A} and this work. 
  
  {\bf HD~36512 ($\upsilon$ Ori).} In this star, we obtained log~$\epsilon_{\rm C}$=8.32$\pm$0.13 from 23 lines of three ionization stages, C\ii, and C\iii, and C\iv. 
  \citet{2018A&A...615A...4C} derived a carbon abundance of log~$\epsilon_{\rm C}$=8.25$\pm$0.22 with NLTE atmosphere code \textsc{fastwind}, that is slightly lower than our result, 
  although still consistent within the error bars. Our result is consistent with \citet{2015AA...575A..34M}, who obtained log~$\epsilon_{\rm C}$=8.38$\pm$0.12 with the code \textsc{cmfgen} 
  that computes NLTE and spherical models. Our result is in line with \citet{2012AA...539A.143N}, in which they obtained 8.35$\pm$0.14 from 19 lines of three ionization stages. 

  {\bf HD~47839 (15 Mon AaAb).} 15 Monocerotis is a bright O-type star (m$_V$ = 4.64) in a spectroscopic binary system 
  that belongs to an open cluster NGC~2264 located at about $d$ $\approx$ 720 pc from the Sun \citep{2010NewA...15..302C, 2019arXiv190802040M}.
  An orbital period of 25 years was estimated by \citet{1993AJ....106.2072G}. Recent astrometric calculations based on the GAIA DR2 indicate a period of 108 years \citep{2019arXiv190802040M}. The difference  in magnitude between Aa and Ab is 1.2 mag \citep{2018A&A...615A.161M}, which indicates that Ab is likely to be a late-O or an early-B star, but there is no spatially separated spectroscopy to confirm that. 
  \citet{1993AJ....106.2072G} showed that the broad component of He\ione\ 5876~\AA\ line in the spectrum of the primary component can be potentially associated with the companion. For the secondary component, they estimated spectral type 09.5 V and quite a large projected rotational velocity $v$~sin~$i$ = 350$\pm$40~\kms\ based on the shape of the wings. The domination of the primary component to the line flux, in combination with such a large rotational velocity, complicates the detection of the secondary by the spectral observations. 
  In the case of 15 Mon AaAb, we do not notice any broad features originating from the secondary in the wings of the investigated carbon lines, even in the strong C\iii\ 5695 and C\iv\ 5801, 5811~\AA\ (Fig. \ref{C3profile}). Due to the high rotational velocity of the secondary, other weak lines would be broad and shallow and almost indistinguishable from the continuum level. Even then, abundance results can be affected by the spectroscopic binarity. When the spectrum of the primary is diluted by that of the rapidly rotating secondary, spectral lines on the normalized continuum might be apparently weakened. Then, abundances obtained by assuming a single star may contain some errors. However, we are presently unable to quantify the effect of the secondary star to the normalized continuum and to estimate the resulting errors quantitatively.
  We obtained a carbon abundance in this star and our mean value, log~$\epsilon_{\rm C}$=8.27$\pm$0.11, is lower by 0.16 dex than the solar value. 
  We find no previous study on the carbon abundance in this star.
    
  {\bf HD~42088.} It is a Blue Straggler and a member of Gem OB1 stellar association \citep{1986A&A...162..369S}. It also is considered as a peculiar star with nitrogen overabundance \citep{1986A&A...162..369S}. 
   The derived carbon abundance for HD~42088 (\Teff\ = 38\,000, log~$g$ = 4.0), log~$\epsilon_{\rm C}$=8.31$\pm$0.11, is consistent with \citet{2015AA...575A..34M}, who obtained log~$\epsilon_{\rm C}$=8.30$\pm$0.06 using the code \textsc{cmfgen} with NLTE model including wind, line-blanketing and spherical geometry. 
   Our line-by-line scatter is higher, because we used a different set of lines. 
   For example, such lines as C\iii\ 4647, 8500, 5695, 7037, 9705$-$9717~\AA, and C\iv\ 5801, 5811~\AA, were not included in analysis of \citet{2015AA...575A..34M}.

\section{Conclusions}\label{Sect:Conclusions}
   
   We present a model atom for C\ione--C\ii--C\iii--C\iv\ using the most up-to-date atomic
  data and evaluated the non-local thermodynamic equilibrium (NLTE) line formation
  in classical 1D atmospheric models of O-B-type stars with the code \textsc{detail}. 
   The model atom allows the absorption and emission lines of C\ione, C\ii, C\iii, and C\iv\ to be analyzed in the atmospheres of O-B type stars with
  effective temperature 15\,800 $\leq$ \Teff $\leq$ 38\,000~K and surface gravity 3.60 $\leq$ log~$g$ $\leq$ 4.30.

    Our modeling predicts the emission lines of C\ii\ 9903 and 18\,535~\AA\ that appear at effective temperature of \Teff $\geq$ 17\,500~K,
   those of C\ii\ 6151 and 6461~\AA\ that appear at effective temperature of \Teff $>$ 25\,000~K,
   and the emission lines of C\iii\ 5695, 6728--44, 9701--17~\AA\ that appear at effective temperature of \Teff $\geq$ 35\,000~K (log~$g$=4.0).
   These emission lines form in the stellar atmosphere, and NLTE effects are responsible for the formation of emission.
   A pre-requisite of the emission is the depletion of population in a minority species, where the photoionization-recombination mechanism provides an inversion of populations of energy levels.
   Our detailed analysis shows that the upper levels of the transitions at C\ii\ 9903 and 18\,535~\AA\ are mainly populated from C\iii\ reservoir through the Rydberg states.
  Photon losses in UV transitions at 885, 1308, and 1426--28~\AA\ that become optically thin in the photosphere drive the extra depopulation of the lower levels of transitions at C\iii\ 5695, 6728--44~\AA.
  The upper levels of transitions at C\iii\ 9701--17~\AA\ are mainly populated due to spontaneous deexcitation at C\iii\ 1577 and 1923~\AA.
   The emission lines of C\ii\ 6151, 6461, 9903, 18\,535~\AA\ and C\iii\ 5695, 6728--44, 9701--17~\AA\ in the spectra of main-sequence O-B-type stars is the result of atomic processes taking place between levels. We confirm that the emission can be obtained without the external mechanisms of excitation, the hypothesis of extended envelopes, atmospheric expansion and supplemental non-radiative influxes of energy.  
   
   We analysed the lines of C\ione, C\ii, C\iii, and C\iv\ in twenty-two O-B-type stars with temperature range 15\,800 $\leq$ \Teff $\leq$ 38\,000~K taking the advantages of their observed
   high-resolution, high SN ratio, and broad wavelength coverage spectra, where SN ratio is higher than 1000 for some of them. 
   Stellar parameters of our stars were reliably determined by \citet{2012AA...539A.143N, 2011AA...532A...2N, 2004AA...413..693M, 2015AA...575A..34M}.
   The emission line C\ii\ 9903~\AA\ was predicted theoretically and well-reproduced in the observations of B-stars for the first time. 
   
    We examine the reliability of carbon abundances determined from the absorption C\ii\ line at 4267~\AA\ and the emission line of C\ii\ at 9903~\AA\ in B-stars and the line C\iii\ at 5695~\AA\ in O-type stars. 
   There is no hint of a trend between the abundances from C\ii\ 4267 and C\iii\ 5695~\AA\ lines and effective temperature in the investigated temperature range. 
   Although, we found a hint on the trend between the abundances from emission C\ii\ 9903~\AA\ line and effective temperature, where 
   the mean abundance from C\ii\ 9903~\AA\ line in the stars with \Teff $<$25\,000~K is 8.48$\pm$0.12, while in the stars with \Teff$>$25\,000~K is 8.16$\pm$0.10. 
   Though the agreement is within 3$\sigma$, we might attribute the apparent discrepancy to be due to the shortcoming of our model atom, where in the $R$-matrix calculations 
   for effective collision strengths are not available in the literature, except for the 30 lowest fine-structure levels in C\ii.
   The C\ii\ 4267~\AA\ line provides reliable abundances at the temperature range from 22\,900~K to 30\,000~K only, while for cooler stars, with \Teff~$\leq$~22\,000~K, the abundances from C\ii\ 4267~\AA\ line are systematically lower by 0.20$-$0.30~dex compared to other C\ii\ lines.

    We conclude that the abundances from emission lines of C\ione, C\ii\ and C\iii\ do not differ significantly from absorption ones, that makes them potential abundance indicator as well. 
    Consistent abundances from emission lines of C\ione\ and absorption lines of C\ii\ have been found for $\iota$~Her, HIP~26000 and $\chi$~Cen, while for other stars, HR~1820 and $\zeta$ Cas, the agreement is still within 1$\sigma$. Consistent abundances within the error bars from C\ii\ emission and C\ii\ absorption lines were found in $\iota$~Her, HR~1820, $\zeta$~Cas, $\chi$~Cen, $\nu$~Eri, $\gamma$~Peg, $\alpha$~Pix, HR~1781, $\lambda$~Lep,
    $\tau$~Sco, and $\upsilon$~Ori. Consistent abundances within the error bars from C\iii\ emission and C\iii\ absorption lines were found in 15~Mon and HD~42088,
    however the abundance from C\iii\ emission lineas at 9705$-$17 are systematically lower by about 0.20~dex.

   For each star, NLTE leads to smaller line-to-line scatter. The NLTE must be taken into account since the deviations from LTE are strong in the investigated temperature regime.
   For example, in C\iv\ lines at 5801 and 5811~\AA\ among O-type stars, the difference between NLTE and LTE abundances, $\Delta_{\rm NLTE}$, can reach up to $-$1.49~dex. 
   In the LTE assumption, we only can recommend the C\ii\ 5132--5151~\AA\ lines as carbon abundance indicator in the stars with 15\,000 $\leq$ \Teff $\leq$ 21\,000~K, due to small NLTE effects for these lines. 
   For stars with 26\,000 $\leq$ \Teff $\leq$ 32\,000~K, the C\iii\ 4056~\AA\ line can be used in a LTE analysis. We obtained log~$\epsilon_{\rm C}$=8.36$\pm$0.08 from twenty B-type stars in the solar vicinity, which is in a good agreement with a present-day Cosmic abundance standard \citep{2012AA...539A.143N}.
   The obtained carbon abundances in 15~Mon and HD~42088 are 8.27$\pm$0.11 and 8.31$\pm$0.11, that is lower by 0.16~dex and 0.12~dex than the solar value, respectively.

\software{DETAIL \citep{detail}, SynthV\_NLTE \citep{2016MNRAS.456.1221R}, BINMAG \citep{binmag3,2018ascl.soft05015K}}.

\acknowledgments
  
   We thank the anonymous referee for valuable suggestions and comments. 
  This work was supported by the National Natural Science Foundation of China and Chinese Academy of Sciences joint fund on astronomy under grants No. U1331102, U1631105 and by the Sino-German Science Foundation under project No. GZ1183. This work is also partly supported by the Young Scholars Program of Shandong University (No.~20820162003).
  S.A. is grateful to the China Postdoctoral international exchange program (ISS-SDU) for financial support.
  This research is based on observations obtained with MegaPrime/MegaCam, a joint project of CFHT and CEA/IRFU, at the Canada-France-Hawaii Telescope (CFHT) which is operated by the National Research Council (NRC) of Canada, the Institut National des Science de l Univers of the Centre National de la Recherch Scientifique (CNRS) of France, and the University of Hawaii.
  We made use of the NORAD-Atomic-Data, NIST, SIMBAD, VALD, and UK APAP Network databases.

\bibliography{Carbon}
\bibliographystyle{aasjournal}

\end{document}